\documentclass[%
 reprint,
 amsmath,amssymb,
 aps,
]{revtex4-2}

\usepackage{graphicx}
\usepackage{dcolumn}
\usepackage{bm}
\usepackage{braket}
\usepackage{xcolor}
\definecolor{NEWSEC}{RGB}{255,137,21}
\usepackage{bbold}
\usepackage{hyperref}
\usepackage{CJK}

\begin{document}
\begin{CJK*}{UTF8}{gbsn}

\title{The robustness condition for general disordered discrete time crystals, and subspace-thermal DTCs from phase transitions between different $n$-tuple DTCs}

\author{Hongye Yu(余泓烨)}
\affiliation{C. N. Yang Institute for Theoretical Physics,
State University of New York at
Stony Brook, Stony Brook, NY 11794-3840, USA}
\affiliation{Department of Physics and Astronomy, State University of New York at
Stony Brook, Stony Brook, NY 11794-3800, USA}
\author{Tzu-Chieh Wei}
\affiliation{C. N. Yang Institute for Theoretical Physics,
State University of New York at
Stony Brook, Stony Brook, NY 11794-3840, USA}
\affiliation{Department of Physics and Astronomy, State University of New York at
Stony Brook, Stony Brook, NY 11794-3800, USA}


\begin{abstract}
We propose a new Floquet time crystal model that responds in arbitrary multiples of the driving period. Such an $n$-tuple discrete time crystal is theoretically constructed by permutations in a disordered spin chain and is well suited for experimental implementations. Transitions between these time crystals with different periods give rise to a novel phase of matter that we call subspace-thermal discrete time crystals, where states within subspaces of definite charges are fully thermalized at an early time. However, the whole system still robustly responds to the periodic driving subharmonically, with a period being the greatest common divisor of the original two periods. Existing theoretical analysis from many-body localization cannot be used to understand the rigidity of such subspace-thermal time crystal phases. To resolve this, we develop a new theoretical framework for the robustness of DTCs from the perspective of the robust $2\pi/n$ quasi-energy gap. Its robustness is rigorously proved if the system satisfies a certain condition where the mixing length, defined by the Hamming distance of the symmetry charges, does not exceed a global threshold. Although whether the condition is satisfied in generic disordered systems is unclear, the rigorous proof for DTC properties applies beyond the models considered here to other existing DTCs realized by kicking disordered MBL systems, where the condition is automatically satisfied and conventional MBL-DTCs can be regarded as a special case of the subspace-thermal DTC with the subspace dimension being one, thus offering a systematic way to construct new discrete time crystal models. We also introduce the notion of DTC-charges that allow us to probe the observables that spontaneously break the time-translation symmetry in both the regular discrete time crystals and subspace-thermal discrete time crystals. Moreover, our discrete time crystal models can be generalized to systems with higher spin magnitudes or qudits, as well as to higher spatial dimensions. 
\end{abstract}

\maketitle
\end{CJK*}

\section{Introduction}

A time crystal was originally proposed~\cite{wilczek2012quantum} by Wilczek as a phase of matter that spontaneously breaks the continuous time translation symmetry ($\tau$). Despite that breaking it in equilibrium was later proved~\cite{bruno2013impossibility,watanabe2015absence} not possible, various models~\cite{else2016floquet,khemani2016phase,von2016absolute,yao2017discrete,else2017prethermal,russomanno2017floquet,chinzei2020time,gong2018discrete,surace2019floquet,machado2020long,fan2020discrete,pizzi2021higher,nie2023mode,liu2023discrete,gargiulo2024swapping} that spontaneously break the {\it discrete} time translation symmetry have been proposed. These discrete time crystals (DTC) are realized in periodic-driven (Floquet) systems, with some observables responding with periods that are multiples of the driving period. Comprehensive reviews of ime crystals can be found in Refs.~\cite{khemani2019brief,else2020discrete,zaletel2023colloquium}.

A major feature and challenge of a DTC system is its stability of subharmonic response against perturbations. A typical Floquet system's state can be thermalized to infinite temperature~\cite{d2014long}, whereas a DTC system contains a mechanism that robustly breaks ergodicity, thus preventing thermalization. The robust ergodicity breaking can be induced in several different settings, such as many-body localization~\cite{else2016floquet,yao2017discrete,surace2019floquet,liu2023discrete,abanin2019colloquium} (MBL), Floquet prethermalization~\cite{abanin2015exponentially,mori2016rigorous,abanin2017effective,machado2019exponentially,else2017prethermal,machado2020long,fan2020discrete,luitz2020prethermalization,peng2021floquet}, and others~\cite{russomanno2017floquet,pizzi2021higher,gong2018discrete,chinzei2020time,nie2023mode}.

Recently, various DTC phases have been observed in experiments~\cite{choi2017observation,zhang2017observation,randall2021many,mi2022time,frey2022realization,kyprianidis2021observation,stasiuk2023observation}. Despite several higher $n$-tuple DTC models~\cite{pizzi2019period,surace2019floquet,pizzi2021higher} having been theoretically proposed, there are rarely experimentally realized DTC models that exhibit periods higher than doubling. In this work, we propose a new model for an arbitrary integer $n$-tuple DTC feasible for experimental realizations, based on permutations on disordered one-dimensional spin-$\frac{1}{2}$ chains. For $n=2$, our model is reduced to a recently proposed swapping Floquet DTC model~\cite{gargiulo2024swapping}. As our $n$-DTC model is based on ``kicking" a disordered spin chain, it shares similar properties to existing MBL-DTC models, such as kicked Ising models~\cite{else2016floquet,yao2017discrete}.

A remarkable property of an MBL-DTC system is the robust
$2\pi/n$ gap in its quasi-spectrum (note that we absorb the gap dependence on driving frequency into the effective Hamiltonian for simplicity). For typical MBL-DTC systems~\cite{else2016floquet,yao2017discrete} with size $L$, when local period-$T$ perturbations are added, the deviations from $2\pi/n$ are argued to be exponentially~\cite{von2016absolute} small $e^{-O(L)}$. Most proofs for this property rely on corollaries from the MBL~\cite{imbrie2016many,von2016absolute,else2016floquet,yao2017discrete,abanin2019colloquium}, such as the existence of a quasi-local unitary relating the perturbed eigenstates and the unperturbed ones. In addition, all the existing proofs are for one-dimensional systems. In this work, we offer a new proof for arbitrary $n$ and arbitrary dimensions that does not rely on the system being MBL. In our proof, the exponentially robust $2\pi/n$ gap against extensive $K$-local period-$T$ perturbations (with $K$ finite) can be rigorously proved, if the system satisfies the condition where local perturbation cannot perturb the eigenstates too globally. Although we are not able to rigorously prove the validity of the condition in generic disordered systems, we give an intuitive argument from perturbative analysis with numerical evidences that such a condition should hold for generic disordered Floquet systems, not just in MBL systems, hinting that MBL may not be a necessary condition for this type of DTC.

Indeed, during the phase transition among our $n$-DTCs, we identify a new form of DTC that is fully thermalized (in subspace) but still exhibits robust subharmonic oscillations for some observables. The thermalization saturates quickly but is restricted in subspaces of the full Hilbert space and remains there for an exponentially long time $\sim e^{O(L)}$. Such restriction is due to the symmetries of our model in the unperturbed case, but is robust against perturbations even if they break the symmetries. The subharmonic oscillations can be viewed as a thermal state restricted in one subspace, cyclically jumping to another subspace without mixing with other subspaces, which can be observed by measuring the symmetry charges that divide those subspaces. We call this new phase a \textit{subspace-thermal DTC} (ST-DTC). 

The ST-DTC is seemingly controversial to existing results in two aspects. Firstly, as shown in \cite{watanabe2015absence}, no time crystal can exist in thermal equilibrium; in ST-DTC, the system quickly reaches thermal equilibrium within each subspace. However, since the thermalization is restricted within those subspaces, it does not violate the no-go theorem proved in \cite{watanabe2015absence}. Secondly, in normal Floquet systems, thermalization can be restricted in subspaces due to the symmetry. However, once the symmetry is broken by perturbations, normal Floquet systems can absorb energies from the driving and will be heated to infinite temperature~\cite{d2014long}, and the thermalization will quickly saturate to the full Hilbert space. In ST-DTC, as the MBL is destroyed by the full-thermalization in the subspace, it seems that no other mechanism can prevent the system from being heated to infinite temperature when the symmetries are broken. In this work, we prove that such heating requires an exponentially (with respect to the system size) long time in ST-DTC even if all the symmetries are broken, due to the robust $2\pi/n$ quasi-spectral gaps. Thus, the time-crystalline order is preserved forever in the thermodynamic limit.

The ST-DTC is different from MBL-DTC: In MBL-DTC~\cite{else2016floquet,yao2017discrete}, the system retains the memory of its initial state forever, whereas, in ST-DTC, the memory is lost at the early time due to the thermalization in subspaces. The ST-DTC is also different from the prethermal-DTC: in the prethermal-DTC, the lifetime scales exponentially~\cite{else2017prethermal,machado2020long} with the driving frequency $\omega_0$ via $e^{O(\omega_0/J)}$ ($J$ is the local energy scale) instead of the system size, whereas the lifetime of ST-DTC is not directly related to $\omega_0$, but with the system size via $e^{O(L)}$ similar to MBL-DTC. In addition, the robust subharmonic response in the prethermal-DTC only holds for low-temperature initial states~\cite{else2017prethermal,machado2020long}, whereas in the ST-DTC it holds for almost every initial state. 

Our proof for the robustness of $2\pi/n$ gap can explain the existence of the ST-DTC well. In addition, the proof also works for existing MBL-DTCs, which can be viewed as the ST-DTC with subspace dimension one (and hence no subspace thermalization). Thus, it offers a systematic way to construct potential $n$-DTC models, which is much broader than MBL-DTC systems, and is not limited by the dimension or interaction range.

Moreover, as shown in previous works~\cite{surace2019floquet,pizzi2021higher},  a non-DTC phase is inevitable during the phase transition between different DTCs. In this work, we show that the transition between two $n$-DTC can be done through an ST-DTC instead of a fully thermalized phase.

The rest of the paper is organized as follows. In Sec.~\ref{sec:theorem}, we briefly outline the proof for robust $2\pi/n$ gap in DTC systems and leave details to Appendix.~\ref{app:perturbGap}. In Sec.~\ref{sec:n-DTC}, we present the $n$-tuple DTC model by permuting a disordered 1-D spin-$\frac{1}{2}$ chain. In Sec~\ref{sec:ST-DTC}, we show that the subspace-thermal DTC emerges, oscillating with period $n_G\equiv \gcd(n_1,n_2)$, during the phase transition between $n_1$-DTC and $n_2$-DTC. With perturbation, the thermalization is robustly confined in the perturbed subspaces due to the emergent symmetries, and one can observe the robust subharmonic oscillations by measuring DTC-charges. In Sec.~\ref{sec:ST-DTCmore}, we provide more evident numerical simulations for ST-DTCs with larger scales and larger perturbation strength, and show that they are qualitatively different from prethermalization results. In Sec.~\ref{sec:imp}, we briefly discuss the potential experimental realization of our models, such as in current noisy intermediate-scale quantum (NISQ) computers. We conclude in Sec.~\ref{sec:conlusion}.

\section{Sketch of proving the robust $2\pi/n$ gap of DTCs}
\label{sec:theorem}
In this section, we briefly outline the proof for the robustness of $2\pi/n$ gap of disordered DTCs against period-$T$ local perturbations, and refer the readers to Appendices~\ref{app:ProofsOfLemma12} and~\ref{app:perturbGap} for further details.

For a Floquet system driven by a periodic Hamiltonian $H_0(t+T)=H_0(t)$, the evolution of the system is governed by $U_0(t)\equiv\mathcal{T} e^{-i \int_0^t H_0(t') \mathrm{d} t'}$ (where $\mathcal{T}$ denotes time ordering), whose stroboscopic properties can be fully described by the Floquet operator $U_F^0\equiv U_0(T)$. We define the quasi-energy $\varepsilon$ of the $U_F^0$ by writing the eigenvalues of $U_F^0$ as $e^{-i \varepsilon}$. We are interested in how the quasi-spectrum of $U_F^0$ changes in the presence of perturbations.

Supposing a periodic perturbation $\lambda \hat{V}(t+T)=\lambda \hat{V}(t)$ is added, the new Floquet operator becomes
\begin{equation}
U_{F}(\lambda)=\mathcal{T} \exp \left(-i \int_0^T\left(H_0(t)+\lambda \hat{V}(t)\right) \mathrm{d} t\right).
\end{equation}
The perturbed Floquet operator $U_{F}(\lambda)$ can be alternatively~\cite{else2016floquet} written as
$U_{F}(\lambda)\equiv U_F^0 U_\lambda$,
where $U_\lambda= \mathcal{T} \exp \left(-i \int_0^T U_0^{\dagger}(t) \lambda \hat{V}(t) U_0(t)\mathrm{d} t\right)=: e^{-i\lambda V}$ and we have defined an effective Hermitian term $\lambda V$ for $U_\lambda$. We note that the new $\lambda$ in $e^{-i\lambda V}$ is different from the original one in the $\lambda \hat{V}(t)$, but they are of the similar scaling.

\subsection{Setup of the model}
\label{sec:setupT}
In this section, we discuss a class of DTC systems with solvable points, where the Floquet operator $U_F$ exactly (up to a global phase) transforms one state $\ket{\chi_{m}}$ to another globally different orthogonal state $\ket{\chi_{m+1}}$, in a cyclic way. 
A paradigmatic example is the period-2 MBL-DTC (also known as the kicked-Ising model)~\cite{else2016floquet,yao2017discrete}, where $U_F= \big(\prod X_i)e^{-i H_{\rm MBL}T }$ and $H_{\rm MBL}$ is a static MBL Hamiltonian. In the solvable point, the globally different orthogonal states connected by $U_F$ are $\ket{z}$ and $\ket{\bar{z}}$, where $\bar{z}$ is obtained by flipping all sites in the $Z$-basis state $\ket{z}$, and there are an exponential number of such pairs.

For an $n$-DTC, the evolving state returns to itself after $n$ periods of driving. One can verify that if $U_F$ acts in this way for the whole Hilbert space, all $\{\ket{\chi_{m}}\}$ form an orthogonal basis, and thus, we can divide the Hilbert space into dynamically disjoint subspaces. Within the $\alpha$-th subspace sector that has period $n$, the action of $U_F$ is
\begin{equation}
    U_F\ket{\chi_{\alpha,m}}= e^{-i \varphi_{\alpha,m+1}}\ket{\chi_{\alpha,m+1}},
\end{equation}
where $1\le m \le n$ for $n$-DTC, and $\ket{\chi_{\alpha,n+1}}\equiv \ket{\chi_{\alpha,1}}$. In such a case, the quasi-energies $\varepsilon_{\alpha,j}$ and eigenstates $\ket{\Phi_{\alpha,j}}$ can be exactly solved as follows, 
\begin{equation}
\label{eq:eStatesForNDTC}
\begin{aligned}
    \ket{\Phi_{\alpha,j}}&=\frac{1}{\sqrt{n}}\sum_{m=1}^n e^{i (\theta_{\alpha,m}+m j\frac{2\pi}{n} )} \ket{\chi_{\alpha,m}},\\
    \varepsilon_{\alpha,j}&=\frac{1}{n}\sum_{m=1}^n \varphi_{\alpha,m} + j \frac{2\pi}{n},
\end{aligned}
\end{equation}
where $\theta_{\alpha,m}=\frac{1}{n}\sum_{i=1}^n (m-i\mod n) \varphi_{\alpha,i}$ and  $U_F|\Phi\rangle=e^{-i\varepsilon}|\Phi\rangle$. Thus, the quasi-energies are separated with exact $\frac{2\pi}{n}$ spacing, and the eigenstates are all composed of an equal-weight superposition of $\ket{\chi_{\alpha,m}}$'s, with different relative phases among them for different eigenstates. When $\ket{\chi_{\alpha,m}}$'s are globally different from each other, the eigenstates $\{\ket{\Phi_{\alpha,j}}\}$ are generalized Schr\"odinger's cat states.

In addition, we require that $\ket{\chi_{\alpha,m}}$'s are eigenstates of some spatially disjoint symmetry charges $\hat{Q}_i$'s, 
\begin{equation}
    \hat{Q}_i \ket{\chi_{\alpha,m}}=q_{\alpha,m}^i \ket{\chi_{\alpha,m}},
\end{equation}
where each $\hat{Q}_i$ only acts on a finite number of sites (denoted as $i$-th spatial unit) and are spatially disjoint from each other, and $q^i$'s can only take finite numbers (e.g. $\sigma^z_i$'s in the kicked-Ising model; see Sec.~\ref{sec:ST-DTC} for another explicit construction for ST-DTC). Then we can write $\ket{\chi_{\alpha,m}}$ as tensor products of local $\hat{Q}_i$-basis states
\begin{equation}
    \ket{\chi_{\alpha,m}}=\left(\bigotimes_{i}\ket{x_i(\hat{Q_i}=q^i_{\alpha,m})}\right)\otimes\ket{R},
\end{equation}
where $\ket{x_i}$ is a local eigenstate of $\hat{Q}_i$ with eigenvalue $q^i_{\alpha,m}$ and $\ket{R}$ is the state on remainder sites not covered by $\hat{Q}_i$'s. In such a setting, we can define the distance between two states according to their $Q$-charges. Specifically, we define the {\it unit Hamming distance} between any two $Q$-basis states: $D^{Q}_u(\ket{Q_\alpha},\ket{Q_\beta})\equiv D^{Q}_u(Q_\alpha,Q_\beta)$, which counts how many spatial units $i$'s having different $Q_i$ values between the two states. We then define $l_\alpha$ to be the minimum unit Hamming distance among states in the $\alpha$-th subspace and call $\ket{\chi_{\alpha,m}}$'s globally different if $l_\alpha$ is $O(L)$.  In DTC systems, $l_\alpha$ is typically $O(L)$. In other words, in one cycle of a DTC evolution, the $Q$'s are changed globally. In the kicked-Ising model, for example, we have $l_\alpha=L$, and hence the two $|\chi\rangle$'s in  any $\alpha$-subspace are globally different. We will show shortly that the stability of DTCs depends only on the distance defined in this form.

In a similar way, we can also define the distance $D_{\alpha,\beta}^Q$ between two different sectors $\alpha,\beta$ as 
\begin{equation}
D_{\alpha,\beta}^Q\equiv\min_{m,m'}D^Q_u(\ket{\chi_{\alpha,m}},\ket{\chi_{\beta,m'}}).
\end{equation}
For DTC models discussed in this work, we also require that the unit Hamming distance defined above is invariant \footnote{This requirement is quite natural as one can verify in the kicked-Ising DTC example that the difference in $\sigma_z$'s values for two states remain the same after flipping all qubits.} under the action of $U_F$,
\begin{equation}
    D^{Q}_u(\ket{Q_\alpha},\ket{Q_\beta})= D^{Q}_u(U_F\ket{Q_\alpha},U_F\ket{Q_\beta}).
\end{equation}
Since the number of different $Q$-configurations is finite, the isometry also implies that $U_F$ acts bijectively on $Q$. We note that the models in this work and other prior works~\cite{else2016floquet,yao2017discrete,surace2019floquet} all satisfy this condition. In general, the condition can be loosened as long as $D^{Q}_u(U_F\ket{Q_\alpha}, U_F\ket{Q_\beta})$ is larger than a finite portion of $D^{Q}_u(\ket{Q_\alpha},\ket{Q_\beta})$.

Compared to the conventional definition of state distance, where the distance between $\ket{\chi_{\alpha,m}}$ and $\ket{\chi_{\alpha,m'}}$ is $l$ if $\braket{\chi_{\alpha,m}|\mathcal{V}|\chi_{\alpha,m'}}$ is nonzero only if $\mathcal{V}$ contains interactions involving at least $l$ sites, the unit Hamming distance defined in this work ignores the difference in the subspace spanned by states with the same $Q$ value. In particular, states with $D^{Q}_u=0$ can still be globally different from the conventional definition. We will show shortly that such differences do not directly contribute to the changes in the $2\pi/n$ gap.

\subsection{The LPCPG condition}
\label{sec:LPCPG-condition}

Before proceeding to the proof, we introduce a sufficient condition for a system to be a robust DTC. 

\smallskip \noindent {\bf The LPCPG Condition}. Supposing an extensive finite-body period-$T$ perturbation is added to a DTC system defined above, the perturbation can be effectively written as $e^{-i\lambda V}$ with the operator 2-norm $\|V\|_2\leq L$. The original eigenstate $\ket{\Phi_{\alpha,j}}$ is then perturbed to $\ket{\tilde{\Phi}_{\alpha,j}(\lambda)}$. Define $\mathbb{V}$ as the subspace spanned by all unperturbed eigenstates $\ket{\Phi_{\beta,j'}}$'s with $D^Q_{\alpha,\beta}<\frac{l_\alpha}{4}$,  where $l_\alpha\geq \gamma L$ and $\gamma$ is a constant. We can define the remainder norm $\epsilon_{\alpha,j}$ outside $\mathbb{V}$ as
\begin{equation}\label{eq:condition}
    \epsilon_{\alpha,j}\equiv\sqrt{1-\braket{\tilde{\Phi}_{\alpha,j}(\lambda)|P_\mathbb{V}|\tilde{\Phi}_{\alpha,j}(\lambda)}},
\end{equation}
where $P_\mathbb{V}$ is the projector to the subspace $\mathbb{V}$ and $\ket{\tilde{\Phi}_{\alpha,j}(\lambda)}$ is normalized. Then the LPCPG condition requires that for sufficiently small but non-vanishing $\lambda$, $\epsilon_{\alpha,j}$ satisfies
\begin{equation}\label{eq:Epsiloncondition}
    \epsilon_{\alpha,j}\leq e^{-\mu L},
\end{equation}
where $\mu>0$ is a scale-independent constant. 

We remark that the condition can be roughly interpreted as ``local perturbations cannot perturb \textit{too} globally" (LPCPG), where the global difference $\frac{l_\alpha}{4}$ here is measured in $Q$-values. Although we are not able to rigorously prove that the condition is satisfied in generic disordered systems, we will give an intuitive argument (with numerical evidence in Fig.~\ref{fig:e-D-STDTC}) for its validity in disordered system shortly, and its consequence on the perturbed quasi-spectrum is also confirmed by our numerical simulations in Figs.~\ref{fig:0.5gap},~\ref{fig:1gapSTDTC} and \ref{fig:gapSTDTC-1}.

Here, we offer an intuitive argument for the validity of this condition in generic disordered Floquet systems. The perturbed eigenstate $\ket{\tilde{\Phi}_{\alpha,i}(\lambda)}$ can be expanded in the original eigenbasis
\begin{equation}
    \ket{\tilde{\Phi}_{\alpha,j}(\lambda)}=\sum_{\beta,j'} a_{\alpha,j;\beta,j'} \ket{\Phi_{\beta,j'}}.
\end{equation}
From the first-order perturbation theory, for those non-resonant components $|a_{\alpha,j;\beta,j'}|\ll 1$, the coefficient is roughly 
\begin{equation}\label{eq:apropVij/dE}
    |a_{\alpha,j;\beta,j'}| \propto \left|\frac{\braket{\Phi_{\alpha,j}|e^{-i\lambda V}|\Phi_{\beta,j'}}}{e^{-i \varepsilon_{\alpha,j}}-e^{-i \varepsilon_{\beta,j'}}}\right|,
\end{equation}
where the quantity on the RHS can also be regarded as a criterion for resonance in the static MBL case~\cite{imbrie2016multi}. Due to the sufficiently strong disorder, the quasi-energies of the two states are roughly uniformly distributed in $[0,2\pi]$, so the expectation of the minimum quasi-spectral gap of $\ket{\Phi_{\alpha,j}}$'s is $\frac{\pi}{2^L}$ for qubit systems. This is also confirmed by numerical simulations (in this work and also in some previous works~\cite{von2016absolute,surace2019floquet}) for disordered Floquet systems, where the average quasi-spectral gap for disordered qubit systems is roughly $O(2^{-L})$. If we consider $V$ to be $K_u$-local (see the explicit definition of $K_u$ in Appendix.~\ref{app:perturbGap}) for simplicity, the numerator is $O(\lambda^{D_{\alpha,\beta}^Q/K_u})$. The whole term is roughly $e^{O(L \ln\lambda)}$ if $\lambda^{D_{\alpha,\beta}^Q/K_u}\ll 2^{-L}$. More precisely, for $\beta$ outside the subspace $\mathbb{V}$, we have $D_{\alpha,\beta}^Q\geq \frac{l_\alpha}{4}$ by definition. As $l_\alpha\geq \gamma L$, the numerator $\lambda^{D_{\alpha,\beta}^Q/K_u}$ is less than $e^{-\frac{\gamma \ln(1/\lambda)}{4K_u} L}$. Thus, the whole term for $\ket{\Phi_{\beta,j'}}$ outside $\mathbb{V}$ is bounded by 
\begin{equation}\label{eq:LPCGP-lambda}
\left|\frac{\braket{\Phi_{\alpha,j}|e^{-i\lambda V}|\Phi_{\beta,j'}}}{e^{-i \varepsilon_{\alpha,j}}-e^{-i \varepsilon_{\beta,j'}}}\right|\leq O(e^{-L(\frac{\gamma \ln(1/\lambda)}{4K_u}-\ln 2)}).
\end{equation}
Thus, as long as $\lambda<2^{-\frac{4K_u}{\gamma}}$, each $|a_{\alpha,j;\beta,j'}|$ can be exponentially small. Since there are less than $2^L$ eigenstates outside $\mathbb{V}$ in qubit systems, if the $\lambda$ further satisfies
\begin{equation}\label{eq:lambdaBound1}
    \lambda<2^{-\frac{8K_u}{\gamma}},
\end{equation}
the whole contribution $\epsilon_{\alpha,j}$ to the perturbed eigenstate $\ket{\tilde{\Phi}_{\alpha,j}(\lambda)}$ from original eigenstates with $D^Q_{\alpha,\beta}\geq \frac{l_\alpha}{4}$ is exponentially small, which validates the LPCPG condition.

Although disorder may not be the only choice to achieve the LPCPG condition, translation invariant models cannot satisfy this condition. In models with translational invariance, globally different states can have exactly the same quasi-energy if they can be connected by translations, which causes the term in Eq.~\eqref{eq:apropVij/dE} divergent. In this case, contributions from globally different states will not vanish, so the condition no longer holds for translation invariant models.

We emphasize that the LPCPG condition is much weaker than the usual assumption ``local perturbations perturb locally'' (LPPL)~\cite{else2016floquet},  an analog of a similar property in gapped systems~\cite{hastings2005quasiadiabatic,bravyi2010topological,de2015local}, which is usually satisfied and well accepted~\cite{von2016absolute,else2016floquet,surace2019floquet,abanin2019colloquium} in MBL-DTC systems with local spectral gaps and served as a crucial stepping stone to prove their stability. Firstly, the distance in LPPL is defined from whether two states can be coupled by spatially local interactions, while the distance in LPCPG is measured in $Q$-values, which has no geometrical restriction and ignores the differences in the same $Q$ subspace. Thus, our distance measure is weaker in this sense. Secondly, let us denote $\mathbb{V}^{(D)}_{\alpha,j}$ as the unperturbed eigenspace with $D_{\alpha,\beta}<D$ from the $\ket{\Phi_{\alpha,j}}$, and the corresponding $\epsilon$ as $\epsilon^{(D)}_{\alpha,j}$. In the LPPL case, $\epsilon^{(D)}_{\alpha,j}\sim \lambda^D$ if $D$ is greater than a finite number. In the LPCPG condition, we only require that $\epsilon^{(D)}_{\alpha,j}\sim \lambda^D$ holds for $D$ greater than a global threshold $l_\alpha/4$. This indicates that systems satisfying the LPCPG condition allow mixing of globally different $Q$-states, as long as such a mixing length $l_{\rm mix}\sim O(L)$ (measured in $D^Q_u$) is still smaller by a global distance than $l_\alpha/4$, i.e.,
\begin{equation}\label{eq:lmix-condition}
    \frac{l_\alpha}{4}-l_{\rm mix}\sim O(L).
\end{equation}
From the criterion $\left|\frac{\braket{\Phi_{\alpha,j}|e^{-i\lambda V}|\Phi_{\beta,j'}}}{e^{-i \varepsilon_{\alpha,j}}-e^{-i \varepsilon_{\beta,j'}}}\right|\sim O(1)$, we can roughly estimate $l_{\rm mix}\sim O(\lambda L)$ in generic disordered Floquet systems by regarding the quasi-spectrum as uniform distributions in $[0,2\pi]$. Here we use the prototype kicked-Ising DTC~\cite{else2016floquet, von2016absolute,yao2017discrete} as an example, where we assume that $V$ is a 1-local perturbation. In this case, we have $|\braket{\Phi_{\alpha,j}|e^{-i\lambda V}|\Phi_{\beta,j'}}|\sim \lambda^{D}$ for $D_{\alpha,\beta}=D$, and $\min_{D_{\alpha,\beta}=D}|e^{-i \varepsilon_{\alpha,j}}-e^{-i \varepsilon_{\beta,j'}}|\sim {L \choose D}^{-1}$. One can verify that $\left|\frac{\braket{\Phi_{\alpha,j}|e^{-i\lambda V}|\Phi_{\beta,j'}}}{e^{-i \varepsilon_{\alpha,j}}-e^{-i \varepsilon_{\beta,j'}}}\right|$ decays exponentially when $D>\lambda e L$. Thus, $l_{\rm mix}\lesssim \lambda e L$. We note that due to MBL in the kicked-Ising DTC, the actual $l_{\rm mix}$ can be much shorter, but the estimation procedure here works for generic disordered Floquet models that may not necessarily exhibit MBL. We remark that $l_{\rm mix}$ of different disordered Floquet models can be different by a factor, but the overall scaling is roughly $O(\lambda L)$. This estimation of the mixing length can also be applied to $Q$-basis states $\ket{\chi_{\alpha,m}}$'s, which can be obtained by the inverse transformation of Eq.~\eqref{eq:eStatesForNDTC} inside the $\alpha$-th subspace. Therefore, condition Eq.~\eqref{eq:lmix-condition} can generally be satisfied by Floquet systems of strong disorder with sufficiently small $\lambda$. In summary, our LPCPG condition is much weaker than the well-accepted LPPL condition (for the MBL-DTC), and systems satisfying the latter automatically satisfy the former. On the other hand, generic disordered Floquet systems in the absence of MBL may not always satisfy the LPPL condition, but usually satisfy the LPCPG condition at small $\lambda$. Our numerics also support that the models we will propose in Sec.~\ref{sec:n-DTC} and \ref{sec:ST-DTC} are also examples that LPPL breaks down but LPCPG condition can still hold.

Finally, we comment that the above arguments for the validity of the LPCPG condition in generic disordered systems are not a rigorous proof but rather a rough estimation. Here, we briefly discuss the weakness of the above arguments. Firstly, all the arguments above are based on first-order perturbative analysis, which may not be sufficient for many-body systems, where many subtle effects can occur such as percolation of resonances to larger-distance states in higher-order perturbation theory. Secondly, the argument for the LPCPG condition can be applied beyond the DTC models to any (static and Floquet) integrable systems with small perturbations. If the condition holds, it may indicate a generic weak violation of the eigenstate thermalization hypothesis (ETH)~\cite{deutsch1991quantum,srednicki1994chaos,rigol2008thermalization} on a long timescale, which is less evident from existing works. These issues require further investigation.

\subsection{\textcolor{black}{Exponentially small change in $2\pi/n$ gap under local perturbations}}
To show that the change of $2\pi/n$ gap under $T$-periodic local perturbations is exponentially small if the LPCPG condition is satisfied, we first introduce two useful lemmas:

\smallskip\noindent{\bf Lemma 1}.
Suppose $\ket{\psi_j}$ is an eigenvector of a unitary matrix $U$ acting in the Hilbert space $\mathbb{H}$, with the corresponding eigenvalue $e^{-i E_j}$. $P_\mathbb{V}$ is a projector that projects a state to a subspace $\mathbb{V}$ of $\mathbb{H}$, and we denote $\epsilon=\sqrt{1-\braket{\psi_j|P_\mathbb{V}|\psi_j}}$ and $\ket{\eta_j}\equiv P_\mathbb{V}\ket{\psi_j}$. Then we have
\begin{equation}
e^{-iE_j}\ket{\eta_j}=P_\mathbb{V}UP_\mathbb{V}\ket{\eta_j}+\ket{\xi},
\end{equation}
where the 2-norm of $\ket{\xi}$ satisfies $\|\ket{\xi}\|_2\leq \epsilon$.

\smallskip\noindent{\bf Lemma 2}.
Let $U$ be a unitary matrix acting in the Hilbert space $\mathbb{H}$, $\mathbb{V}$ be a subspace of $\mathbb{H}$ with corresponding projector $P_\mathbb{V}$, and $\ket{\eta}$ be a nonzero vector that is zero outside the subspace $\mathbb{V}$. Suppose we have
\begin{equation}
e^{-iE}\ket{\eta}=P_\mathbb{V}UP_\mathbb{V}\ket{\eta}+\ket{\xi}
\end{equation}
where the 2-norm $\|\ket{\xi}\|_2= \epsilon\ll\|\eta\|_2$ is small. Then $U$ has an eigenvalue $e^{-iE_j}$ satisfying $|e^{-iE_j}-e^{-iE}|\leq \sqrt{\frac{2\epsilon}{\|\ket{\eta}\|_2}}$.

We show the proofs of these two lemmas in Appendix.~\ref{app:ProofsOfLemma12}. Intuitively, Lemma 1 states that if an eigenstate $\ket{\psi_j}$ of a unitary $U$ resides almost within a subspace $\mathbb{V}$, then $\ket{\psi_j}$ and $e^{-iE_j}$ is approximately the eigenstate and eigenvalue of the diagonal block $\mathbb{V}\times\mathbb{V}$ of $U$.  Lemma 2 states that if one can find an approximate eigenstate $\ket{\eta}$ with an approximate eigenvalue $e^{-iE}$ of a diagonal submatrix of a unitary $U$, then one can always find an exact eigenvalue of $U$ close to $e^{-iE}$. Therefore, we can reduce the task of diagonalizing $U$ to that in subspace $\mathbb{V}$ with a bounded error, as long as the leakage of the perturbed eigenstate from $\mathbb{V}$ is sufficiently small. For $U_F(\lambda)$ of DTC systems, when we consider the perturbed eigenstates $\ket{\tilde{\Phi}_{\alpha,j}(\lambda)}$'s we set $\mathbb{V}$ to be the unperturbed eigenspace spanned by states $\ket{\Phi_{\beta,j'}}$ satisfying $D_{\alpha,\beta}^Q<\frac{l_\alpha}{4}$, the same as in the LPCPG condition.

From now on, we write all the matrices in the unperturbed eigenbasis. When the perturbation $e^{-i\lambda V}$ is added, we assume that $V$ is $K_u$-local (see the explicit definition of $K_u$ in Appendix.~\ref{app:perturbGap}) with its operator 2-norm no more than an extensive number $\|V\|_2\leq L$. It is important to note that the submatrix $A\equiv P_\mathbb{V}U_F(\lambda)P_\mathbb{V}$ is closely related to a unitarily transformed submatrix $\tilde{A}_c\equiv P_\mathbb{V}\tilde{U}_F(\lambda,c)P_\mathbb{V}$, where $c$ is an integer $0\le c <n$ and $\tilde{U}_F(\lambda,c)\equiv U_c^\dagger\Pi_c^\dagger U_F(\lambda) \Pi_c U_c$, with $\Pi_c$ being a permutation matrix that permutes $\ket{\Phi_{\beta,m}}$ to $\ket{\Phi_{\beta,m+c \pmod n}}$, and $U_c$ a diagonal unitary matrix. Specifically, if we denote 
\begin{equation}
    \Delta A_c\equiv A-e^{ic \frac{2\pi}{n}}\tilde{A}_c,
\end{equation}
where the $e^{ic \frac{2\pi}{n}}$ comes from $\Pi_c^\dagger U_F(0) \Pi_c=e^{-ic \frac{2\pi}{n}}U_F(0)$, then we can show that the operator 2-norm $\|\Delta A_c\|_2$ is exponentially small in the system size $L$ for sufficiently small but non-vanishing $\lambda$. In Appendix.~\ref{app:perturbGap}, we give the explicit definition of $\tilde{A}_c$ and prove that the $\|\Delta A_c\|_2$ is bounded by
\begin{equation}
    \|\Delta A_c\|_2\leq 2N_{\mathbb{V}} R_\lambda\Big(\Big\lceil \frac{l_\alpha}{2K_u} \Big\rceil\Big),
\end{equation}
where $N_{\mathbb{V}}$ is the size of the subspace $\mathbb{V}$, $R_\lambda(k)$ is defined by $R_\lambda(k)\equiv\frac{(\lambda L)^k}{k!}$, and $\lceil\cdot\rceil$ is the ceiling function. For qubit systems, we have $N_{\mathbb{V}}<2^L$, so that
\begin{equation}
    \|\Delta A_c\|_2<2e^{-L(\frac{\gamma}{2K_u}\ln\frac{\gamma}{2eK_u\lambda}-\ln 2)}.
\end{equation} 
One can verify that the norm decays exponentially $\|\Delta A_c\|_2\leq e^{-\nu L}$ as long as 
\begin{equation}\label{eq:lambdaBound}
\lambda<\frac{\gamma}{2e K_u }2^{-\frac{2 K_u }{\gamma}},
\end{equation}
where $e$ is the base of the natural logarithm and $\gamma \leq \frac{l_\alpha}{L}$ is the constant defined in the LPCPG condition. We note that to obtain a robust $2\pi/n$ gap, this bound should be combined with the $\lambda$ bound Eq.~\eqref{eq:lambdaBound1} in the rough estimation of the LPCPG condition, and use whichever is tighter.

With these preparations, the rest of the proof is then straightforward. Let $\ket{\tilde{\Phi}_{\alpha,j}(\lambda)}$ be a perturbed eigenstate of $U_F(\lambda)$ satisfying the LPCPG condition with eigenvalue $e^{-iE_{\alpha,j}(\lambda)}$. From Lemma 1, we know that they are also an approximate eigensolution of the submatrix $A$. As $e^{ic \frac{2\pi}{n}}\tilde{A}_c$ is close to $A$, $\ket{\tilde{\Phi}_{\alpha,j}(\lambda)}$ is also an approximate eigenstate of $\tilde{A}_c$, with the corresponding approximate eigenvalue $e^{-i(E_{\alpha,j}(\lambda)+c\frac{2\pi}{n})}$. As $\tilde{A}_c$ is a diagonal submatrix of $\tilde{U}_F(\lambda,c)$ whose eigenvalues are identical with those of $U_F(\lambda)$, from Lemma 2 we can find an exact eigenvalue $e^{-iE_{\alpha,j+c}(\lambda)}$ of $U_F(\lambda)$ that is close to $e^{-i(E_{\alpha,j}(\lambda)+c\frac{2\pi}{n})}$. Then we can bound the change of the $2\pi/n$ gap by
\begin{equation}\label{eq:DeltaBound}
    |\Delta^{(n,c)}_{\alpha,j}|<\frac{\pi}{2} \sqrt{2\left(\|\Delta A_c\|_2+\frac{\epsilon_{\alpha,j}}{\sqrt{1-(\epsilon_{\alpha,j})^2}}\right)},
\end{equation}
where $\Delta^{(n,c)}_{\alpha,j}\equiv|E_{\alpha,j}(\lambda)+c\frac{2\pi}{n}-E_{\alpha,j+c}(\lambda)|$. As we already proved that $\|\Delta A_c\|_2$ is exponentially small, if the state further satisfies the LPCPG condition, then $\epsilon_{\alpha,j}$ is also exponentially small. Thus, the total change of the $2\pi/n$ gap remains exponentially small. We leave the rigorous proofs and details of the above procedure in Appendix~\ref{app:perturbGap}. In summary, we arrive at the following theorem:

\smallskip\noindent{\bf Theorem 1}. Suppose an extensive $K_u$-local period-$T$ perturbation $e^{-i\lambda V}$ with $\|V\|_2\leq L$ is added to the DTC systems defined in Sec.~\ref{sec:setupT}. If the perturbed eigenstate satisfies the LPCPG condition \eqref{eq:Epsiloncondition} and the perturbation strength $\lambda$ is bounded by Eq.~\eqref{eq:lambdaBound}, then the change of the $2\pi/n$ gap induced by the perturbation is exponentially small, with the bound explicitly given by Eq.~\eqref{eq:DeltaBound}.

We note that the whole proof is non-perturbative and ignores the differences in states with the same $Q$ values. Therefore, even when the system is thermalized within the same $Q$ subspaces, the DTC properties are still exponentially robust, which justifies the existence of ST-DTC. Another evidence supporting this is from considering perturbations satisfying $[V,Q_i]=0$, where one can verify that the $2\pi/n$ gap is exactly preserved.

Moreover, the proof also works for many existing models (see the discussions in Appendix.~\ref{app:KIsing}), and offers a systematic way to construct new DTC models. To do this, one can first consider a solvable model where the Floquet operator $U_F$ exactly transforms one state $\ket{\chi_{m}}$ to another globally different orthogonal state $\ket{\chi_{m+1}}$ in a cyclical way with period $n>1$, where each state can be labeled by symmetry charges $Q$ satisfying the properties in Sec.~\ref{sec:setupT}. Then one can add disorders that do not mix $Q_i$'s to break quasispectral degeneracies. Note that such disorders may not always lead to MBL, but are usually sufficient to satisfy the LPCPG condition. According to our proof, the $2\pi/n$ gap for such models is exponentially robust against local perturbations, leading to a DTC phase. In the following, we provide concrete model constructions of $n$-DTC.

\begin{figure}[t]
\begin{tabular}{l}
(a) \\
\includegraphics[width=0.44\textwidth]{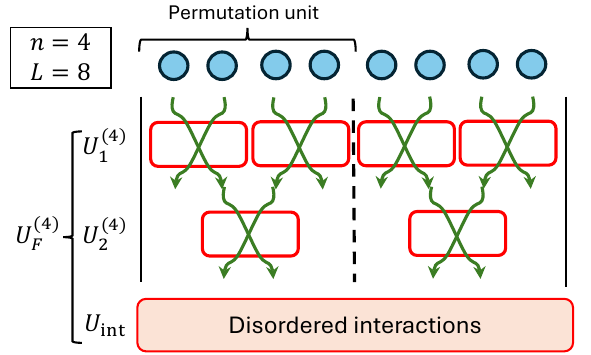}\\
(b) \\
\includegraphics[width=0.428\textwidth]{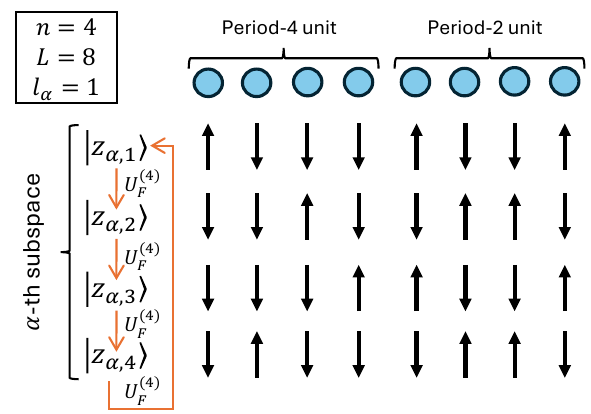}
\end{tabular}
    \caption{Illustration for the unperturbed $n$-DTC model with $n=4$ from the perspective of (a) permutation on spins and (b) transformation on $Z$-basis states. The red blocks and green arrows in (a) represent the swap gates and their actions. Each $n$-spin block forms a permutation unit. As $Z$-basis states may oscillate at a lower period in some units, we use $l_\alpha$ to count the number of period-$n$ units in $\alpha$-th subspace. }
    \label{fig:nDTC}
\end{figure}

\section{$n$-tuple DTC model from permutation}
\label{sec:n-DTC}
In this section, we define the period-$nT$ discrete time crystal on a 1-D length-$L$ spin-$\frac{1}{2}$ chain with open boundary condition~\footnote{different boundary conditions will not bring qualitative differences in this model}, whose evolution is governed by a period-$T$ Hamiltonian $H(t+T)=H(t)$. Within one period ($T=t_1+t_2+t_3$), the Hamiltonian is (see Fig.~\ref{fig:nDTC})
\begin{gather}
    H^{(n)}(t)= \begin{cases}
    H_{1}^{(n)}, & \text { for } 0 \leq t<t_1 ,\\ 
    H_{2}^{(n)}, & \text { for } t_1 \leq t<t_1+t_2,\\
    H_{\rm int}, & \text { for } t_1+t_2 \leq t<t_1+t_2+t_3,
    \end{cases}
\end{gather}
where $H_{1}^{(n)}$ and $H_{2}^{(n)}$ both consist of swap gates acting as permutations and the interaction Hamiltonian $H_{\rm int}$ includes disordered couplings and fields 
\begin{gather}
 H_{\rm int}=\sum_{i<j} J_{ij} \sigma_i^z \sigma_{j}^z+\sum_{i=1}^L h_i^z \sigma_i^z+\sum_{i=1}^L \epsilon_i^x \sigma_i^x,
\end{gather}
where the parameters $h_i^z\in [0,2\bar{h^z}]$ are randomly chosen from uniform distributions, and $\epsilon_i^x\in [-\epsilon^x,\epsilon^x]$'s are small uniformly random perturbations. We choose the coupling $J_{ij}$ to be a disordered power-law interaction obtained by
\begin{equation}
J_{ij}=\frac{1}{\mathcal{L}_{L, \kappa}} \frac{\tilde{J}_{i, j}}{|i-j|^\kappa},
\end{equation}
where $\tilde{J}_{ij}\in [\frac{1}{2}\bar{J},\frac{3}{2}\bar{J}]$ is uniformly and randomly chosen, and $\mathcal{L}_{L, \kappa}$ is a coefficient to make the energy extensive~\cite{kac1963van}
\begin{equation}
\mathcal{L}_{L, \kappa}= \begin{cases}1 & \text { if } \kappa \geqslant 1, \\ \ln L & \text { if } \kappa=1, \\ L^{1-\kappa} & \text { if } \kappa<1.
\end{cases}
\end{equation}
We remark that power-law interaction is not the only choice~\footnote{For example, one can randomly choose $J_{ij}$ uniformly for $|i-j|\leq 2n$, and set $J_{ij}=0$ for $|i-j| > 2n$, where $n$ is the length of the spatial permutation units.} for our model being DTC; however, nearest-neighbor interactions are not sufficient~\cite{gargiulo2024swapping} to break all the degeneracies in the quasi-spectrum. For the even period $n\ge 2$, the $H_{1}^{(n)}$ and $H_{2}^{(n)}$ are
\begin{eqnarray}
    H_{1}^{(n)}&=&\frac{\pi}{2 t_1}(1-\epsilon_1)\sum_{i=1}^{L/2-1} \frac{1}{2}(\hat{\bm{\sigma}}_{2i-1}\cdot \hat{\bm{\sigma}}_{2i}-1),\\\nonumber
    H_{2}^{(n)}&=&\frac{\pi}{2 t_2}(1-\epsilon_2)\left( \sum_{i=1}^{L/2-1} \frac{1}{2} (\hat{\bm{\sigma}}_{2i}\cdot \hat{\bm{\sigma}}_{2i+1}-1)\right.\\
    &&+\left.\sum_{i=1}^{\frac{L}{n}-1} \frac{1}{2}(1-\hat{\bm{\sigma}}_{ni}\cdot \hat{\bm{\sigma}}_{ni+1}) \right),
\end{eqnarray}
where $\hat{\bm{\sigma}}_{i}\cdot \hat{\bm{\sigma}}_{j}=\sigma_{i}^x \sigma_{j}^x+\sigma_{i}^y \sigma_{j}^y+\sigma_{i}^z \sigma_{j}^z$ is the Heisenberg type of interaction. For the odd period $n\ge 3$, the $H_{1}^{(n)}$ and $H_{2}^{(n)}$ are
\begin{widetext}
\begin{eqnarray}
    H_{1}^{(n)}&=&\frac{\pi}{2 t_1}(1-\epsilon_1)\left( \sum_{i=1}^{L/2-1} \frac{1}{2}(\hat{\bm{\sigma}}_{2i-1}\cdot \hat{\bm{\sigma}}_{2i}-1)+\sum_{i=1}^{L/(2n)-1}\frac{1}{2}(1-\hat{\bm{\sigma}}_{(2i-1)n}\cdot \hat{\bm{\sigma}}_{(2i-1)n+1}) \right),\\
    H_{2}^{(n)}&=&\frac{\pi}{2 t_2}(1-\epsilon_2)\left( \sum_{i=1}^{L/2-1} \frac{1}{2}( \hat{\bm{\sigma}}_{2i}\cdot \hat{\bm{\sigma}}_{2i+1}-1)+\sum_{i=1}^{L/(2n)-1} \frac{1}{2}(1- \hat{\bm{\sigma}}_{2ni}\cdot \hat{\bm{\sigma}}_{2ni+1}) \right),
\end{eqnarray}
\end{widetext}
where $\epsilon_1$ and $\epsilon_2$ are close to 0, implying small imperfections. For convenience, in all the following discussions, we assume that the system length $L$ is divisible~\footnote{This assumption is for convenience and is not essential for DTC properties. If $L$ is not divisible by $n$ or $2n$, one can always set $H_1^{(n)}$ and $H_2^{(n)}$ to be zero at the remainder sites.} by $n$ for even $L$ or $2n$  for odd $L$, respectively.  The Floquet evolution operator for a single period is then $U_F^{(n)}(\hat{\bm{\epsilon}})=U_{\rm int}U_2^{(n)}U_1^{(n)}\equiv e^{-i H_{\rm int} t_3}e^{-i H_{2}^{(n)}t_2}e^{-i H_{1}^{(n)}t_1}$.

To grasp the $nT$ periods of such evolution, it is helpful to look into the solvable, unperturbed case ($\epsilon_i^x=\epsilon_1=\epsilon_2=0$). We define the unperturbed Floquet operator as $U_F^{(n)}\equiv U_F^{(n)}(\hat{\bm{\epsilon}}=0)$ for simplicity. Firstly, we notice that the unitaries $U_1^{(n)}, U_2^{(n)}$ are groups of swap gate, as $\mathrm{SWAP_{i,j}}\equiv \frac{1}{2}(I+\hat{\bm{\sigma}}_{i}\cdot \hat{\bm{\sigma}}_{j})=e^{-i \frac{\pi}{4} (\hat{\bm{\sigma}}_{i}\cdot \hat{\bm{\sigma}}_{j}-1) }$. Thus, $U_1^{(n)}$ and $U_2^{(n)}$ effectively perform swap gates on odd and even bonds, respectively (see Fig.~\ref{fig:nDTC}). The gate structure is periodic in space, with each smallest spatial permutation unit consisting of $n$ spins~\footnote{For odd period $n$, the minimum permutation unit consisting of $2n$ spins, but the left half and the right half are symmetrical (see Fig.~\ref{fig:3-DTC}) and do not have qualitative differences in analytical analysis.}. Within each unit, it can be proved (see Appendix~\ref{app:n-period}) that such spin permutation has a period of $n$. In addition, the $H_{\rm int}$ only adds a phase for $Z$-basis states. Thus, the whole system follows an $nT$-period evolution for $Z$-basis initial states. One may notice that some $Z$-basis states can oscillate with a smaller period $k<n$, where $k$ divides $n$, i.e., $k\vert n$. Take $n=L=4$, for example. While all of the states satisfy $(U_F^{(4)})^4 \ket{\psi^{(4)}}=\ket{\psi^{(4)}}$, some of the states (e.g., $\ket{\uparrow\downarrow\downarrow\uparrow}$) also satisfy $(U_F^{(4)})^2\ket{\psi^{(2)}}=\ket{\psi^{(2)}}$ and the ferromagnetic states (e.g., $\ket{\uparrow\uparrow\uparrow\uparrow}$) satisfy $U_F^{(4)}\ket{\psi^{(1)}}=\ket{\psi^{(1)}}$, where we label the $k$-period state as $\ket{\psi^{(k)}}$. We prove that such lower-period states are exponentially rare, and most of them also have robust $k$-period subharmonic oscillation and will not be mixed with $n$-period states even if perturbations are added (see discussions in Appendices~\ref{app:n-period} and~\ref{app:perturbGap}). Thus, we will mostly consider $n$-period states in the rest of the discussion in the main text.

When local perturbations are added, we expect that the subharmonic response will be preserved. However, different initial states may have different robustness. As shown in Sec.~\ref{sec:theorem}, the robustness is determined by the minimum distance among states in the same subspace. If an $n$-period state has a $k$-period unit (with $k<n$), then this unit does not contribute to the distance. Thus, the robustness only relies on the number of $n$-period units for states in $\alpha$-th subspace, which is denoted by $l_\alpha$ in our model. Note that the $l_\alpha$ here may differ from the definition for general DTC models in Sec.~\ref{sec:theorem} by a factor, but will not change the overall scaling.  In Appendix~\ref{app:l=O(L)}, we prove that for the vast majority of the Hilbert space, $l_\alpha$ is $O(L/n)$. Thus, for most states, the subharmonic oscillation will maintain up to an $e^{O(L/n)}$ time scale. In the thermodynamic limit, the time scale approaches infinity.

In the following discussions on numerics, we set, for convenience,  $\bar{J}=4,~\bar{h^z}=12,~\kappa=0.5$, and fix the evolution time for each Hamiltonian $t_1=t_2=1/2$ and $t_3=1$. For simplicity, we set all $\epsilon$ proportional to a parameter $\lambda$, controlling the overall perturbation strength, where $\epsilon^x=\lambda$, $ \epsilon_1=0.9\lambda$ and $ \epsilon_2=1.1\lambda$.

\subsection{Symmetries and DTC-charges in solvable case}

In the unperturbed case (all $\epsilon=0$), the Floquet operator $U_F^{(n)}$ exactly permutes spins to different sites. For convenience, we relabel them as $\sigma_{i,1}\equiv\sigma_{(i-1)n+1}$ and
\begin{equation}
   \sigma^z_{i,j} \equiv \left(U_F^{(n)}\right)^{j-1}\circ (\sigma^z_{i,1}),\\
\end{equation}
where $U\circ \sigma\equiv U^\dagger \sigma U$ and we use $U_F^{(n)}$ to denote the unperturbed Floquet operator for $n$-DTC throughout the paper. The actions of $U_F^{(n)}$ on them are 
\begin{gather}
\label{eq:ufnZ}
\begin{aligned}
    U_F^{(n)} \circ (\sigma^z_{i,j})=\sigma^z_{i,j+1},\\
    [U_F^{(n)}]^n \circ (\sigma^z_{i,j})=\sigma^z_{i,j}.
\end{aligned}
\end{gather}

A slightly different picture arises from the perspective of states (see Fig.~\ref{fig:nDTC}b). For $Z$-basis states , the Floquet operator cyclically transforms one $Z$-basis state $\ket{z}$ to another 
$U_F^{(n)}\ket{z_{m}}=e^{-i E_{\rm int}(z_{m+1})}\ket{z_{m+1}}$, where $E_{\rm int}(z)$ is the energy of the $H_{\rm int}$ for the state $\ket{z}$. Thus, the whole Hilbert space is divided into exponentially many dynamically disjoint fragments, with each subspace consisting of $k$ $Z$-basis states, where $k$ (with $ k| n$) is the period of them. For $n$-period states in the $\alpha$-th subspace, the corresponding eigenstates $\ket{\phi_{\alpha,j}^{(n)}}$ and quasi-energies $\varepsilon_{\alpha,j}^{(n)}$ are (see Eq.~\eqref{eq:eStatesForNDTC} and Appendix~\ref{app:ExactE} for more details)
\begin{equation}
\begin{aligned}
    \ket{\phi_{\alpha,j}^{(n)}}&=\frac{1}{\sqrt{n}}\sum_{m=1}^n e^{i (\theta_{\alpha,m}^{(n)}+m j\frac{2\pi}{n} )} \ket{z^{(n)}_{\alpha,m}},\\
    \varepsilon_{\alpha,j}^{(n)}&=\frac{1}{n}\sum_{m=1}^n E_{\rm int}(z^{(n)}_{\alpha,m}) + j \frac{2\pi}{n},    
\end{aligned}
\end{equation}
where $\theta_{\alpha,m}^{(n)}=\frac{1}{n}\sum_{i=1}^n (m-i\mod n) E_{\rm int}(z_{\alpha, i}^{(n)})$. We thus see that the quasi-energies within each subspace are separated by exactly $\frac{2\pi}{n}$ spacing. Similarly, one can obtain $\frac{2\pi}{k}$ spacing for $k$-period states.

With the exact $\frac{2\pi}{k}$ (where $k|n$) quasi-spectral separations, one can identify a symmetry group $\mathbb{Z}_n$ for the system using the group generator
\begin{eqnarray}
    S&\equiv&\sum_{\alpha}\sum_{j=1}^{k_\alpha} e^{-i j\frac{2\pi}{k_\alpha}} \ket{\phi^{(k_\alpha)}_{\alpha,j}}\bra{\phi^{(k_\alpha)}_{\alpha,j}},
\end{eqnarray}
where $k_\alpha$ (which divides $n$) is the dimension of the subspace $\alpha$.
By construction, it is easy to see that $S^n=1$ and that $S$ commutes with $U_F^{(n)}$. The non-trivial relation of the $U_F^{(n)}$ and $S$ is 
\begin{equation}\label{eq:U_F=S_inMainText}
    \left.U_F^{(n)}\right|_\alpha= \left.e^{-i \frac{E_\alpha}{k_\alpha}} S\right|_\alpha,
\end{equation}
where $E_\alpha\equiv \sum_{m=1}^{k_\alpha} E_{\rm int}(z^{(k_\alpha)}_{\alpha,m})$ and $U|_\alpha$ is the operator $U$ restricted in the $\alpha$-th subspace. This indicates that, when acting within the $\alpha$-th subspace, the Floquet operator $U_F^{(n)}$ has exactly the same effects as the $\mathbb{Z}_n$ symmetry generator $S$, up to a global phase $e^{-i E_\alpha/k_\alpha}$. This immediately leads to $[U_F^{(n)}]^n \bigr{|}_\alpha = e^{-i E_\alpha \frac{n}{k_\alpha}}$ being a pure phase in the $\alpha$-th sector.

Thus, all $Z$-basis states within the same sector have the same quasi-energies w.r.t. the $n$-th power of $U_F^{(n)}$, i.e., $[U_F^{(n)}]^n$, which gives rise to more symmetries than $U_F^{(n)}$; see Eq.~\eqref{eq:ufnZ}. We are interested in symmetries that only appear in $[U_F^{(n)}]^n$ but not in $U_F^{(n)}$ so that they naturally exhibit subharmonic oscillations under $U_F^{(n)}$. As expectation values of symmetry operators are conserved charges for $[U_F^{(n)}]^n$, but they oscillate under $U_F^{(n)}$ within each $n$-period evolution,  so we also refer to them as {\it DTC-charges}. One can verify that all $\sigma^z$'s, along with most their linear combinations, are DTC-charges of $U_F^{(n)}$. By measuring those DTC-charges, we can observe the subharmonic oscillations, which can be seen by Eq.~\eqref{eq:ufnZ}.

\begin{figure}[ht]
    \centering
    \includegraphics[width=0.9\linewidth]{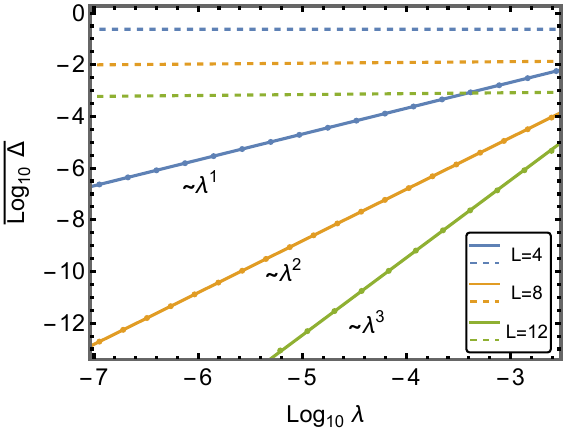}
    \caption{The scaling of the gaps for 4-DTC. The dashed lines and solid lines are for $\overline{\log_{10} \Delta^{(0)}}$ and $\overline{\log_{10} \Delta^{(4)}}$, respectively. All the spectral data is obtained by exactly diagonalizing the $U_F^{(4)}(\lambda)$ in full Hilbert space, whereas the $\overline{\log_{10} \Delta^{(4)}}$ is picked within a subspace for the consistency of the scaling. The average data is obtained from 4000, 1000, and 80 random field samples for $L=4,8,12$, respectively. The solid line is fitted by fixing the slope to be exactly $1,2,3$ for $L=4,8,12$, respectively. One can clearly see the agreement with our theoretical prediction that the $\pi/2$ gap is perturbed at the order of $O(\lambda^{L/4})$.
    }
    \label{fig:0.5gap}
\end{figure}

\subsection{Emergent symmetries and robust subharmonic oscillation against local perturbations}

\begin{figure*}[t]
\begin{tabular}{lll}
(a) & (b) & (c)  \\
\includegraphics[width=0.282\textwidth]{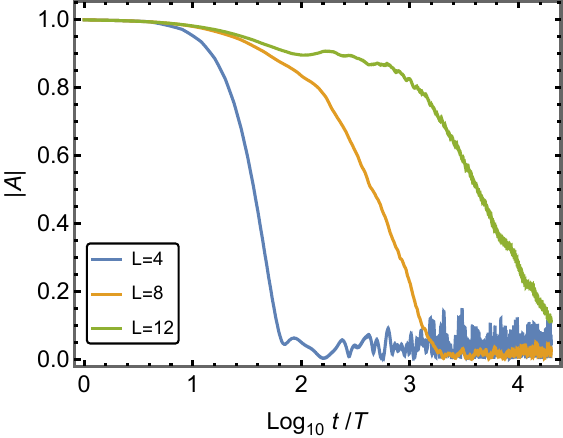}&
\includegraphics[width=0.3\textwidth]{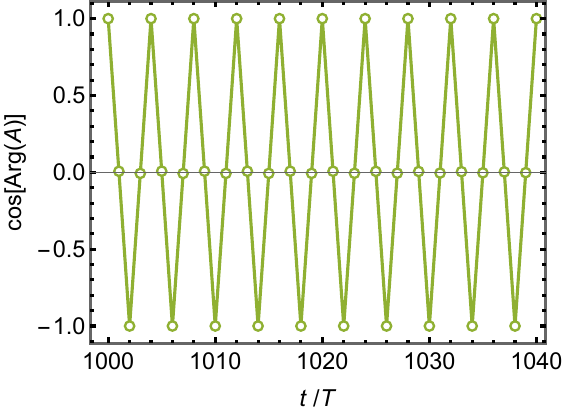}&
\includegraphics[width=0.296\textwidth]{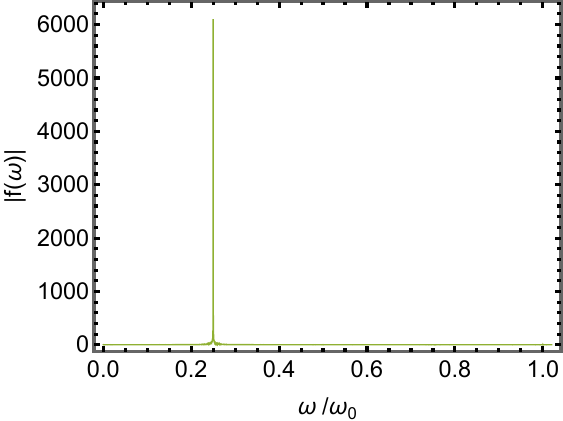}\\
(d) & (e) & (f)  \\
\includegraphics[width=0.282\textwidth]{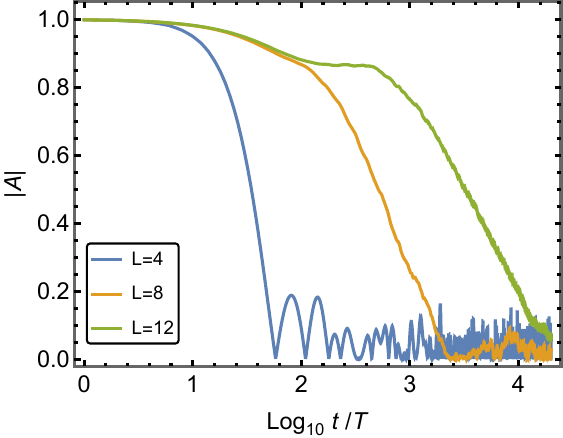}&
\includegraphics[width=0.3\textwidth]{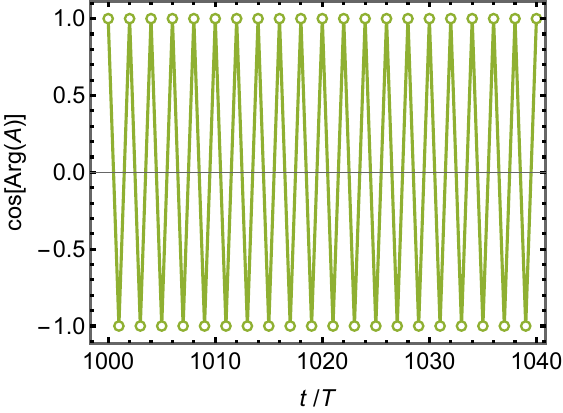}&
\includegraphics[width=0.296\textwidth]{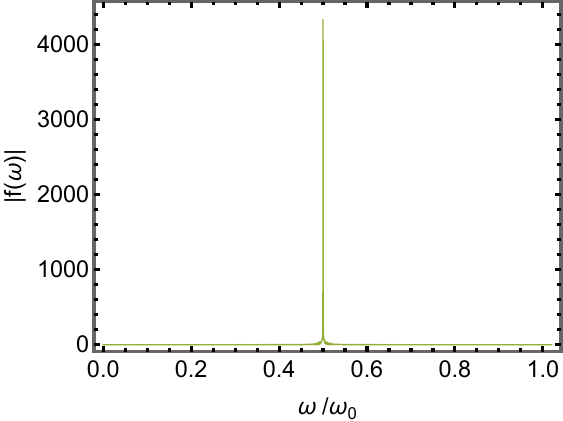}
\end{tabular}
    \caption{The amplitude, phase, and Fourier transformation of the stroboscopic observable $A(t)$ at $t=NT$, respectively, for (a)-(c): 4-DTC and (d)-(f): period doubling subspace-thermal-DTC during the phase transition from 2-DTC to 4-DTC at $s=0.6$. We set the perturbation strength $\lambda=0.02$ for all cases. To visualize the oscillation of the phase of $A(NT)$, (b) and (e) show the cosine of the argument of $A(NT)$ for $L=12$ at the late time $t\approx 1000T$. Diagrams (c) and (f) are Fourier transformations for $A(NT)$ for $L=12$ compared to the driving frequency $\omega_0$, where one can clearly see delta-function-like peaks at (c) ${\omega_0}/{4}$ for 4-DTC, and (f) ${\omega_0}/{2}$ for the ST-DTC.
    }
    \label{fig:AandFT}
\end{figure*}

We now show that the subharmonic oscillations remain robust against local perturbations. In the unperturbed case, for $Z$-basis states in the $\alpha$-th subspace, their components within one spatial unit are distinct if they oscillate with period-$n$ in the unit. We use $l_\alpha$ to count the number of period-$n$ units in the whole chain. From the discussion in Sec.~\ref{sec:theorem}, the correction for the $2\pi/n$ quasi-spectral separation is $\Delta^{(n)} \sim \lambda^{O(l_\alpha)}$. Thus, the time-crystalline structure is exponentially robust to local perturbations and becomes exact in the thermodynamic limit. We have also numerically simulated the quasi-spectral gap dependence with $\lambda$ for 4-DTC (see Fig.~\ref{fig:0.5gap}). With sorted quasi-energies in full space $\{\varepsilon_{i}\}$ and $\alpha$-th subspace $\{\varepsilon_{\alpha,i}\}$, we define the quasi-spectral spacing and the deviation from $\frac{2\pi}{n}$ quasi-spectral separation, respectively
\begin{gather}
\begin{aligned}\label{eq:Delta-0n}
\Delta^{(0)}_i&=\varepsilon_{i+1}-\varepsilon_{i}, \\
\Delta^{(n)}_{\alpha,i}&=\left|\varepsilon_{\alpha,i+1}-\varepsilon_{\alpha,i}-2 \pi / n\right|.
\end{aligned}
\end{gather}
We then average them as 
\begin{gather}
\begin{aligned}
\overline{\log_{10} \Delta^{(0)}}&\equiv\mathbf{avg.}\left[\frac{1}{\mathcal{N}}\sum_i\log_{10} \Delta^{(0)}_i\right], \\
\overline{\log_{10} \Delta^{(n)}}&\equiv\mathbf{avg.}\left[\max_{\alpha,i}(\log_{10} \Delta^{(n)}_{\alpha,i})\right], 
\end{aligned}
\end{gather}
where the symbol $\mathbf{avg.}[\cdots]$ means averaging over disorders, $\mathcal{N}$ counts the number of gaps, and the maximum is only picked from $n$-period sectors. Since $l_\alpha$ affects the scaling behavior of $2\pi/n$ gap, in numerical simulations, we average the $\Delta^{(0)}_i$ over the whole Hilbert space, but only pick the maximum of $\Delta^{(n)}_{\alpha,i}$ in a subspace with the same $l_\alpha$'s for consistency of the scaling. In the unperturbed case $\lambda=0$, the subspace is chosen in the $Z$-basis, where each spatial unit contains exactly one spin down. For $\lambda>0$, the subspace is induced by applying $\mathcal{V}(\lambda)$ to the unperturbed subspace, where $\mathcal{V}(\lambda)$ relates the unperturbed eigenstates to perturbed ones. Suppose that the original eigenstates of $U_F^{(n)}(0)$ are $\{\ket{\phi_{\alpha,j}}\}$ and the perturbed eigenstates of $U_F^{(n)}(\lambda)$ are $\{\ket{\psi_{\alpha,j}(\lambda)}\}$, respectively. We have $\mathcal{V}(\lambda)\ket{\phi_{\alpha,j}}=\ket{\psi_{\alpha,j}(\lambda)}$. For $n=4$, we see from Fig.~\ref{fig:0.5gap} that the average quasi-spectral spacing is almost fixed $\Delta^{(0)} \sim O(1/2^L)$,  whereas the average changes of the $2\pi/n=\pi/2$ (for $n=4$) quasi-spectral separation show a clear exponential scaling $\Delta^{(4)}\sim O(\lambda^{L/4})$. The scaling is consistent with the analytical results.

With robust $2\pi/n$ gaps, the symmetry $\mathbb{Z}_n$ and the DTC-charges $\sigma^z$'s are perturbed into approximate symmetries and DTC-charges, where the deviations from their exact definitions are exponentially small. Defining $\ket{\tilde{z}(\lambda)}\equiv\mathcal{V}(\lambda)\ket{z}$, one can verify that (see Appendix~\ref{app:SymmetryS} for details)
\begin{equation}\label{eq:UflambdaOnZtildeInMainText}
    U_F^{(n)}(\lambda)\ket{\tilde{z}_{\alpha,m}(\lambda)}=e^{-i \tilde{E}_{\alpha,m+1}}\ket{\tilde{z}_{\alpha,{m+1}}(\lambda)}+\lambda^{O(l_\alpha)}.
\end{equation}
The $\mathbb{Z}_n$ symmetry is then perturbed to $\tilde{S}(\lambda)\equiv\mathcal{V}(\lambda)S\mathcal{V}^\dagger(\lambda)$.
One can verify that $\tilde{S}(\lambda)$ is still an exact $\mathbb{Z}_n$ generator satisfying $\tilde{S}^n=1$, and remains an exact symmetry of $U_F^{(n)}(\lambda)$. However, the relation Eq.~\eqref{eq:U_F=S_inMainText} no longer holds exactly; instead, it becomes an approximate relation
\begin{equation}\label{eq:UfSsameinAlphaWithLambda}
    \left.U_F^{(n)}(\lambda)\right|_\alpha= \left.e^{-i \frac{\tilde{E}_\alpha}{k_\alpha}} \tilde{S}\right|_\alpha + \lambda^{O(l_\alpha)},
\end{equation}
where the $\alpha$-th subspace is induced by the perturbed eigenstates $\ket{\psi_{\alpha}}$'s. These results are a direct consequence of the fact that the $2\pi/k_\alpha$ quasi-energy separation is perturbed at the order of $\lambda^{O(l_\alpha)}$ (see Appendix~\ref{app:SymmetryS} for details). Therefore, $U_F^{(n)}(\lambda)$ approximately acts like a $\mathbb{Z}_n$ symmetry generator in the new basis, up to a $\lambda^{O(l_\alpha)}$ correction. We also show that the $\mathbb{Z}_n$ symmetry can be spontaneously broken in the thermodynamic limit in Appendix.~\ref{app:STauB}.

This leads to the perturbed DTC-charges $\tau_i^z$'s of $U_F^{(n)}(\lambda)$, which can be obtained from similar transformations on $\sigma^z$'s. We can define $\tau_i^a\equiv\mathcal{V}(\lambda)\sigma_i^a\mathcal{V}^\dagger(\lambda)$ with $a \in \{x,y,z\}$. It is easy to check that $ \ket{\tilde{z}}$'s are eigenstates of $\tau_i^z$'s. From Eq.~\eqref{eq:UflambdaOnZtildeInMainText}, we see that $\tau^z$'s become approximate emergent symmetries of $[U_F^{(n)}(\lambda)]^n$, and act like exact DTC-charges in the thermodynamic limit, i.e., 
\begin{equation}
\begin{aligned}
    \left[\left(U_F^{(n)}(\lambda)\right)^n,\tau^z\right]\Bigr|_\alpha&\sim \lambda^{O(l_\alpha)},\\
    U_F^{(n)}(\lambda) \circ (\tau^z_{i,j})\Bigr|_\alpha&=\tau^z_{i,j+1} \Bigr|_\alpha+\lambda^{O(l_\alpha)}.
\end{aligned}
\end{equation}
In addition, we expect $\mathcal{V}(\lambda)$ to be close to identity when $\lambda$ is small; we can then expand $\sigma^z$'s in $\tau$-basis
\begin{equation}
    \sigma^z_{i,j}=\mathcal{V}(\lambda)^\dagger\tau^z_{i,j}\mathcal{V}(\lambda)=[1-O(\lambda)]\tau^z_{i,j}+O(\lambda).
\end{equation}
Thus, by measuring $\sigma^z$'s, we will observe that their major components (dominated by $\tau^z$'s) have robust subharmonic oscillations, up to a time scale $e^{O(L/n)}$, which tends to infinity in the thermodynamic limit.

\begin{figure*}[t]
\begin{tabular}{lll}
(a) & (b) & (c)  \\
\includegraphics[width=0.3\textwidth]{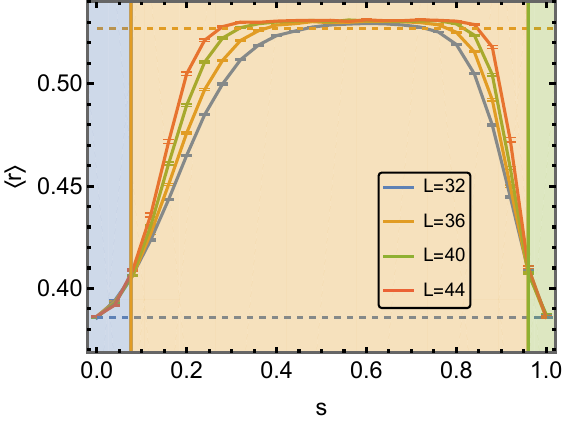}&
\includegraphics[width=0.3\textwidth]{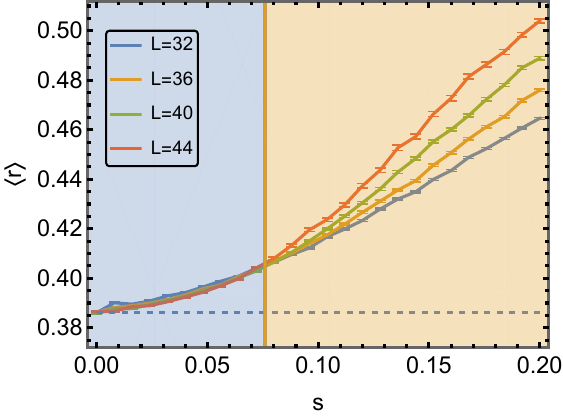}&
\includegraphics[width=0.3\textwidth]{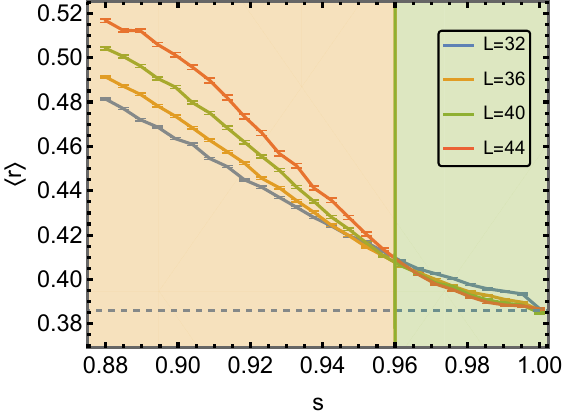}
\end{tabular}
    \caption{Disorder averaged quasi-energy level statistics ratio $\braket{r}$ in the subspace with respect to $s$. The blue, orange, and green regions indicate $2$-DTC, periodic doubling subspace-thermal DTC, and $4$-DTC, respectively. The yellow and blue dashed lines are the COE limit 0.5269 and the Poisson limit 0.3863, respectively.  Diagrams (b) and (c) are zoomed-in views near the phase transition points, where the two critical values  $s^\star $ are $s^\star_1=0.076\pm0.008$ and $s^\star_2=0.960\pm0.005$, respectively. The data is obtained from average over 4000,2000,1000,400 disorder realizations for $L=32,36,40,44$, respectively.
    } 
    \label{fig:rWiths}
\end{figure*}

To evaluate the overall behavior of the subharmonic oscillations, we introduce an observable for each spatial permutation unit
\begin{equation}
    \Lambda_j\equiv \sum_{k=1}^n e^{-ik\frac{2\pi}{n}} \sigma^z_{j,k}.
\end{equation}
One can easily verify that $U_F^{(n)}(0) \circ \Lambda_j=e^{i \frac{2\pi}{n}} \Lambda_j$, and $\braket{z|\Lambda_j|z}$ is non-zero only if the state $\ket{z}$ is period-$n$ in the $j$-th spatial unit. We then normalize and average it over all possible $Z$-basis states with non-zero expectations
\begin{equation}\label{eq:LambdaJTilde}
 \tilde{A}_j(t)\equiv\frac{1}{\mathcal{N}} \sum_{z,\braket{z|\Lambda_j(0)|z}\neq 0}\frac{\braket{z|\Lambda_j(t)|z}}{\braket{z|\Lambda_j(0)|z}},
\end{equation}
where $\Lambda_j(t)=U^\dagger(t)\Lambda_j U(t)$ and $\mathcal{N}$ counts the total number of states. We remark that the quantity can also be regarded as overall evaluation for autocorrelation functions of $\sigma^z_{i,j}$'s for $Z$-basis initial states, as $\Lambda_j$ is diagonal in $Z$-basis. Finally, we further average it over all spatial units and different disorders $J_{ij}$'s and $h_z$'s,
\begin{equation}\label{eq:A(t)forNDTC}
    A(t) \equiv \mathbf{avg.}\Big[\frac{1}{L/n}\sum_{j=1}^{L/n} \tilde{A}_j(t)\Big].
\end{equation}
In the solvable case, one can easily see that $A(NT)=e^{i N\frac{2\pi}{n}}$ and can use it to probe the subharmonic oscillations. For perturbed cases, we expect that the subharmonic oscillation of $A(t)$ persists up to $e^{O(L/n)}$ time. We have simulated ($n$=4)-DTC with $\lambda=0.02$ for $L=4,8,12$, averaging over 100, 100, 48 realizations of disorder, respectively. To maintain the scaling consistency, we only randomly choose 16 $Z$-basis states with one down spin in each unit as initial states, so that all sectors $\alpha_j$ considered have $l_{\alpha_j}=L/4$. We then measure the stroboscopic $\sigma^z$'s for 20000 periods of evolution to obtain $A(NT)$. The results are shown in Fig.~\ref{fig:AandFT}(a)-(c), where we see clearly that the subharmonic oscillations maintain for a long time, with the time scale growing exponentially with $L$, and the Fourier transformation shows an extremely sharp peak at $\omega_0/4$.

\section{Phase transitions among different $n$-DTCs thorough subspace-thermal DTCs}
\label{sec:ST-DTC}

\begin{figure}[ht]
\begin{tabular}{l}
(a) \\
\includegraphics[width=0.39\textwidth]{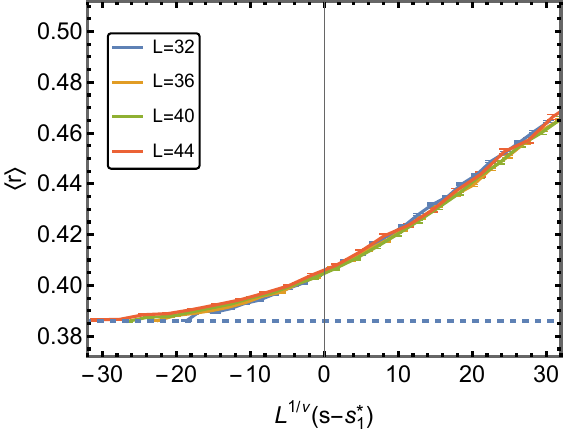}\\
(b) \\
\includegraphics[width=0.4\textwidth]{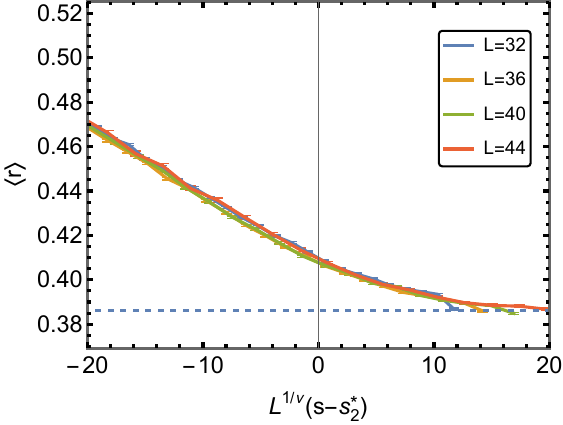}
\end{tabular}
    \caption{Finite size scaling of the $\braket{r}$ near the two phase transition points, for which we choose (a) $\nu_1=0.63$ and (b) $\nu_2=0.61$. In both cases, we see clear data collapse.}
    \label{fig:FiniteSizeScaling}
\end{figure}

Our construction of $n$-DTC offers a natural path for exploring the transition between different $n$-DTC's. We design the potential phase transition between $n_1$-DTC and $n_2$-DTC controlled by a parameter $s$ that goes from 0 to 1, where the changes in the periodic evolution are reflected in the following Hamiltonian terms, 
\begin{gather}\label{eq:HforST-DTC}
    H(s,t)= \begin{cases}
    H_{1}^{(n_1\rightarrow n_2)}(s), & \text { for } 0 \leq t<t_1, \\ 
    H_{2}^{(n_1\rightarrow n_2)}(s), & \text { for } t_1 \leq t<t_1+t_2,\\
    H_{\rm int}, & \text { for } t_1+t_2 \leq t<T,
    \end{cases}
\end{gather}
where $T=t_1+t_2+t_3$, and
\begin{gather}
\begin{aligned}
    H_{1}^{(n_1\rightarrow n_2)}(s)\equiv(1-s)H_{1}^{(n_1)}+sH_{1}^{(n_2)},\\
    H_{2}^{(n_1\rightarrow n_2)}(s)\equiv(1-s)H_{2}^{(n_1)}+sH_{2}^{(n_2)},
\end{aligned}
\end{gather}
and we set $t_1=t_2=\frac{1}{2}$ for both $H^{(n_1)}_{1(2)}$ and $H^{(n_2)}_{1(2)}$, and $H_{\rm int}$ with $t_3=1$ unchanged during the transition. When studying the properties of the intermediate phase $0<s<1$, we fix an $s$ and choose an initial state, then let it evolve under $U_F^{(n_1\rightarrow n_2)}(s)$ to see how observables respond to the periodic driving.

\subsection{Emergent subspace-thermal DTC during the phase transition in solvable case}
We first consider the phase transition between  $n_1$-DTC and $n_2$-DTC in unperturbed cases,  where all $\epsilon$'s are set to 0. When $s$ is not close to 0 and 1, the Heisenberg interactions in $H_{1}^{(n_1\rightarrow n_2)}(s)$ and $H_{2}^{(n_1\rightarrow n_2)}(s)$ are no longer swap gates, and will take $Z$-basis states to superpositions of them. At the first glance, one may think that such periodic driving will quickly thermalize the system. However, due to the symmetry $[\hat{\bm{\sigma}}_{i}\cdot \hat{\bm{\sigma}}_{j},\sigma_{i}^z+ \sigma_{j}^z]=0$ of Heisenberg interactions, some combinations of $\sigma^z$'s can be preserved after several periods of evolution. Take $L=n_2=4$ and $n_1=2$ as an example, one can easily verify that $\sigma_1^z+\sigma_4^z$ is preserved after two periods of driving for all $s$, which can be viewed as a symmetry of $[U_{F}^{(2\rightarrow 4)}(s)]^2$. Thus, the thermalization is restricted in subspaces divided and characterized by those symmetries. 

To verify the above argument, for the unperturbed case, we calculate the quasi-energy level statistics ratio $r\equiv\frac{\min(\delta_i,\delta_{i+1})}{\max(\delta_i,\delta_{i+1})}$ for $U_F^{(2\rightarrow 4)}(s)$ with different $s$, where $\delta_i\equiv \varepsilon_{i+1}-\varepsilon_i$. In a thermal phase, one expects $r$ close to the circular orthogonal ensemble (COE) limit of 0.5269, whereas the value for an integrable phase should approach the Poisson limit of 0.3863. As the $U_{F}^{(2\rightarrow 4)}(s)$ is fragmentized due to the symmetries, if one calculates $r$ in the full Hilbert space, one can {\it only} observe the Poisson limit~\cite{giraud2022probing}. Thus, we calculate $r$ only in the subspace spanned by states dynamically coupled with $\ket{\downarrow \uparrow\uparrow\uparrow,\downarrow \uparrow\uparrow\uparrow,...}$. The results are shown in Fig.~\ref{fig:rWiths}, and one clearly sees a transition from an integrable to a thermal phase, exhibited by the change of $\langle r\rangle$ from approximately 0.386 to 0.527 and back to 0.386 as $s$ varies. Near the critical points $s^\star$, we assume the correlation length scales as $\xi\propto (s-s^\star)^{-\nu}$ so that $\braket{r}$ can be written as a function of $\frac{L}{\xi}$, i.e., $\braket{r}=f(L/\xi)$. With finite-size scaling (see Fig.~\ref{fig:FiniteSizeScaling}), we estimate the critical component $\nu \approx 0.6\pm 0.1$ for both transition points.

Therefore, $U^{(n_1\rightarrow n_2)}_F(s)$ thermalizes the system when $s$ is not close to 0 or 1, but the thermalization is restricted in subspaces characterized by some combinations of $\sigma^z$'s. We will show shortly that such combinations can be systematically constructed for general $n_1$ and $n_2$. Despite that the subspaces are already thermalized, these combinations of $\sigma^z$'s can still exhibit subharmonic oscillation. Furthermore, we will show later that such subharmonic oscillation persists even with perturbations that break the symmetry of Heisenberg interactions. The robust subharmonic response in the presence of fully thermalized subspaces is distinct from all existing DTC models, where the systems are either not thermalized or prethermalized in the latter cases. Therefore, we call this new type of phase \textit{subspace-thermal DTC} (ST-DTC).

To continue, we first set the notation. For general nonzero natural numbers $n_1$ and $n_2$, we denote their greatest common divisor and least common multiplier as  $n_G\equiv \gcd (n_1,n_2)$ and  $n_L\equiv \mathrm{lcm} (n_1,n_2)$, respectively. Note that for every $\mathbb{Z}_n$ group, one can find a $\mathbb{Z}_k$ subgroup where $k|n$. As the $U_F^{(n)}$ transforms one DTC-charge to another in our model, we can linearly combine DTC-charges $\sigma^z$'s to construct an observable oscillating in a smaller period $k$, where $k$ can be any factor of $n$. In this way, by constructions of our $n$-DTC model, we can find common combinations of DTC-charges for {\it both\/} $n_1$-DTC and $n_2$-DTC in an $n_L$-site block, oscillating with $n_G$ period, that retain their forms during the phase transition from $s=0$ to 1.  In Appendix~\ref{app:ExactU(s)}, we show that such new DTC-charges, denoted as $Q^{(n_1\rightarrow n_2)}_{i,j}$'s, can be constructed as follows, 
\begin{gather}
    Q_{i,j}^{(n_1\rightarrow n_2)}\equiv\sum_{m=1}^{n_L/n_G} \left(U_F^{(n_L)}\right)^{n_G (m-1)+j-1}\circ (\sigma^z_{i,1}),
\end{gather}
where we have relabeled the first spin in $i$-th $n_L$-site unit as $\sigma_{i,1}\equiv\sigma_{(i-1)n_L+1}$, and $i$ denotes the labeling of which unit and $j \in[1,n_G]$ denotes the $j$-th group of sites connected by $[U_F^{(n_L)}]^{n_G}$ within each unit. When $n_1=n_2$, the DTC-charges $Q$'s reduce to $\sigma^z$'s. We may omit the superscript ${(n_1\rightarrow n_2)}$ of $Q_{i,j}$ occasionally for simplicity. One can verify that (see Appendix~\ref{app:ExactU(s)} for details) all $Q_{i,j}$'s commute with $[U_F^{(n_1\rightarrow n_2)}(s)]^{n_G}$. With a similar procedure, we obtain
\begin{equation}
\begin{aligned}
    U_{F}^{(n_1\rightarrow n_2)}(s) \circ (Q^{(n_1\rightarrow n_2)}_{i,j})=Q^{(n_1\rightarrow n_2)}_{i,j+1},\\
    [U_{F}^{(n_1\rightarrow n_2)}(s)]^{n_G} \circ (Q^{(n_1\rightarrow n_2)}_{i,j})=Q^{(n_1\rightarrow n_2)}_{i,j},
\end{aligned}
\end{equation}
where $Q^{(n_1\rightarrow n_2)}_{i,n_G+1}\equiv Q^{(n_1\rightarrow n_2)}_{i,1}$. We can also define one eigenstate  of $[U_F^{(n_1\rightarrow n_2)}(s)]^{n_G}$ in a certain $\{Q_{i,j}\}$-subspace as $\ket{\chi_{\alpha,1}(\{Q_{i,j}=q_{i,j}\})}$ with eigenvalue $e^{-i E_\alpha}$, and obtain the relation
\begin{multline}
    U_F^{(n_1\rightarrow n_2)}(s)\ket{\chi_{\alpha,m}(\{Q^{(n_1\rightarrow n_2)}_{i,j}=q_{i,j}\})}\equiv\\
    \ket{\chi_{\alpha,m+1}(\{Q^{(n_1\rightarrow n_2)}_{i,j+1}=q_{i,j}\})},
\end{multline}
for $1\leq m<n_G$, where we absorb the possible phase ambiguity into the definition of  $\ket{\chi_{\alpha,m+1}}$. By definition $[U_F^{(n_1\rightarrow n_2)}(s)]^{n_G}\ket{\chi_{\alpha,1}}=e^{-i E_\alpha} \ket{\chi_{\alpha,1}}$, we have $U_F^{(n_1\rightarrow n_2)}(s)\ket{\chi_{\alpha,n_G}}= e^{-i E_\alpha} \ket{\chi_{\alpha,1}}$. Similarly, depending on the configuration of $\{q_{i,j}\}$'s, some states may have smaller periods $k$'s than $n_G$. For $n_G$ period states, the eigenstates and quasi-energies of $U_F^{(n_1\rightarrow n_2)}(s)$ are thus obtained, similarly to previous Floquet eigenstates and quasi-energies, as follows,
\begin{eqnarray}\label{eq:ST-eState}
\begin{aligned}
    \ket{\Phi_{\alpha,j}}&=\frac{1}{\sqrt{n_G}}\sum_{m=1}^{n_G} e^{i (\theta_{\alpha,m}+mj\frac{2\pi}{n_G}) } \ket{\chi_{\alpha,m}},\\
    \varepsilon_{\alpha,j}&= \frac{E_\alpha}{n_G} + j \frac{2\pi}{n_G}.
\end{aligned}
\end{eqnarray}

To probe the subharmonic oscillation of $Q$'s, we can simply prepare an $n_G$-period $Z$-basis initial states $\ket{z(\{Q_{i,j}=q_{i,j}\})}$, which is a superposition of eigenstates with the same $Q_{i,j}$ configurations
\begin{equation}
    \ket{z(\{Q_{i,j}=q_{i,j}\})}=\sum_{k} a_k \ket{\chi_{\alpha_k,m_k}(\{Q_{i,j}=q_{i,j}\})},
\end{equation}
where we have used the fact that each $\alpha_k$-th sector contains only one $\ket{\chi}$ that has exactly the same $Q_{i,j}$ configurations as the $\ket{z(\{Q_{i,j}=q_{i,j}\})}$. After $n_G$ periods of evolution, the state evolves to
\begin{multline}
    [U_F^{(n_1\rightarrow n_2)}(s)]^{n_G}\ket{z(\{Q_{i,j}=q_{i,j}\})}=\\
    \sum_{k} a_k e^{-i E_{\alpha_k}} \ket{\chi_{\alpha_k,m_k}(\{Q_{i,j}=q_{i,j}\})} .
\end{multline}
Due to the random phases $e^{-iE_{\alpha_k}}$'s, $[U^{(n_1\rightarrow n_2)}_F(s)]^{n_G}\ket{z}$ gets thermalized in the $\{Q_{i,j}=q_{i,j}\}$-subspace after sufficient steps and is no longer a $Z$-basis state. However, as all components of the outcome state share the same $\{Q_{i,j}\}$ values, the expectations of $\{Q_{i,j}\}$ evolve back to themselves after $n_G$ periods. The whole evolution can be viewed as a thermal state restricted in one $\{Q_{i,j}\}$-subspace, oscillating cyclically among different $\{Q_{i,j}\}$-subspaces. We will show shortly that such subharmonic oscillation is robust even if the symmetry dividing $\{Q_{i,j}\}$-subspaces is broken. Since $Q_{i,j}$'s are combinations of $\sigma^z$'s, if one measures the expectations of single $\sigma^z$'s, one will find that they quickly become evenly distributed within each $Q_{i,j}=q_{i,j}$. As $Q_{i,j}$'s are oscillating with period-$n_G$, all $\sigma^z$'s will eventually oscillate synchronously with corresponding $Q_{i,j}$'s, with amplitudes reduced by a factor $n_G/n_L$. Therefore, even if $\sigma^z$'s have diffused, $Q_{i,j}$'s remain localized in each $n_L$-site permutation unit.

\subsection{Robust subharmonic oscillations against local perturbations}

\begin{figure}[ht]
    \centering
    \includegraphics[width=0.9\linewidth]{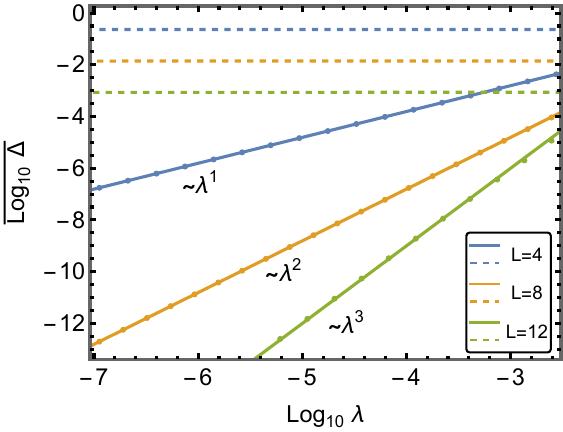}
    \caption{The scaling of the gap for the emergent period doubling subspace-thermal DTC during the phase transition $s=0.6$ from 2-DTC to 4-DTC. The dashed lines and solid lines are for $\overline{\log_{10} \Delta^{(0)}}$ and $\overline{\log_{10} \Delta^{(2)}}$, respectively. All the spectral data is obtained by exactly diagonalizing the $U_F^{(2\rightarrow4)}(s=0.6,\lambda)$ in full Hilbert space, whereas the $\overline{\log_{10} \Delta^{(2)}}$ is picked within a subspace for the consistency of the scaling. The average data is obtained from 4000, 1000, and 80 random field samples for $L=4,8,12$, respectively. The solid line is fitted by fixing the slope to be exactly $1,2,3$ for $L=4,8,12$, respectively. One can clearly see the $\pi$ gap is perturbed at the order of $O(\lambda^{L/4})$.
    }
    \label{fig:1gapSTDTC}
\end{figure}

As $n_1$-DTC and $n_2$-DTC are both robust against local perturbations, we expect the system will remain $n_1$-DTC or $n_2$-DTC when $s$ is close to 0 or 1. When $s$ is far from 0 and 1, the system will become thermalized in the $\{Q_{i,j}\}$-subspaces. However, as such thermalization does not close the quasi-spectral gap for states with different $Q_{i,j}$ configurations, the LPCPG condition still holds, so that $2\pi/n_G$ quasi-spectral separation remains robust against local perturbations. From the discussion in Sec.~\ref{sec:theorem} and Appendix~\ref{app:perturbGap}, the correction for the $2\pi/n_G$ quasi-spectral separation is approximately $\Delta^{(n_G)} \sim \lambda^{O(l_\alpha)}$, where the $l_\alpha$ counts the number of $n_G$-period $n_L$-sites units in the whole chain. Thus, the $2\pi/n$ quasi-spectral gap is exponentially robust to local perturbations and becomes exact in the thermodynamic limit. As an illustration, we have also numerically simulated the quasi-spectral gap dependence with $\lambda$ for period doubling ST-DTC $U_F^{(2\rightarrow 4)}(s)$ at fully thermalized point $s=0.6$, during the phase transition from 2-DTC to 4-DTC (see Fig.~\ref{fig:1gapSTDTC}), where we see a clear exponential scaling $O(\lambda^{L/4})$. For general $n_1$ and $n_2$, based on the analysis in Appendix~\ref{app:perturbGap}, the perturbation on the $2\pi/n_G$ spacing should scale with $\lambda^{O(L/n_L)}$.

With robust $2\pi/n_G$ quasi-spectral spacing, we can define $\tilde{Q}_{i,j}\equiv\mathcal{V}(\lambda)Q^{(n_1\rightarrow n_2)}_{i,j}\mathcal{V}^\dagger(\lambda)$, in a similar way to Sec.~\ref{sec:n-DTC}. These $\tilde{Q}_{i,j}$'s are emergent approximate symmetries of $[U_F(s,\lambda)]^{n_G}$, and $U_F(s,\lambda)$ approximately acts like a $\mathbb{Z}_{n_G}$ symmetry generator on $\tilde{Q}_{i,j}$ (see discussions in Appendix.~\ref{app:SymmetryS}), i.e.,
\begin{equation}
\begin{aligned}
&\left[\left(U_F^{(n_1\rightarrow n_2)}(s,\lambda)\right)^{n_G},\tilde{Q}_{i,j}\right]\Bigr|_\alpha\sim \lambda^{O(l_\alpha)},\\
&U_F^{(n_1\rightarrow n_2)}(s,\lambda) \circ (\tilde{Q}_{i,j})\Bigr|_\alpha=\tilde{Q}_{i,j+1}\Bigr|_\alpha+\lambda^{O(l_\alpha)}.
\end{aligned}
\end{equation}

Thus, the thermalizations are still restricted within subspaces characterized by $\tilde{Q}_{i,j}$'s, stable within a time scale $e^{O(L)}$ approaching infinity in the thermodynamic limit.
In addition, we expect $\mathcal{V}(\lambda)$ is close to identity when $\lambda$ is small. We can expand $Q_{i,j}$'s in $\tau$-basis,
\begin{equation}
    Q_{i,j}=\mathcal{V}(\lambda)^\dagger \tilde{Q}_{i,j}\mathcal{V}(\lambda)=[1-O(\lambda)] \tilde{Q}_{i,j}+O(\lambda),
\end{equation}
where we use the fact that $\tilde{Q}_{i,j}$'s are combinations of $\tau^z$'s. Thus, by measuring $Q_{i,j}$'s, we can observe that their major components $\tilde{Q}_{i,j}$'s have robust subharmonic oscillations, up to a time scale $e^{O(L)}$ going to infinity in the thermodynamic limit.

To probe the overall behavior of the subharmonic oscillations, we define a similar observable to the $n$-DTC case for each $n_L$-site spatial permutation unit
\begin{equation}\label{eq:Lambda-STDTC}
    \Lambda_j\equiv \sum_{k=1}^{n_G} e^{-ik\frac{2\pi}{n_G}} Q^{(n_1\rightarrow n_2)}_{j,k}.
\end{equation}
By averaging over $Z$-basis states, spatial units, and disorder realizations, in a similar way to Eq.~\eqref{eq:LambdaJTilde} and~\eqref{eq:A(t)forNDTC}, we get the stroboscopic observable $A(NT)$, which is $e^{i NT}$ in the unperturbed case. We simulated period doubling ST-DTC, emerging at the phase transition from 2-DTC to 4-DTC at $s=0.6$, with perturbation strength $\lambda=0.02$ for $L=4,8,12$, averaging over 100, 100, 48 realizations of disorder, respectively. In the simulations, to maintain the consistency of scaling,  we only randomly choose 16 $Z$-basis states with one down spin in each unit as initial states, so that all sectors $\alpha_j$ considered have $l_{\alpha_j}=L/4$. We then measure the stroboscopic $\sigma^z$'s for 20000 periods of evolution and calculate $A(NT)$. The results are in Fig.~\ref{fig:AandFT}(d)-(f), where we see clearly that the subharmonic oscillations maintain for a long time, with the time scale growing exponentially with $L$, and the Fourier transformation shows a clean, sharp peak at $\omega_0/2$.

\section{More numerics on ST-DTC}
\label{sec:ST-DTCmore}

In this section, we perform more thorough numerical simulations for the ST-DTC, with larger scales and larger perturbations. We will firstly show the validity of the intuitive arguments we give for the LPCPG condition in Sec.~\ref{sec:theorem}, and verify that the $2\pi/n$ gap is indeed exponentially robust for larger scales. Then, we calculate the stroboscopic order parameter $A(NT)$ of the ST-DTC, and show that the exponential growth of the DTC lifetime proceeds to $L=24$ and $\lambda=0.2$. Finally, to rule out the prethermal effect in our ST-DTC, we simulate the translation invariant version of our model, where only prethermalization governs, and show that the stability from prethermalization is qualitatively different from that in ST-DTCs.

To extend the numerical simulation of the ST-DTCs to larger sizes, we modify the $Z$-$Z$ couplings and perturbation terms in $H_{\rm int}$ so that the size of the dynamical Hilbert space can be reduced. The evolution of the modified ST-DTC model is defined similarly to Eq.~\eqref{eq:HforST-DTC}:
\begin{gather}
    H(s,t)= \begin{cases}
    H_{1}^{(n_1\rightarrow n_2)}(s), & \text { for } 0 \leq t<t_1, \\ 
    H_{2}^{(n_1\rightarrow n_2)}(s), & \text { for } t_1 \leq t<t_1+t_2,\\
    \tilde{H}_{\rm int}, & \text { for } t_1+t_2 \leq t<T.
    \end{cases}
\end{gather}
We focus on numerics for $n_1=2$ and $n_2=4$ in this section, whose Floquet operator is denoted as $\tilde{U}_F^{(2\rightarrow4)}(s,\lambda)$. Specifically, the $U_1$ and $U_2$ of $\tilde{U}_F^{(2\rightarrow4)}(s,\lambda)$ are the same as $U_F^{(2\rightarrow4)}(s,\lambda)$ defined in Sec.~\ref{sec:ST-DTC}, where we set $s=0.6$ so that the $Q$-subspaces are fully thermalized. We replace the $H_{\rm int}$ in $U_F^{(2\rightarrow4)}(s,\lambda)$ with the following Hamiltonian:
\begin{equation}\label{eq:new-H-int}
 \tilde{H}_{\rm int}=\sum_{i=1}^{L/4} J_{i} \Lambda_i \Lambda_{i+1}+\sum_{i=1}^L h_i^z \sigma_i^z+\lambda \hat{V},
\end{equation}
where we choose the periodic boundary condition and the $\Lambda_i$ is defined~\footnote{Here all the $\Lambda_i$ for $n_1=2$ and $n_2=4$ is Hermitian, since $n_G=2$ and all the phase factors are either +1 or -1. In general, $\Lambda_i$ may not be Hermitian due to the phase factor in its definition, but one can choose the following form $\Lambda^\dagger_i \Lambda_{i+1}+\Lambda^\dagger_{i+1}\Lambda_i$ in $H_{\rm int}$.} from Eq.~\eqref{eq:Lambda-STDTC}. Note that $\Lambda$'s are linear combinations of $\sigma^z$'s, so the interactions are still $Z$-$Z$ types, similar to the previous model. By setting the interaction in this way, unperturbed eigenstates with locally different $Q$'s will have larger quasi-spectral gaps compared to the power-law interactions, which enhances the robustness of the system against perturbations. We note that the system is no longer a robust 2- or 4-DTC defined in Sec.~\ref{sec:n-DTC} at $s=0$ or 1 due to the degeneracies in the $Q$-subspaces (but are still robust 2-ST-DTCs). However, this does not matter since our focus here is to give an example to strongly display subspace-thermal DTC and further strengthen the claim that the ST-DTC phase exists. In addition, we set $s=0.6$ and this will lead to the thermalization of the $Q$-subspaces anyway, as shown in Fig.~\ref{fig:rWiths} for the previous model. The perturbation $\hat{V}$ is defined as
\begin{equation}
    \hat{V}=\sum_{i=1,4\nmid i}^L j_i^{xy}( \sigma_i^x\sigma_{i+1}^x+\sigma_i^y\sigma_{i+1}^y)+\sum_{i=1}^Lj_i^{z}\sigma_i^z\sigma_{i+1}^z.
\end{equation}
The form of $\hat{V}$ ensures that the $Q$-symmetries are broken, but the particle number in each spatial unit is still conserved. In addition, we set $\epsilon_1=\epsilon_2=0$ in $H_1^{(2\rightarrow4)}$ and $H_2^{(2\rightarrow4)}$ so that the particle number in each spatial unit is preserved throughout the period of $\tilde{U}_F^{(2\rightarrow4)}(s,\lambda)$. 

In the numerical simulations in this section, we set $J_i\in [0.5\bar{J},1.5\bar{J}]$, $h^z_i\in [0,2\bar{h^z}]$, $j^{xy}\in [0.8\bar{j^{xy}},1.2\bar{j^{xy}}]$ and $j^z_i\in [0.8\bar{j^z},1.2\bar{j^z}]$, with $\bar{J}=4$, $\bar{h^z}=12$ and $\bar{j^{xy}}=\bar{j^z}=1$. We also set the same values for $t_1=t_2=1/2$ and $t_3=1$ as before. In addition, we reduce the numerical simulations in the subspace spanned by $Z$-basis states with one down spin in each unit, so that all states in the subspace have $l_\alpha=L/4$. As $\tilde{U}_F^{(2\rightarrow4)}(s,\lambda)$ preserves the particle number in each unit, the subspace reduction brings no error for initial states with one down spin in each unit.

\subsection{Verification of the LPCPG condition}

\begin{figure}[ht]
    \centering
    \includegraphics[width=0.9\linewidth]{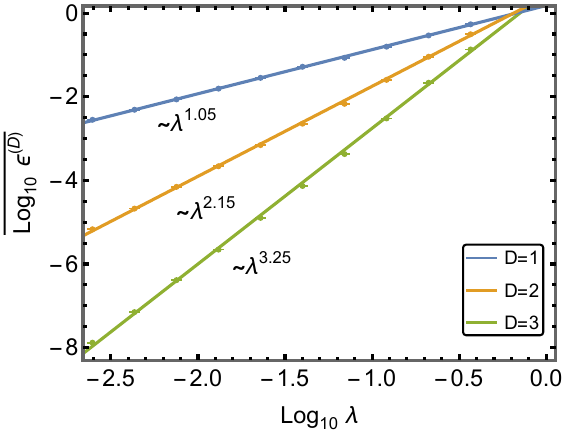}
    \caption{The averaged $\epsilon^{(D)}$ vs. $\lambda$ at different $D$ for the ST-DTC at $L=24, s=0.6$. The $\epsilon^{(D)}$ roughly scales with $\lambda^D$, which is consistent with our intuitive arguments in Sec.~\ref{sec:LPCPG-condition}. We note that $\epsilon^{(D\geq4)}=0$ since $\max (D_{\alpha,\beta})=3$ for the subspace we considered at $L=24$.
    }
    \label{fig:e-D-STDTC}
\end{figure}

Directly verifying the LPCPG condition requires numerical simulations on a large scale, which suffers from the exponential growth of the Hilbert space size. In this section, we try to verify the intuitive argument in Sec.~\ref{sec:LPCPG-condition} and provide numerical data for a decent system size. The argument states that, for a perturbed eigenstate $\ket{\tilde{\Phi}_{\alpha,j}(\lambda)}$ of a strongly disordered Floquet system, the contribution from states distanced from $\ket{\Phi_{\alpha,j}}$ by $D$ is roughly $O(\lambda^{D/K_u}2^L)$, which can be exponentially small in $L$ if $D\gtrsim O(L)$ and $\lambda$ is sufficiently small (but still finite).

For an unperturbed eigenstate $\ket{\Phi_{\alpha,j}}$, we define $\mathbb{V}^{(D)}_{\alpha,j}$ as the subspace spanned by unperturbed eigenstates $\ket{\Phi_{\beta,j'}}$ with $D_{\alpha,\beta}<D$. Similarly, we can define the leakage $\epsilon^{(D)}_{\alpha,j}$ outside the subspace $\mathbb{V}^{(D)}_{\alpha,j}$, according to Eq.~\eqref{eq:condition}, for the perturbed eigenstate $\ket{\tilde{\Phi}_{\alpha,j}(\lambda)}$. From the intuitive arguments in Sec.~\ref{sec:theorem}, $\epsilon^{(D)}_{\alpha,j}$ should scale like $\lambda^D$ for small $\lambda$, given $K_u=1$ for the model defined in this section. This is indeed verified by our numerical simulations presented in Fig.~\ref{fig:e-D-STDTC} for $L=24$, where we exactly diagonalize the perturbed  Floquet operator $\tilde{U}_F^{(2\rightarrow4)}(s,\lambda)$ and obtain the perturbed eigenstate $\ket{\tilde{\Phi}_{\alpha,j}(\lambda)}$, which has the maximum overlap with the unperturbed eigenstate $\ket{\Phi_{\alpha,j}}$. In this way, we can calculate $\epsilon^{(D)}_{\alpha,j}$ for all perturbed eigenstates, and then obtain the averaged $\epsilon^{(D)}$ by log-averaging them in all eigenstates. This quantity is further averaged with 20 random realizations of the disorder. In Fig.~\ref{fig:e-D-STDTC}, one can clearly see that the leakage $\epsilon^{(D)}$ roughly scales with $\lambda^D$, which confirms the intuitive arguments in Sec.~\ref{sec:LPCPG-condition}.

\begin{figure}[ht]
    \centering
\begin{tabular}{l}
(a) \\
\includegraphics[width=0.41\textwidth]{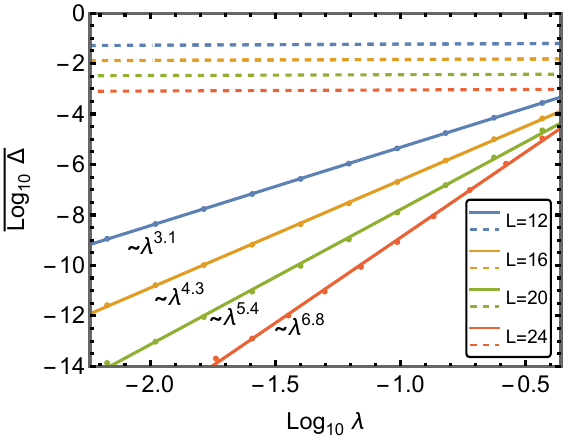}\\
(b) \\
~~~~\includegraphics[width=0.38\textwidth]{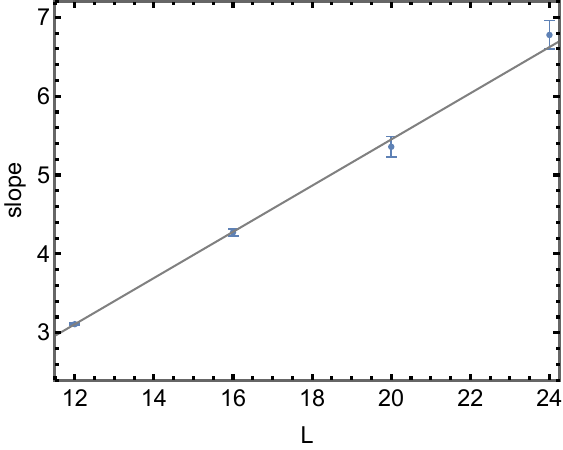}
\end{tabular}
    \caption{(a) The scaling of the deviation from $\pi$ gap for the ST-DTC $\tilde{U}_F^{(2\rightarrow4)}(s,\lambda)$ at $s=0.6$. The dashed lines and solid lines are for $\overline{\log_{10} \Delta^{(0)}}$ and $\overline{\log_{10} \Delta^{(2)}}$, respectively. All the spectral data is obtained by exactly diagonalizing the $\tilde{U}_F^{(2\rightarrow4)}(s,\lambda)$ in the particle number preserving subspace. The average data is obtained from 2000, 1000, 400, and 80 random field samples for $L=12,16,20,24$. (b) The slope of the solid lines in (a) with respect to the system size $L$. One can clearly see the agreement with our theoretical prediction that the $\pi$ gap is perturbed at the order of $e^{-O(L\ln \frac{1}{\lambda})}$.
    }
    \label{fig:gapSTDTC-1}
\end{figure}

\subsection{Verification of the robust $\pi$ gap}

Here, we numerically verify the exponentially robust $\pi$ gap of the ST-DTC defined earlier in this section for a larger scale than in Sec.~\ref{sec:n-DTC}. We can define the quasi-spectral spacing $\Delta^{(0)}$ and the deviation from $\pi$ gap $\Delta^{(2)}$ in the same way as Eq.~\eqref{eq:Delta-0n}. Since all states in the subspace we choose here have period-doubling oscillations with the same $l_\alpha=L/4$, we can directly average both of them as 
\begin{gather}
\begin{aligned}
\overline{\log_{10} \Delta^{(0)}}&\equiv\mathbf{avg.}\left[\frac{1}{\mathcal{N}}\sum_i\log_{10} \Delta^{(0)}_i\right], \\
\overline{\log_{10} \Delta^{(2)}}&\equiv\mathbf{avg.}\left[\frac{1}{\mathcal{N}}\sum_{\alpha,i}\log_{10} \Delta^{(2)}_{\alpha,i}\right], 
\end{aligned}
\end{gather}
where the symbol $\mathbf{avg.}[\cdots]$ means averaging over disorders, and $\mathcal{N}$ counts the number of gaps. We show the results for $\tilde{U}_F^{(2\rightarrow4)}(s,\lambda)$ in Fig.~\ref{fig:gapSTDTC-1} that the average quasi-spectral spacing is almost fixed $\Delta^{(0)} \sim O(2^{-L/2})$,  whereas the average change in the $\pi$ gap show a clear exponential scaling $e^{-O\left(L\ln\frac{1}{\lambda}\right)}$ with $L$. The scaling is consistent with the analytical results in Sec.~\ref{sec:theorem}.

\subsection{Scaling of the lifetime}

\begin{figure}[ht]
    \centering
    \includegraphics[width=0.9\linewidth]{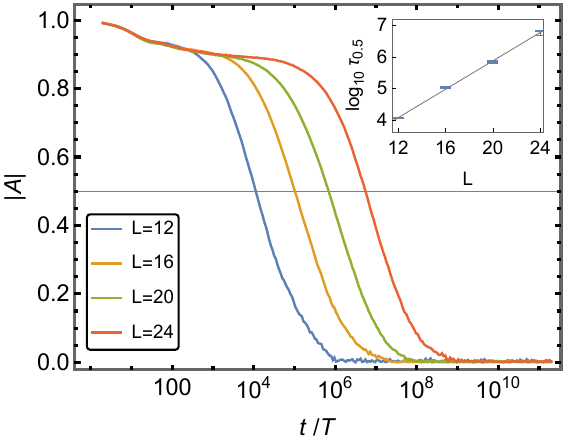}
    \caption{The stroboscopic observable $|A(t)|$ at $t=NT$ for the ST-DTC $\tilde{U}_F^{(2\rightarrow4)}(s,\lambda)$ at different system size with $s=0.6$ and $\lambda=0.2$, averaged over 32 random initial $Z$-basis states and random realizations of disorders. One can clearly see the exponential growth of the lifetime of the ST-DTC.
    }
    \label{fig:A-STDTC-1}
\end{figure}

In this section, we present numerical simulations of the lifetime of the ST-DTC for larger scales by calculating the amplitude of $A(t)$ defined in Sec.~\ref{sec:ST-DTC}. We show the results for $\tilde{U}_F^{(2\rightarrow4)}(s,\lambda)$ at $s=0.6$ in Fig.~\ref{fig:A-STDTC-1}, where we set the perturbation strength to be $\lambda=0.2$ and average the result over 32 random initial $Z$-basis states and 1000, 800, 400, 80 realizations of disorder for $L=12,16,20,24$, respectively. We then calculate the stroboscopic $\sigma^z$'s by exact diagonalization and calculate $A(t=NT)$. In Fig.~\ref{fig:A-STDTC-1}, we clearly see that the subharmonic oscillations maintain for a long time. We define the lifetime $\tau_{0.5}$ (in the unit of $T$) of the DTCs as the evolution time when the amplitude of the order parameter $|A|$ decays by a half, i.e.,
\begin{equation}
    |A(\tau_{0.5}T)|=\frac{1}{2}.
\end{equation} 
In the inset of Fig.~\ref{fig:A-STDTC-1}, we can clearly see that the lifetime grows exponentially with $L$.

\subsection{Effects of prethermalization}

\begin{figure}[ht]
    \centering
    \includegraphics[width=0.9\linewidth]{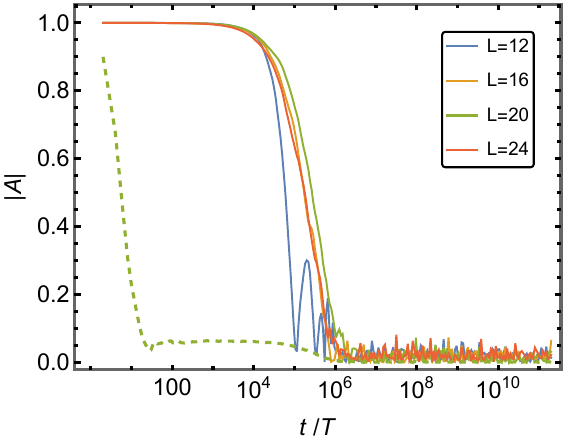}
    \caption{The stroboscopic observable $|A(t)|$ at $t=NT$ for the translation-invariant prethermal-DTC at different size with $s=0.6$ and $\lambda=0.2$, where we use fixed couplings $J_i$ and local fields $h^z_i$, with random perturbations. The initial state of the solid lines is fixed to be the ground state of unperturbed $\tilde{H}_{\rm int}$ so that the prethermal time can be maximized, and one can see clearly that the lifetime does not grow with $L$ for $L\geq 16$. For the green dashed line, we fix $L=20$ and average over 32 randomly chosen $Z$-basis initial states. In this case, most of them get thermalized quickly, and only a small portion of them can have the prethermal plateau.
    }
    \label{fig:A-STDTC-pre-1}
\end{figure}

As prethermalization always~\cite{else2017prethermal} exists for local Floquet systems at $\lambda T\ll 1$, we use the translation-invariant version of our ST-DTC model to examine its effects. Specifically, the model, which we called prethermal-DTC in this section, is defined by setting $J_i=4$ and $h_i^z=12$ to be fixed values in the Eq.~(\ref{eq:new-H-int}). As the system is now translation invariant at $\lambda=0$, the majority of the perturbed eigenstates do not satisfy the LPCPG condition. Thus, we expect that the $\pi$ gap is not robust against perturbations. However, evolutions from low-temperature initial states can be protected by the prethermalization mechanism and still display robust subharmonic oscillations for a long time.

In Fig.~\ref{fig:A-STDTC-pre-1}, we show the simulation results for the translation invariant prethermal-DTC at $s=0.6$, where we set the perturbation strength $\lambda=0.2$ and average over 200, 200, 200, 100 realizations of random perturbations for $L=12,16,20,24$, respectively. When setting the initial state as the ground state of $\tilde{H}_{\rm int}$ at $\lambda=0$ in Eq.~\eqref{eq:new-H-int} (solid lines in Fig.~\ref{fig:A-STDTC-pre-1}), we can observe the prethermal plateau, but the lifetimes almost remain roughly the same when $L$ grows, while in the ST-DTC, the lifetime grows exponentially with $L$. In addition, the lifetime is shorter for initial states other than the ground state. To demonstrate this, we also plot the $|A(NT)|$ averaged over 32 randomly chosen initial $Z$-basis states at $L=20$ and $\lambda=0.2$ with 400 random perturbations (the dashed line in Fig.~\ref{fig:A-STDTC-pre-1}). From the figure, we can see that most initial states quickly get thermalized at short times. Despite the fact that a small portion of them can have prethermal plateaus, the overall $|A|$ is much more suppressed than that in the ST-DTC. Therefore, the results from prethermalization are qualitatively different from those of ST-DTC.

\section{Prospect of Experimental Implementations}
\label{sec:imp}

Since our model for $n$-DTC and ST-DTC only contains Z-Z interactions and Heisenberg interactions, it can be easily implemented on current noisy quantum computers, such as on superconducting qubits, ion traps, and perhaps Rydberg atom systems. Although the power-law terms of $H_{\rm int}$ in our simulation require long-range couplings, which are relatively hard to implement in many systems such as superconducting qubits, one can choose short-range interactions instead. For example, one can randomly choose $J_{ij}$'s independently from some distribution for $|i-j|\leq 2n$ and set $J_{ij}=0$ for $|i-j| > 2n$, and this is sufficient to break the degeneracies as well. In addition, one can also replace the $\hat{\bm{\sigma}}_i \cdot \hat{\bm{\sigma}}_{i+1} $ with $\sigma^x_i\sigma^x_{i+1}+\sigma^y_i\sigma^y_{i+1}$ to further reduce the circuit depth. Since $\sigma^x_i\sigma^x_{i+1}+\sigma^y_i\sigma^y_{i+1}$ can be obtained from the dipolar couplings in NMR systems, the model can also be potentially realized on NMR systems \cite{rovny2018observation,stasiuk2023observation}.

While simulating ST-DTCs is feasible for current noisy quantum computers, it is hard for classical computers. Ordinary methods such as exact diagonalization suffer from exponentially growing complexity. Other efficient numerical methods, such as matrix product state~\cite{vidal2004efficient,verstraete2008matrix}, are not able to accurately simulate the dynamics of large-scale ST-DTC even for short-time evolution, due to the rapid entanglement growth from the subspace thermalization in the ST-DTC. Therefore, experimental realization of ST-DTCs can explore their large-scale behavior beyond classical simulations.

In our perturbative analysis, with strength $\lambda$ perturbations, the $2\pi/n$ quasi-energy gaps are perturbed at $\Delta^{(n)}\sim O(\lambda^{L/n})$. On the other hand, the average quasi-energy gap $\Delta^{(0)}$ scales like $O(\frac{1}{2^L})$. To maintain a time-crystalline phase, a rough bound~\cite{von2016absolute} for the perturbation strength $\lambda$ is $\Delta^{(n)} \lesssim \Delta^{(0)}$, which leads to $\lambda \lesssim \frac{1}{2^n}$. This is also consistent with our theoretical bound in Eq.~\eqref{eq:lambdaBound}. Thus, larger-$n$ DTCs are less robust and harder to implement. In addition, the noise is usually not $T$-periodic in open systems coupling to environments. As $n$-DTCs are likely unstable against non-periodic noise, this gives further challenges to realize a stable $n$-DTC in experiments, but these are beyond the scope of this work.

\section{Conclusion and discussion}
\label{sec:conlusion}
In this work, we have constructed models to realize $n$-tuple DTCs via permutations on disordered spin-$\frac{1}{2}$ chains. The construction of the model has enabled investigation into the nature of transitions between two different DTCs, and has led to a new phenomenon that we call subspace-thermal DTCs. In the phase transition between $n_1$- and $n_2$-DTC, we have identified the DTC-charges $\{Q_{i,j}\}$ responding subharmonically, with a period $n_G=\gcd (n_1,n_2)$, whereas the states within subspaces characterized by $\{Q_{i,j}\}$ are thermalized. This property is distinct from all existing DTC models. We also numerically verified that the robustness of the ST-DTC is qualitatively different from prethermal-DTCs.

To understand the $n$-DTC and ST-DTC, we have developed a new theoretical framework to analyze the robust subharmonic responses, from the perspective of robust $2\pi/n$ gap instead of MBL. We also provided a novel rigorous proof for the stability of the $2\pi/n$ quasi-energy separation, and gave the explicit condition (LPCPG condition) that a system can be a robust DTC. The proof of stability works for models in this work, but can also apply to generic disordered DTCs, including many existing MBL-DTCs, and thus provides a systematic way to construct new DTC models. We have also introduced the notion of DTC-charges that allow us to probe the response that spontaneously breaks the time-translation symmetry in both the regular DTCs and subspace thermal DTCs. 

In addition, we believe our construction of $n$-tuple DTCs and their phase transitions can be straightforwardly realized in experiments, such as in recent NISQ devices. Our results can also be generalized to higher spin magnitudes or qudits, as well as higher spatial dimensions (see discussion in Appendix~\ref{app:higherSandD}).

However, in our proof, we still require that the LPCPG condition (``local perturbations cannot perturb too globally'') holds in the ST-DTC, which is not rigorously proved in this work. Although our numerics showed its validity for small systems, whether it can persist to thermodynamic limit is still unclear. In future works, one may explore such properties with more thorough perturbative analysis, or construct models that can be rigorously proved to satisfy the LPCPG condition. Either of them can fill the final gap for rigorously proving the existence of ST-DTC.

Finally, we emphasize that our results soften the condition for the stability of general DTCs from the LPPL (``local perturbations perturb locally'') to the LPCPG. Although the LPPL condition can be rigorously proved in static MBL~\cite{imbrie2016many}, whether it rigorously holds in Floquet systems is still unknown, despite the fact that it was invoked in prior works to argue for the stability of DTCs (see, e.g., Ref.~\cite{else2016floquet,von2016absolute}). The weaker LPCPG condition already enables us to go beyond the MBL region to construct potential new stable DTCs. Intuitively, our proof states that if one can find a Floquet system that weakly breaks the ergodicity and such breaking is robust against local perturbations (i.e., it satisfies the LPCPG condition), then one can easily construct a stable DTC model upon it.

\section{Acknowledgements}

The numerical simulations for the time evolution of $A(NT)$ in Fig.~\ref{fig:AandFT} is performed using dynamite package~\cite{gregory_d_kahanamoku_meyer_2024_10906046}.
This work was partly supported by the U.S. Department of Energy, Office
of Science, National Quantum Information Science Research
Centers, Co-design Center for Quantum Advantage (C2QA)
under Contract No. DE-SC0012704, in
particularly on the part of the algorithmic design of spin swapping, and by the National Science Foundation under Grant No. PHY 2310614, in particular, on the part of the stability of discrete time crystals.

\bibliography{nDTC}

\clearpage
\appendix

\section{The $n$-period permutation in the unperturbed case}\label{app:n-period}

\begin{figure}[h]
    \centering
    \includegraphics[width=0.9\linewidth]{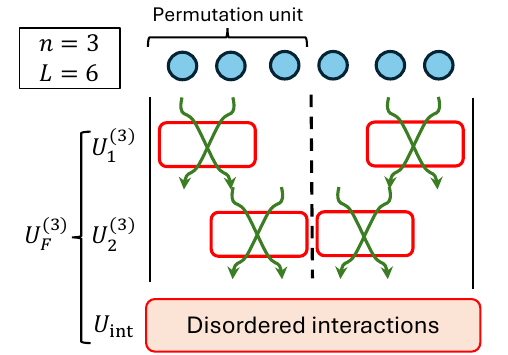}
    \caption{Illustration for the unperturbed $3$-DTC.
    }
    \label{fig:3-DTC}
\end{figure}

Here we prove that for unperturbed Floquet driving $U_F^{(n)}$ (all $\epsilon^x,\epsilon_1,\epsilon_2$ being equal to 0), all states $\ket{z}\equiv\ket{z_1 z_2...z_L}$ in $\sigma^z$ basis evolve back to itself after $n$ periods $(U_F^{(n)})^n \ket{z}= e^{i\varphi_z}\ket{z}$, up to a negligible global phase. We note that $U_{\rm int}$ only adds a phase to the $Z$-basis states, and the SWAP gates $U_1^{(n)}$ and $U_2^{(n)}$ only transform one $Z$-basis state to another. Thus, we need only consider the effects of SWAP gates on $Z$-basis states. For a spin chain with length $L$, the SWAP gates have the same effect for all $n$-site spatial permutation units (up to a reverse order of the spins for odd $n$). Therefore, we can further reduce the problem of finding the period of $U_F^{(n)}$ to the period of SWAP gates acting on a single $n$-site spatial permutation unit.

We use an even $n$ as an example (the proof for odd $n$ is quite similar). To start with, we consider the effects of the SWAP gates as permuting the spatial positions of spins instead of transforming among states. The whole effects of the two-layer SWAP gates $U_1^{(n)}$ and $U_2^{(n)}$ are the following:
\begin{quote}
    1. For an odd site spin $q_i$ ($i$ labeling the position inside a unit beginning with 1 and ending with $n$) with $i+2\le n$, it is permuted to the new site $q_{i+2}$;\\
    2. For an even site spin $q_i$ with $i> 2$, it is permuted to the new site $q_{i-2}$;\\
    3. If the permutation from site $i$ to $i\pm 2$ is out of the unit boundary (i.e., $i+2>n$ or $i-2<1$), then the $q_i$ is permuted to the boundary (i.e., $q_n$ or $q_1$, respectively).
\end{quote}
For an odd site spin $q_i$, it will firstly take $\frac{n+1-i}{2}$ steps to go to the boundary $q_n$, then take $\frac{n}{2}$ steps to go to the boundary $q_1$, and finally take $\frac{i-1}{2}$ steps to go back to itself $q_i$. The total number of steps is then $n$. Similarly, one can find that it takes $n$ steps as well for even site spin $q_i$ to return to itself. Thus, we can conclude that, for even $n$, $(U_F^{(n)})^n \ket{z}= e^{i\varphi_z}\ket{z}$. With similar arguments, one can confirm the statement is true for odd $n$ as well (see Fig.~\ref{fig:3-DTC} for $3$-DTC as an example).

We remark that, although all $Z$-basis states satisfy $(U_F^{(n)})^n \ket{z}= e^{i\varphi_z}\ket{z}$, some of the states may also satisfy $(U_F^{(n)})^k \ket{z}= e^{i\varphi_z'}\ket{z}$, where $k$ can be any divisor of $n$. Take $k=n/2$ for even $n$ as an example, where the $\frac{n}{2}$-step evolution takes spin $q_i$ to $q_{n+1-i}$. If we choose the initial state $\ket{z}\equiv\ket{z_1 z_2...z_L}$ satisfying $z_i=z_{n+1-i}$ for any $i$ inside a unit, then the state can return to itself after $n/2$ steps of evolution $(U_F^{(n)})^{n/2} \ket{z}= e^{i\varphi_z'}\ket{z}$. For each pair of $z_i=z_{n+1-i}$, we can choose either $\uparrow \uparrow$ or $\downarrow \downarrow$ for both spins. Thus, there are in total $2^{n/2}$ states inside a unit satisfying $(U_F^{(n)})^{n/2} \ket{z}= e^{i\varphi_z'}\ket{z}$.

For arbitrary $k|n$, in order to satisfy $(U_F^{(n)})^{k} \ket{z}= e^{i\varphi_z'}\ket{z}$, one requires similar equal-spin chains for the initial states (e.g., $z_i=z_{i+2k}=z_{i+4k}=...$ for odd $i$ and even $n$). All chains contain $n/k$ spins, and there are in total $k$ number of such chains. Within each chain, one can set spins to be all up or all down. Thus, the total number of the states satisfying $(U_F^{(n)})^{k} \ket{z}= e^{i\varphi_z'}\ket{z}$ is $2^k$. We remark that the number $2^k$ also contains the states with smallest period $k'$ less than $k$, where $k'$ divides $k$. We label the states $\ket{z}$ with smallest period $k$ as $\ket{z^{(k)}}$.

Within one single spatial unit, the number $C(k)$ of states $\ket{z^{(k)}}$ with smallest period $k$ under the evolution of $U_F^{(n)}$ can be obtained recursively
\begin{equation}
    C(k)=2^k-\sum_{k'<k,k'|k} C(k'),
\end{equation}
where $k$ must divide $n$. Note that $C(k)$ value remains the same even if the Floquet period $n$ changes, as long as the condition $k|n$ is satisfied.  One can easily check that $C(1)=2$ and $C(p)=2^p-2$ for prime number $p$. We can then estimate the lower bound on the number of states $\ket{z^{(n)}}$:
\begin{equation}\label{eq:Cn>}
    C(n)\geq 2^n-\sum_{k'<n,k'|n} 2^{n/2}\geq 2^n-\frac{n}{2}2^{n/2}.
\end{equation}
Thus, the density of the $n$-period states for one single permutation unit is exponentially close to 1 when $n$ becomes sufficiently large. For the whole $L$-site chain with fixed $n$, the states $\ket{z}$ is $n$-period as long as any of the spatial permutation units is $n$-period. Therefore, we can conclude that the density of states with a period less than $n$ is $(1-C(n)/2^n)^{L/n}$, which becomes exponentially small as $L$ grows. Next, we further prove that almost every $Z$-basis state contains at least $O (L)$ number of $n$-period permutation unit, with exponentially small exceptions.

\subsection{Almost every $Z$-basis state contains at least $\gamma L$ number of $n$-period permutation unit}\label{app:l=O(L)}

The number of states containing $l$ number of $n$-period permutation units on an $L$-site spin chain is $\binom{L/n}{l} C(n)^l [2^n-C(n)]^{\frac{L}{n}-l}$, then the density of states containing less than $\gamma L$ (with an apparent constraint $0<\gamma < 1/n$) number $\mathcal{N}$ of $n$-period permutation units is 
\begin{eqnarray}\label{eq:gamma=app}
\nonumber
    \frac{\mathcal{N}(l\leq \gamma L)}{2^L}&=&\sum_{l=0}^{\gamma L}\binom{L/n}{l} \left(\frac{C(n)}{2^n}\right)^l \left(1-\frac{C(n)}{2^n}\right)^{\frac{L}{n}-l}\\
    &\leq& \exp\left(-2\frac{L}{n}\left(\frac{C(n)}{2^n}-n\gamma \right)^2\right),
\end{eqnarray}
where the inequality comes from the Hoeffding's inequality for binomial distribution, which is valid only for $n\gamma<\frac{C(n)}{2^n}$. Since we know $\frac{C(n)}{2^n}\geq \frac{1}{2}$ from Eq.~\eqref{eq:Cn>}, if we choose $\gamma=\frac{1}{3 n}$ (or any constant smaller than $\frac{C(n)}{n2^n}$), the density of states containing less than $\gamma L$ number of $n$-period permutation units decays exponentially with $L$. In other words, when $L$ is sufficiently large, almost every state contains $l\sim O(L)$ number of $n$-period permutation units.

\section{Brief review of the quantum Floquet system and analytical results for solvable cases}
\label{app:ExactE}

\subsection{Brief review of the quantum Floquet system}
For a Floquet system driven by a period $T$ Hamiltonian $H(t)=H(t+T)$, the states evolve according to the time-dependent Schr\"odinger Equation
\begin{equation}\label{eq:shrdg}
i \frac{\partial}{\partial t}\ket{\psi(t)}=H(t)\ket{\psi(t)},
\end{equation}
or equivalently according to the evolution operator $\hat{U}(t', t) \ket{\psi(t)}=\ket{\psi(t')}$, with 
\begin{equation}
\hat{U}(t', t)=\mathcal{T}\mathrm{exp}\left(-i \int_{t}^{t'} H(s) \mathrm{d} s\right),
\end{equation}
where $\mathcal{T}$ denotes the integral is time-ordered. The Floquet theorem states that~\cite{shirley1965solution} the eigenstate solution $\ket{\psi_j(t)}$ of the Eq.~\eqref{eq:shrdg} takes the form of $\ket{\psi_j(t)}\equiv e^{-i \varepsilon_j t}\ket{\phi_j(t)}$, where $\varepsilon_j$ is the quasi-energy and the Floquet mode $\ket{\phi_j(t)}$ is a periodic wavefunction $\ket{\phi_j(t+T)}=\ket{\phi_j(t)}$. The evolution of $\ket{\phi_j (t)}$ is then
\begin{equation}
\left(H(t)-i \frac{\partial}{\partial t}\right)\ket{\phi_j (t)}=\varepsilon_j\ket{\phi_j(t)}.
\end{equation}
Suppose one has an initial state $\ket{\psi(0)}=\sum_j a_j \ket{\phi_j(0)}$, then the evolving state can be written as the combined evolution of eigenstates
\begin{equation}
    \ket{\psi(t)}=\sum_j a_j e^{-i\varepsilon_j t}\ket{\phi_j(t)},
\end{equation}
which is similar to the time-independent Hamiltonian case.

If one only cares about the stroboscopic properties of the states $\ket{\psi(t)}$ at $t=N T$, one can use an effective Hamiltonian $H_{\rm eff}$ according to $e^{-i H_{\rm eff} T}=U_F\equiv\hat{U}(T, 0)$ to model the evolution of the state $\ket{\psi(t)}$:
\begin{equation}\label{eq:strobo}
\ket{\psi(NT)}=e^{-i H_{\rm eff} NT}\ket{\psi(0)}=\sum_j a_je^{-i \varepsilon_j NT}\ket{\phi_j(0)},
\end{equation}
where the quasi-energies $\varepsilon_j$'s and the Floquet modes $\ket{\phi_j(0)}$'s become eigenenergies and eigenstates of $H_{\rm eff}$.

\subsection{Analytical results for the unperturbed $n$-DTC}\label{app:ExactUn}
Here, we set the evolution time $t_3=1$ and absorb the period $T$ into the definition of $H_{\rm eff}$ for convenience (i.e., $e^{-i H_{\rm eff}}\equiv U_F$). In the unperturbed case, i.e., when all $\epsilon^x,\epsilon_1,\epsilon_2$ are equal to 0, the eigenstates $\ket{z}\equiv\ket{z_1 z_2...z_L}$ of $H_{\rm int}$ are in $\sigma^z$ basis, with corresponding energies being
\begin{equation}
    E_{\rm int}(z)=\sum_{i<j} J_{ij} z_i z_{j}+\sum_{i=1}^L h_i^z z_i,
\end{equation}
where $z_{L+1}=z_1$ represents the spatial periodic boundary condition (otherwise, the site $L+1$ does not exist for the open boundary condition). Within the $\alpha$-th subspace of size $k$, the Floquet operator will transform one $Z$-basis state to another cyclically as follows
\begin{equation}\label{eq:zm}
    U_F\ket{z^{(k)}_{\alpha,m}}=e^{-i E_{\rm int}(z^{(k)}_{\alpha,{m+1}})}\ket{z^{(k)}_{\alpha,{m+1}}},
\end{equation}
where the index addition $m+1$ means $(m+1 \mod k)$ and the state goes back to itself after $k$ times of evolution $\ket{z^{(k)}_{\alpha,{m}}}=\ket{z^{(k)}_{\alpha,{m+k}}}$. Note that the exponential on the r.h.s. is associated with $z_{\alpha,m+1}$ rather than $z_{\alpha,m}$, as the Floquet operation, in one period of time, starts with the swap operations, followed by the interaction term. Thus, for any state within the sector, $U_F$ only acts nontrivially within the subspace spanned by $\{\ket{z^{(k)}_{\alpha,m}}\}$, and we label the subspace of the full Hilbert space as the $\alpha$-th sector. Diagonalizing the matrix  corresponding 
 to $U_F$ within this subspace exactly, we get eigenstates $\phi_{\alpha,j}^{(k)}$ ($j=1,2,...,k$) of $U_F$,
\begin{eqnarray}\label{eq:eState}
    \ket{\phi_{\alpha,j}^{(k)}}&=&\frac{1}{\sqrt{k}}\sum_{m=1}^k e^{i (\theta^{(k)}_{\alpha,m}+mj\frac{2\pi}{k}) } \ket{z^{(k)}_{\alpha,m}},\\
    \theta^{(k)}_{\alpha,m}&=&\frac{1}{k}\sum_{i=1}^k (m-i\mod k)E_{\rm int}(z^{(k)}_{\alpha,i}),
\end{eqnarray}
with  the corresponding quasi-energy satisfying 
\begin{eqnarray}
    U_F\ket{\phi_{\alpha,j}^{(k)}}&=&e^{-i \varepsilon_{\alpha,j}^{(k)}}\ket{\phi_{\alpha,j}^{(k)}},\\
    \varepsilon_{\alpha,j}^{(k)}&=&\frac{1}{k}\sum_{m=1}^k E_{\rm int}(z^{(k)}_{\alpha,m}) + j \frac{2\pi}{k}.
\end{eqnarray}
We see clearly that the quasi-energies $\varepsilon_{\alpha,j}^{(k)}$ in each sector are separated with an exact quasi-energy gap $2\pi/k$. The above results can be easily verified by applying $U_F$ directly on the eigenstate \eqref{eq:eState} according to \eqref{eq:zm}. For convenience, we also define $E_\alpha\equiv \sum_{m=1}^{k_\alpha} E_{\rm int}(z^{(k_\alpha)}_{\alpha,m})$.

\subsection{Analytical results for subpace-thermal-DTC in the phase transition in the unperturbed case}\label{app:ExactU(s)}

The change of the system during the phase transition from $n_1$-DTC to $n_2$-DTC is governed by the following Hamiltonian:
\begin{gather}
    H(s,t)= \begin{cases}
    H_{1}^{(n_1\rightarrow n_2)}(s), & \text { for } 0 \leq t<t_1, \\ 
    H_{2}^{(n_1\rightarrow n_2)}(s), & \text { for } t_1 \leq t<t_1+t_2,\\
    H_{\rm int}, & \text { for } t_1+t_2 \leq t<T,
    \end{cases}
\end{gather}
where $T=t_1+t_2+t_3$, and
\begin{gather}
    H_{1}^{(n_1\rightarrow n_2)}(s)=(1-s)H_{1}^{(n_1)}+sH_{1}^{(n_2)},\\
    H_{2}^{(n_1\rightarrow n_2)}(s)=(1-s)H_{2}^{(n_1)}+sH_{2}^{(n_2)},
\end{gather}
where $s$ changes from 0 to 1. We define the corresponding Floquet operator at a certain $s$ as $U_F^{(n_1\rightarrow n_2)}(s)\equiv e^{-i H_{\rm int} t_3}e^{-i H_{2}^{(n_1\rightarrow n_2)}(s)t_2}e^{-i H_{1}^{(n_1\rightarrow n_2)}(s)t_1}$. For simplicity, we will omit the superscript and use $U_F(s)$ to refer to the transition Floquet operator in this section when there is no ambiguity.

\smallskip\noindent{\bf I. Case $n_1$ divides $n_2$}. Let us first consider the case where $n_1$ is a proper divisor of $n_2$ (i.e., $n_1 | n_2, \mbox{with}\, n_1\neq n_2$). We first rewrite the $H_{1}^{(n_1\rightarrow n_2)}(s)$ and $H_{2}^{(n_1\rightarrow n_2)}(s)$ as
\begin{gather}
    H_{1}^{(n_1\rightarrow n_2)}(s)=H_{1}^{(n_1)}+sH_{\rm odd}^{(n_1\rightarrow n_2)},\\
    H_{2}^{(n_1\rightarrow n_2)}(s)=H_{2}^{(n_1)}+sH_{\rm even}^{(n_1\rightarrow n_2)}.
\end{gather}
For simplicity, we omit all the superscripts $(n_1\rightarrow n_2)$ in the Hamiltonians from now on in this section. One can verify that, $H_{\rm odd}$ and $H_{\rm even}$ can be written as
\begin{equation}
\begin{aligned}
    H_{\rm odd}&=\frac{\pi}{2 t_1} \sum_{n_1|j,n_2\nmid j,2\nmid j} \hat{\bm{\sigma}}_{j}\cdot \hat{\bm{\sigma}}_{j+1},\\
    H_{\rm even}&=\frac{\pi}{2 t_2} \sum_{n_1|j,n_2\nmid j,2 |j} \hat{\bm{\sigma}}_{j}\cdot \hat{\bm{\sigma}}_{j+1},
\end{aligned}  
\end{equation}
where we see the change in the Hamiltonian lies in the interactions between boundaries of $n_1$-units {\it within} every $n_2$-unit.

As the system changes similarly within each $n_2$ unit, for convenience, we relabel the spins for each $n_2$-unit as follows. The first site in the $j$-th $n_2$-unit is labeled as $\sigma_{i,1}\equiv\sigma_{(i-1)n_2+1}$. As  $U_F^{(n_2)}$ successively permutes this site through the rest (not in the spatial order), we can relabel the rest of the spins in this unit via 
\begin{equation}
   \sigma^z_{i,j} \equiv \left(U_F^{(n_2)}\right)^{j-1}\circ (\sigma^z_{i,1}),
\end{equation}
where $1\leq i\leq \frac{L}{n_2},1\leq j\leq n_2$ and we have defined the unitary transformation by $U$ on $\sigma$ as $U\circ (\sigma)=U \sigma U^{\dagger}$. The index $j$ will then visit all the spins in the $n_2$-unit region according to the rule stated in Appendix~\ref{app:n-period}. 

With the above definition, we can then define the oscillating DTC-charges $Q_{i,m}^{(n_1\rightarrow n_2)}$ within each $n_2$-unit as
\begin{gather}
    Q_{i,m}^{(n_1\rightarrow n_2)}\equiv\sum_{j=1}^{n_2/n_1} \left(U_F^{(n_2)}\right)^{n_1 (j-1)}\circ (\sigma^z_{i,m}),
\end{gather}
where $1\leq m\leq n_1$. We will use the symbol $Q_{i,m}\equiv Q_{i,m}^{(n_1\rightarrow n_2)}$ for simplicity from now on in this section. One can easily verify that $(U_{F}^{(n_1)})^{n_1}$ (which is proportional to a pure phase in each block) and $(U_{F}^{(n_2)})^{n_1}$ both commute with all $Q_{i,m}$'s and that both $U_F$'s transforms one $Q_{i,m}$ to the next $Q_{i,m+1}$, i.e.,
\begin{equation}
    \begin{aligned}
        U_{F}^{(n_1)} \circ (Q_{i,m})=Q_{i,m+1},\\
        U_{F}^{(n_2)} \circ (Q_{i,m})=Q_{i,m+1}.\\
    \end{aligned}
\end{equation}

One can further verify that, if a $n_1$-unit boundary spin $\sigma^z_{k n_1}$ (with $k n_1$ representing the spatial location) appears in some $Q_{i,m}$, so does $\sigma^z_{k n_1+1}$. As $\sigma^z_j+\sigma^z_{j+1}$ commutes with $\hat{\bm{\sigma}}_{j}\cdot \hat{\bm{\sigma}}_{j+1}$, we can conclude that all $Q_{i,m}$'s commute with $H_{\rm odd}$ and $H_{\rm even}$. As $H_{\rm odd}$ and $H_{\rm even}$ also commute with $H_1^{(n_1)}$ and $H_2^{(n_1)}$ (or $H_1^{(n_2)}$ and $H_2^{(n_2)}$), respectively, all $Q_{i,m}$'s are transformed by $U_F^{(n_1\rightarrow n_2)}(s)$, in the same way as $U_F^{(n_1)}$ or $U_F^{(n_2)}$:
\begin{equation}
    U_{F}(s) \circ (Q_{i,m})=Q_{i,m+1}.
\end{equation}

Since all $Q_{i,j}$'s commute with $[U_F(s)]^{n_1}$ for arbitrary $s$, the eigenstates of $[U_F(s)]^{n_1}$ can be written in subspaces characterized by distinct $Q_{i,j}$ values, $\ket{\chi(Q_{1,1}=q_{1,1},...,Q_{L/n_2,n_1}=q_{L/n_2,n_1})}$. We then have the relation
\begin{multline}
    [U_F(s)]^{n_1}\ket{\chi_{\alpha,m}(\{Q_{i,j}=q_{i,j}\})}=\\
    e^{-i E_\alpha(s)}\ket{\chi_{\alpha,m}(\{Q_{i,j}=q_{i,j}\})},
\end{multline}
and we can define
\begin{multline}
    U_F(s)\ket{\chi_{\alpha,m}(\{Q_{i,j}=q_{i,j}\})}\equiv\\
    \ket{\chi_{\alpha,m+1}(\{Q_{i,j+1}=q_{i,j}\})},
\end{multline}
where $1\leq m < n_1$, and we denote the subspace spanned by $\ket{\chi_{\alpha,m}}$'s as the $\alpha$-th sector of the whole Hilbert space. We have absorbed possible phase terms into the definition of $\ket{\chi_{\alpha,m+1}}$, except for $m=n_1$, which is
\begin{multline}
    U_F(s)\ket{\chi_{\alpha,n_1}(\{Q_{i,j}=q_{i,j}\})}=\\
    e^{-i E_\alpha}\ket{\chi_{\alpha,1}(\{Q_{i,j+1}=q_{i,j}\})},
\end{multline}
where we use the fact that $\ket{\chi_{\alpha,1}}$ is an eigenstate of $[U_F(s)]^{n_1}$. One can verify that, all $\ket{\chi_{\alpha,m}}$'s are eigenstates of $[U_F(s)]^{n_1}$, with the same quasi-energy $ E_\alpha$. Furthermore, when $U_F(s)$ acts on the eigenstates: $U_F(s)\ket{\chi_{\alpha,m\ne n_1}}=\ket{\chi_{\alpha,m+1}}$ and $U_F(s)\ket{\chi_{\alpha,n_1}}=e^{-iE_\alpha}\ket{\chi_{\alpha,1}}$, it permutes the DTC-charges by $Q_{i,j}\rightarrow Q_{i,j+1}$, which can be {\it experimentally} observed by measuring $Q_{i,j}$'s. All $Q_{i,j}$'s are permuted back to themselves after $n_1$ periods of driving by $U_F(s)$. We note that some configurations may have a smaller period $k$ than $n_1$, with $k|n_1$, which happens when $Q_{i,j}=Q_{i,j+Nk}$ for any integer $N$. With a similar derivation to Appendix~\ref{app:n-period}, one can verify that densities of those smaller-period states are exponentially small with $n_1$ in a single $n_2$-unit and are exponentially small with $L$ in the whole spin chain.

\smallskip\noindent{\bf II. General $n_1$ and $n_2$}. For general $n_1,n_2$, by considering a similar procedure, one can find DTC-charges $Q_{i,j}$, with $1\leq j\leq \gcd (n_1,n_2)\equiv n_G$, in the $i$-th $n_L\equiv\mathrm{lcm} (n_1,n_2)$-site permutation units. Similar to the above divisible case, we relabel the spins for each $n_L$ unit $\sigma_{i,1}\equiv\sigma_{(i-1)n_L+1}$
and
\begin{equation}
   \sigma^z_{i,j} \equiv \left(U_F^{(n_L)}\right)^{j-1}\circ (\sigma^z_{i,1}),
\end{equation}
where $1\leq i\leq \frac{L}{n_L}$ and $1\leq j\leq n_L$, and the DTC-charges are defined similarly as
\begin{gather}
    Q_{i,m}^{(n_1\rightarrow n_2)}\equiv\sum_{j=1}^{n_L/n_G} \left(U_F^{(n_L)}\right)^{n_G (j-1)}\circ (\sigma^z_{i,m}),
\end{gather}
where $1\leq m\leq n_G$. Notice that when $n_1=n_2$, the DTC charges $Q$' s reduce to $\sigma^z$' s. In addition, as $Q_{i,m}$'s oscillate with period $n_G$, in case $n_1$ and $n_2$ are coprime ($n_G=1$), the charges $Q_{i,m}$'s no longer have subharmonic oscillations but remain constant.

To prove that the $Q_{i,m}$'s are indeed quasi-symmetries, one can verify that, in a similar vein as above, all $Q_{i,m}$'s commute with every $\hat{\bm{\sigma}}_{j}\cdot \hat{\bm{\sigma}}_{j+1}$ interaction in $H_{\rm odd }^{(n_G\rightarrow n_L)}$ and in $H_{\rm even }^{(n_G\rightarrow n_L)}$, respectively, and that $H_{\rm odd }^{(n_G\rightarrow n_L)}$ and $H_{\rm even }^{(n_G\rightarrow n_L)}$ contain all the interaction terms in $H_{\rm odd }^{(n_1\rightarrow n_2)}$ and $H_{\rm even }^{(n_1\rightarrow n_2)}$, respectively. Thus, all $Q_{i,m}$'s transformed by $U_F^{(n_1\rightarrow n_2)}(s)\equiv U_F(s)$, in a similar manner to $U_F^{(n_1)}$ and $U_F^{(n_2)}$
\begin{equation}
    U_{F}(s) \circ (Q_{i,m})=Q_{i,m+1},
\end{equation}
where $Q_{i,n_G+1}\equiv Q_{i,1}$. We can also define the eigenstates of $[U_F(s)]^{n_G}$ in a certain $\{Q_{i,j}\}$-subspace as $\ket{\chi_{\alpha,1}(\{Q_{i,j}=q_{i,j}\})}$, and obtain the relation
\begin{multline}
    U_F(s)\ket{\chi_{\alpha,m}(\{Q_{i,j}=q_{i,j}\})}\equiv\\
    \ket{\chi_{\alpha,m+1}(\{Q_{i,j+1}=q_{i,j}\})},
\end{multline}
where $1\leq m < n_G$, and we denote the subspace spanned by $\ket{\chi_{\alpha,m}}$'s as the $\alpha$-th sector of the whole Hilbert space. We have absorbed possible phase terms into the definition of $\ket{\chi_{\alpha,m+1}}$, except for $m=n_G$, which is
\begin{multline}
    U_F(s)\ket{\chi_{\alpha,n_G}(\{Q_{i,j}=q_{i,j}\})}=\\
    e^{-i E_\alpha(s)}\ket{\chi_{\alpha,1}(\{Q_{i,j+1}=q_{i,j}\})}.
\end{multline}

We have shown that all states in the Hilbert space evolve back to themselves with $n_G$-period evolution, with only an exponentially small portion of smaller-period states. For $n_G$-period states, the configurations of DTC-charges are different for different $\ket{\chi_{\alpha,m}}$'s, and $U_F(s)$ transforms such a state in one $\{Q_{i,j}\}$-subspace to another subspace, which indicates the states are orthogonal to each other. Thus, by a similar derivation to that in Appendix~\ref{app:ExactUn}, we obtain the eigenstates of $U_F(s)$
\begin{eqnarray}\label{eq:ST-eStateInapp}
    \ket{\Phi_{\alpha,j}}&=&\frac{1}{\sqrt{n_G}}\sum_{m=1}^{n_G} e^{i (\theta_{\alpha,m}+mj\frac{2\pi}{n_G}) } \ket{\chi_{\alpha,m}},\\
    \theta_{\alpha,m}&=&\frac{1}{n_G}  (m-1)E_{\alpha},
\end{eqnarray}
with  the corresponding state and quasi-energy satisfying 
\begin{eqnarray}
    U_F(s)\ket{\Phi_{\alpha,j}}&=&e^{-i \varepsilon_{\alpha,j}}\ket{\Phi_{\alpha,j}},\\
    \varepsilon_{\alpha,j}&=& \frac{E_\alpha}{n_G} + j \frac{2\pi}{n_G}.
\end{eqnarray}
We see clearly that the quasi-energies $\varepsilon_{\alpha,j}$ for each sector are separated with an exact energy gap $2\pi/n_G$. Suppose one prepares an initial $n_G$-period $Z$-basis state $\ket{z}$ with $Q_{i,j}=q_{i,j}$, it can be written as a superposition of eigenstates in ($Q_{i,j}=q_{i,j}$)-subspace:
\begin{equation}
    \ket{z(Q_{i,j}=q_{i,j})}=\sum_{k} a_k \ket{\chi_{\alpha_k,m_k}(Q_{i,j}=q_{i,j})},
\end{equation}
where we have used the fact that each $\alpha_k$-th sector contains only one $\ket{\chi}$ that has exactly the same $Q_{i,j}$ configurations as the $\ket{z(\{Q_{i,j}=q_{i,j}\})}$. After one period, the state evolve to
\begin{multline}
    U_F(s)\ket{z(Q_{i,j}=q_{i,j})}=\\
    \sum_{k} a_k e^{-i\varphi_k} \ket{\chi_{\alpha_k,m_k+1}(Q_{i,j+1}=q_{i,j})},
\end{multline}
where one can clearly see the oscillation of $Q_{i,j}$'s. After $n_G$ periods of evolution, the state evolves to
\begin{multline}
    [U_F(s)]^{n_G}\ket{z(Q_{i,j}=q_{i,j})}=\\
    \sum_{k} a_k e^{-i E_{\alpha_k}} \ket{\chi_{\alpha_k,m_k}(Q_{i,j}=q_{i,j})} .
\end{multline}
Due to the randomness in $e^{-iE_{\alpha_k}}$'s, $[U_F(s)]^{n_G}\ket{z}$ gets thermalized and is no longer a $Z$-basis state. However, as all components of the outcome state share the same $Q_{i,j}$ values, the thermalization is confined within the subspaces characterized by different $Q_{i,j}$ values. Since $Q_{i,j}$'s are combinations of $\sigma^z$'s, if one measures the expectations of $\sigma^z$'s, one will find that they quickly become evenly distributed within each $Q_{i,j}=q_{i,j}$. As $Q_{i,j}$'s are oscillating with period-$n_G$, all $\sigma^z$'s will eventually oscillate synchronously with corresponding $Q_{i,j}$'s, with amplitudes reduced by a factor $n_G/n_L$.

\section{\textcolor{black}{Proofs of Lemma 1 and 2}}
\label{app:ProofsOfLemma12}
\subsection{Proof of Lemma 1}
\label{app:proofOflemma-1}

\smallskip\noindent{\bf Lemma 1}.
Suppose $\ket{\psi_j}$ is an eigenvector of a unitary $U$, with corresponding eigenvalue $e^{-i E_j}$. $P_\mathbb{V}$ is a projector that projects a state to a subspace $\mathbb{V}$, and we denote the subspace fidelity as $\braket{\psi_j|P_\mathbb{V}|\psi_j}=1-\epsilon^2$ and $\ket{\eta_j}\equiv P_\mathbb{V}\ket{\psi_j}$. Then we have
\begin{equation}
e^{-iE_j}\ket{\eta_j}=P_\mathbb{V}UP_\mathbb{V}\ket{\eta_j}+\ket{\xi},
\end{equation}
where the 2-norm $\|\ket{\xi}\|_2\leq \epsilon$.

Denote $U=\begin{pmatrix}
    A & B\\
    C & D
\end{pmatrix}$ with $A$ acting in the subspace $\mathbb{V}$, and $\ket{\psi_j}=\begin{pmatrix}
    \eta_j\\
    \tau_j
\end{pmatrix}$. It is easy to see that $\|\tau_j\|_2=\epsilon$.
By definition, we have
\begin{equation}
    A \ket{\eta_j}+B\ket{\tau_j}=e^{-i E_j}\ket{\eta_j}.
\end{equation}
Then we can set $\ket{\xi}=B\ket{\tau_j}$. Note that $B$ is a submatrix of a unitary matrix, thus satisfying $\|B\|_2\leq 1$, so it cannot increase the norm of a state. Thus, we have $\|\ket{\xi}\|_2\leq\epsilon$. By definition, $A=P_\mathbb{V}UP_\mathbb{V}$ is the restriction of $U$ in the subspace $\mathbb{V}$, and the lemma is thus proved.

\subsection{Proof of Lemma 2}
\label{app:proofOflemma-2}

\smallskip\noindent{\bf Lemma 2}.
Let $U$ be a unitary and $\ket{\eta}$ be a nonzero vector that is nonzero only in subspace $\mathbb{V}$. Suppose we have
\begin{equation}
e^{-iE}\ket{\eta}=P_\mathbb{V}UP_\mathbb{V}\ket{\eta}+\ket{\xi},
\end{equation}
where the 2-norm $\|\ket{\xi}\|_2= \epsilon$ is small. Then $U$ has a quasi-energy $E_j$ satisfying $|e^{-iE_j}-e^{-iE}|\leq \sqrt{\frac{2\epsilon}{\|\ket{\eta}\|_2}}$.

Denoting $U=\begin{pmatrix}
    A & B\\
    C & D
\end{pmatrix}$ with $A$ acting in the subspace $\mathbb{V}$, and applying it directly to the state $\begin{pmatrix}
    \eta\\
    0
\end{pmatrix}$, we obtain
\begin{equation}
    U\begin{pmatrix}
    \eta\\
    0
\end{pmatrix}=\begin{pmatrix}
    A\eta\\
    C\eta
\end{pmatrix}=\begin{pmatrix}
    e^{-iE}\eta-\xi\\
    C\eta
\end{pmatrix},
\end{equation}
where the norm of $e^{-iE}\eta-\xi$ is at least $\|\eta\|_2-\epsilon$. Since $U$ does not change the norm of a state, the norm of $C\eta$ is at most $\sqrt{2\epsilon\|\eta\|-\epsilon^2}$. Thus, we have
\begin{equation}
    U\begin{pmatrix}
    \eta\\
    0
\end{pmatrix}=e^{-iE}\begin{pmatrix}
    \eta\\
    0
\end{pmatrix}+\begin{pmatrix}
    -\xi\\
    C\eta
\end{pmatrix}.
\end{equation}
Thus, $e^{-iE}$ and $\begin{pmatrix}
    \eta\\
    0
\end{pmatrix}$ are an approximate eigensolution of $U$. By Bauer-Fike theorem~\cite{bauer1960norms}, we can find an eigenvalue $e^{-i E_j}$ of $U$, such that 
\begin{equation}
    |e^{-i E_j}-e^{-i E}|<\kappa_2(X)\frac{\left\|\begin{pmatrix}
    -\xi\\
    C\eta
\end{pmatrix}\right\|_2}{\|\ket{\eta}\|_2}\leq \sqrt{\frac{2\epsilon}{\|\ket{\eta}\|_2}},
\end{equation}
where $\kappa_2$ is the 2-norm condition number and $X$ is the eigenvector matrix of $U$. Since $X$ is unitary, its 2-norm condition number is always 1. Thus, the lemma is proved.

\section{\textcolor{black}{Proof for the robust $2\pi/n$ gap}}
\label{app:perturbGap}
\subsection{Main proof}

The goal of this section is to prove that the $2\pi/n$ gap for the $n$-DTC and the ST-DTC is perturbed at $\Delta^{(n)}\leq e^{-O(L)}$. Here, we will mainly focus on the ST-DTC, as the $n$-DTC is merely a special case of ST-DTC with the subspace dimension equal to 1. We remark that the proof can also be applied to generic DTC models satisfying the LPCPG condition.

Let us begin with the unperturbed Floquet operator from a periodic Hamiltonian $H_0(t+T)=H_0(t)$,
\begin{equation}
U_0(t)=\mathcal{T} e^{-i \int_0^t H_0(t') \mathrm{d} t'},
\end{equation}
and denote  $U_0\equiv U_0(T)$. When a periodic perturbation $\lambda \hat{V}(t+T)=\lambda \hat{V}(t)$ is added to the unperturbed Hamiltonian, the new Floquet operator becomes
\begin{equation}
U_{F}(\lambda)=\mathcal{T} \exp \left(-i \int_0^T\left(H_0(t)+\lambda \hat{V}(t)\right) \mathrm{d} t\right),
\end{equation}
which can also be effectively written as~\cite{else2016floquet}, 
\begin{equation}
\begin{aligned}
U_{F}(\lambda)&=U_0  \mathcal{T} \exp \left(-i \int_0^T\left(U_0\right)^{\dagger}(t) \lambda \hat{V}(t) U_0(t) \mathrm{d} t\right)\\
&\equiv U_0 U_\lambda,
\end{aligned}
\end{equation}
where $U_\lambda= \mathcal{T} \exp \left(-i \int_0^T\left(U_0\right)^{\dagger}(t) \lambda V(t) U_0(t)\mathrm{d} t\right)$. If we only care about the stroboscopic properties, then the effective Hamiltonian is all we need for the stability analysis. Defining $e^{-i H_F(\lambda) }\equiv U_{F}(\lambda)$, $ e^{-i H_F(0) }\equiv U_{F}(0)$, and $e^{-i \lambda V }\equiv U_\lambda$, we have
\begin{equation}
    e^{-i H_F(\lambda) }=e^{-i H_F(0) } e^{-i \lambda V },
\end{equation}
where we absorb the factor $T$ in the effective Hamiltonian for simplicity. It has been shown in the supplemental material of Ref.~\cite{else2016floquet} that any $T$-periodic local perturbation of the Floquet Hamiltonian can be transformed into the local perturbation on the Floquet operator in the form of $U_{F}(\lambda)=U_0e^{-i\lambda V}\equiv e^{-i H_{\rm eff}}e^{-i\lambda V}$, where the $U_{F}(\lambda)$ is the perturbed Floquet operator and $V$ remains local. The statement is true because $U_0(t)$ consists of a finite time evolution of local Hamiltonians, or equivalently, $U_0$ can be transformed into a finite-depth quantum circuit. In this case, we expect that $U_0$ satisfies some Lieb-Robinson bound. As long as $T$ is finite and not too long, the spreading of $V$ from the original $\hat{V}(t)$ will remain local.

For the ST-DTC discussed in the main text, at $\lambda=0$ the quasi-energies $\varepsilon_{\alpha,j}$ and eigenstates $\ket{\Phi_{\alpha,j}}$ can be exactly solved as follows, 
\begin{equation}
\label{eq:eStatesForSTDTC}
\begin{aligned}
    \ket{\Phi_{\alpha,j}^{(n)}}&=\frac{1}{\sqrt{n}}\sum_{m=1}^n e^{i (\theta^{(n)}_{\alpha,m}+m j\frac{2\pi}{n} )} \ket{\chi^{(n)}_{\alpha,m}},\\
    \varepsilon_{\alpha,j}^{(n)}&=\frac{1}{n}\sum_{m=1}^n \varphi_{\alpha,m} + j \frac{2\pi}{n},
\end{aligned}
\end{equation}
where $\theta^{(n)}_{\alpha,m}=\frac{1}{n}\sum_{i=1}^n (m-i\mod n) \varphi_{\alpha,i}$ and we use superscript $(n)$ to denote its $n$-period oscillation. In addition, each $\ket{\chi_{\alpha,m}}$ is also the eigenstate of $Q$'s,
\begin{equation}
Q_{i,j}\ket{\chi_{\alpha,m}}=q_{i,j}^{\alpha,m}\ket{\chi_{\alpha,m}}.
\end{equation}

To characterize the locality of the $V$ for our $n$-DTC model, it is convenient to introduce the concept of unit locality $K_u$. Suppose $V$ is $K$-local and can be written as
\begin{equation}
    V=\sum_i V_i,
\end{equation}
where each $V_i$ couples at most $K$ spins (without any geometric constraints). $K_u$ is defined as the maximum number of coupled permutation units among all $V_i$. It is easy to see $K_u \leq K$. We also define the corresponding {\it unit Hamming distance} between any two $Q$-basis states: $D^{Q}_u(\ket{Q_\alpha},\ket{Q_\beta})\equiv D^{Q}_u(Q_\alpha,Q_\beta)$, which counts how many permutation units having different $Q$ values between states $\ket{Q_\alpha}$ and $\ket{Q_\beta}$. Note that $i$-th permutation unit can have multiple $Q_{i,j}$'s, and two states are the same in $i$-th permutation unit only if $Q_{i,j}$'s are the same for all $j$. Based on these definitions, one can verify that, for a $K_u$-local $V$, the matrix element $\bra{Q_\alpha}V\ket{Q_\beta}=0$ if $K_u < D^{Q}_u(\ket{Q_\alpha},\ket{Q_\beta})$.

In this formulism, we can further define the distance $D_{\alpha,\beta}$ for two unperturbed eigenstates $\ket{\Phi_{\alpha,j}^{(n)}}$ and $\ket{\Phi_{\beta,j'}^{(n')}}$ as the minimum distance among any pair $\ket{\chi_{\alpha,m}}$ and $\ket{\chi_{\beta,m'}}$ in the respective decompositions
\begin{equation}
    D^{Q}_{\alpha,\beta}\equiv\min_{m,m'} D^{Q}_u(\ket{\chi_{\alpha,m}},\ket{\chi_{\beta,m'}}).
\end{equation}
Note that the $D^{Q}_{\alpha,\beta}$ defined in this way is independent of $j,j'$ and $n,n'$, as different eigenstates with fixed $\alpha$ (or $\beta$) have the same set of $|\chi\rangle$'s in their decomosition. For simplicity, we will occasionally omit the superscript $Q$ from now on in this section. We also remark that the two eigenstates $\ket{\Phi_{\alpha,j}^{(n)}}$ and $\ket{\Phi_{\beta,j'}^{(n')}}$ can be globally different in $Z$ basis even if the $D^Q_{\alpha,\beta}=0$, as the distance $D^Q$ compares only $Q$-charges.

One can also verify that for the ST-DTC introduced in the main text, if $D_u(\chi^{(n)}_{\alpha,m},\chi^{(n')}_{\beta,m'})=D_{\alpha,\beta}$, then $D_u(\chi^{(n)}_{\alpha,m+c},\chi^{(n')}_{\beta,m'+c})=D_{\alpha,\beta}$. Thus, we can define the invariant relative index difference $d_{\alpha,\beta}$ to be
\begin{equation}\label{eq:dab=app}
    d_{\alpha,\beta}\equiv m-m'.
\end{equation}
This quantity will be used in the proof shortly.

For simplicity, we assume $V$ is $K_u$-local. In addition, we assume the operator 2-norm of $V$ is no more than an extensive number, so that we can simply set $\|V\|_2\leq L$. We remark that the results should also apply to the case where $\|V_i\|_2$ decays exponentially with the $K_u$-locality of $V_i$. We also assume the perturbation strength $\lambda$ to be sufficiently small but still non-vanishing, namely,
\begin{equation}\label{eq:lambdaBound=app}
    \lambda<\frac{\gamma}{2e K_u }2^{-\frac{2 K_u }{\gamma}},
\end{equation}
where $\gamma$ can be defined from Eq.~\eqref{eq:gamma=app} and the majority of the states satisfy $l_\alpha/L\geq\gamma$. In this way, we find a relation for matrix elements of  $e^{-i\lambda V}$ in the unperturbed eigenbasis, as stated in the following lemma.

\smallskip\noindent{\bf Lemma \thesection.1}. The matrix elements of the $e^{-i\lambda V}$ in the unperturbed eigenbasis satisfy
\begin{multline}\label{eq:VnPhiOffdiagExp}
\left|e^{-ic\frac{2\pi}{n}d_{\alpha,\beta}}\braket{\Phi^{(n)}_{\alpha,j}|e^{-i\lambda V}|\Phi^{(n)}_{\beta,j'}}-\right.\\
\left.\braket{\Phi^{(n)}_{\alpha,j+c}|e^{-i\lambda V}|\Phi^{(n)}_{\beta,j'+c}}
\right|\leq 2R_\lambda\Bigr(\Bigr\lceil \frac{D^{(2)}_{\alpha,\beta}}{K_u} \Bigr\rceil\Bigr).
\end{multline}
where $D^{(2)}_{\alpha,\beta}$ is the second minimum unit Hamming distance between $\ket{\chi^{(n)}_{\alpha,m}}$'s and $\ket{\chi^{(n)}_{\beta,m'}}$ and $R_\lambda(k)$ is a function defined as
\begin{equation}
    R_\lambda(k)\equiv\frac{(\lambda L)^k}{k!}.
\end{equation}
We present the proof in Appendix.~\ref{app:proofOflemma1}. Note that $D^{(2)}_{\alpha,\beta}\geq \max(l_\alpha,l_\beta)-D_{\alpha,\beta}$ and $D^{(2)}_{\alpha,\beta}\geq D_{\alpha,\beta}$. Thus, we have
\begin{equation}
    D^{(2)}_{\alpha,\beta}\geq \frac{\max(l_\alpha,l_\beta)}{2}\sim O(L).
\end{equation}
On the other hand, for large $k$ we have
\begin{equation}
    R_\lambda(k)  \approx \frac{1}{\sqrt{2 \pi k}}\left(\frac{e \lambda L}{k}\right)^k.
\end{equation}
Thus, as long as $\lambda<\frac{1 }{2e K_u}\frac{l_\alpha}{L}$, the difference in \eqref{eq:VnPhiOffdiagExp} is exponentially small. Since for the majority of states, $l_\alpha\geq \gamma L$, $\lambda$ can be nonvanishing for large $L$.

Following the LPCPG condition, we set $\mathbb{V}$ in the unperturbed eigenbasis, spanned by all states $\ket{\Phi_{\beta,j'}}$ with $D^{Q}_{\alpha,\beta}<\frac{l_\alpha}{4}$. We can then define $A=P_\mathbb{V}U_F(\lambda)P_\mathbb{V}$ as the diagonal block of the perturbed Floquet operator. In addition, we define $A_c$ by reordering the matrix elements in $A$
\begin{equation}\label{eq:Acj=j+c=app}
    (A_c)_{\alpha,j;\beta,j'}\equiv A_{\alpha,j+c;\beta,j'+c}.
\end{equation}
Equivalently, $A_c$ can be obtained by a permutation transformation of $A$
\begin{equation}
    A_c=\Pi_c^TA\Pi_c,
\end{equation}
where $\Pi_c$ is the permutation matrix according to the rule \eqref{eq:Acj=j+c=app}. We can find a relation between the $A_c$ and original $A$ according to the following lemma.

\smallskip\noindent{\bf Lemma \thesection.2}. Define $\tilde{A}_c\equiv U_c^\dagger A_c U_c$, with $U_c$ being a diagonal unitary matrix: $(U_c)_{\beta,j';\beta,j'}=e^{ i c \frac{2\pi}{n}d_{\alpha,\beta}}$, where $d_{\alpha,\beta}$ is defined in Eq.~\eqref{eq:dab=app}. Then the following relation is satisfied:
\begin{equation}
    A=e^{ic\frac{2\pi}{n}}\tilde{A}_c+\Delta A_c,
\end{equation}
with 
\begin{equation}
    \|\Delta A_c\|_2\leq 2N_{\mathbb{V}} R_\lambda\Big(\Big\lceil \frac{l_\alpha}{2K_u} \Big\rceil\Big),
\end{equation}
where the $N_{\mathbb{V}}$ is the size of the subspace $\mathbb{V}$. We present the proof in Appendix.~\ref{app:proofOflemma2}. One can verify that if $\lambda$ satisfies the bound \eqref{eq:lambdaBound=app}, then $\|\Delta A_c\|_2$ decays exponentially with $L$, given that $l_\alpha\geq \gamma L$ and $N_{\mathbb{V}}<2^L$. 

When the perturbation is added, we define the corresponding perturbed state as $\ket{\tilde{\Phi}^{(n)}_{\alpha,j}(\lambda)}$ and the perturbed quasi-energy as $E^{(n)}_{\alpha,j}(\lambda)$. Our goal is equivalently to prove that there exists an $E^{(n)}_{\alpha,j+c}(\lambda)$ such that $|\Delta^{(n,c)}_{\alpha,j}|\equiv| E^{(n)}_{\alpha,j}(\lambda)- E^{(n)}_{\alpha,j+c}(\lambda)+c\frac{2\pi}{n}|\sim e^{-O(L)}$. For $\ket{\tilde{\Phi}_{\alpha,j}(\lambda)}$ satisfying the LPCPG condition, we have
\begin{equation}
    1-\braket{\tilde{\Phi}_{\alpha,j}(\lambda)|P_\mathbb{V}|\tilde{\Phi}_{\alpha,j}(\lambda)}=(\epsilon_{\alpha,j})^2,
\end{equation}
where $\epsilon_{\alpha,j}\sim e^{-O(L)}$. Define $\ket{\eta_{\alpha,j}(\lambda)}=P_\mathbb{V}\ket{\tilde{\Phi}_{\alpha,j}(\lambda)}$, from Lemma 1 in the main text we have
\begin{equation}
    A\ket{\eta_{\alpha,j}(\lambda)}=e^{-iE^{(n)}_{\alpha,j}(\lambda)}\ket{\eta_{\alpha,j}(\lambda)}-\ket{\xi_{\alpha,j}},
\end{equation}
where $\|\ket{\xi_{\alpha,j}}\|_2\leq \epsilon_{\alpha,j}$. We can then obtain

\begin{widetext}
\begin{equation}
\begin{aligned}
    \tilde{A}_c\ket{\eta_{\alpha,j}(\lambda)}
    &=e^{-ic\frac{2\pi}{n}}(A-\Delta A_c)\ket{\eta_{\alpha,j}(\lambda)}\\
    &=e^{-i(E^{(n)}_{\alpha,j}(\lambda)+c\frac{2\pi}{n})}\ket{\eta_{\alpha,j}(\lambda)}-e^{-ic\frac{2\pi}{n}}\Delta A_c\ket{\eta_{\alpha,j}(\lambda)}-e^{-ic\frac{2\pi}{n}}\ket{\xi_{\alpha,j}}.
\end{aligned}
\end{equation}
\end{widetext}
Note that $\tilde{A}_c$ is a submatrix of $ \tilde{U}_{F}(\lambda,c)$, which can be defined as 
\begin{equation}
    \tilde{U}_{F}(\lambda,c)=(U_c\otimes I)\circ[(\Pi_c\otimes I)\circ U_F(\lambda)].
\end{equation}
Since both $U_c\otimes I$ and $\Pi_c\otimes I$ are unitary, $\tilde{U}_{F}(\lambda,c)$ shares exactly the same eigenvalues as $U_F(\lambda)$. Using Lemma 2 in the main text, we can find an eigenvalue of $\tilde{U}_{F}(\lambda,c)$ that is close to $e^{-i(E^{(n)}_{\alpha,j}(\lambda)+c\frac{2\pi}{n})}$. This leads to the following theorem:

\smallskip\noindent{\bf Theorem \thesection.1}. For ST-DTCs, the perturbation of the $2\pi/n$ gap is perturbed at $\Delta^{(n)}_{\alpha,j}$, with
\begin{equation}
    |\Delta^{(n,c)}_{\alpha,j}|\leq\frac{\pi}{2} \sqrt{2\left(\|\Delta A_c\|_2+\frac{|\epsilon_{\alpha,j}|}{\sqrt{1-(\epsilon_{\alpha,j})^2}}\right)},
\end{equation}
where we have used $|a-b|<\frac{\pi}{2}|e^{ia}-e^{ib}|$ when $|a-b|<\pi$. From Lemma \thesection.2, we know that $\|\Delta A_c\|_2$ decays exponentially with $L$ if $\lambda$ satisfies the bound \eqref{eq:lambdaBound=app}. In addition, the LPCPG condition states that $\epsilon_{\alpha,j}$ decays exponentially with $L$. Therefore, the perturbation $|\Delta^{(n,c)}_{\alpha,j}|$ for the $2\pi/n$ gap also decays exponentially with $L$. Thus, the robust $2\pi/n$ gap is proved.

\medskip\noindent{\bf Robustness for smaller periods}. The above proof also works for smaller periods $k<n$ for the permutation DTCs introduced in the main text, with a small modification on the definition of $k$-period unit number $l^{(k)}$. Note that the $2\pi/k$ separation appears not only in the states with the smallest period $k$ but also in any states with period $k'$ containing $k$ as a factor. This can be seen by $\frac{2\pi}{k'}\cdot \frac{k'}{k}=\frac{2\pi}{k}$, where $\frac{k'}{k}$ is an integer as we assume $k|k'$. Thus, to count the number of $k$-period local permutation units, we should not only count units with the smallest period $k$ but also with the smallest period $k'$ satisfying $k|k'$. In particular, $k|n$ is always satisfied in the Floquet operator $U_{F}^{(n)}$. Thus, most of the eigenstates of $U_{F}^{(n)}$ also contain $2\pi/k$ quasi-spectral separation with $k|n$.  For a randomly chosen $Z$-basis state in $\alpha$-th sector, the number $l^{(k)}_\alpha$ of $k$-period local permutation units is almost $O(L)$, with exponentially small exceptions. One can verify that, by repeating all the above derivation with small modifications, the bound for the change of $\frac{2\pi}{k}$ spectral separation can be obtained by replacing $l^{(n)}_\alpha$ with $l^{(k)}_\alpha$, which is also exponentially small.

\subsection{Proof of Lemma \thesection.1}
\label{app:proofOflemma1}
\smallskip\noindent{\bf Lemma \thesection.1}. The matrix elements of the $e^{-i\lambda V}$ in the unperturbed eigenbasis satisfy
\begin{multline}
\left|e^{-ic\frac{2\pi}{n}d_{\alpha,\beta}}\braket{\Phi^{(n)}_{\alpha,j}|e^{-i\lambda V}|\Phi^{(n)}_{\beta,j'}}-\right.\\
\left.\braket{\Phi^{(n)}_{\alpha,j+c}|e^{-i\lambda V}|\Phi^{(n)}_{\beta,j'+c}}
\right|\leq 2R_\lambda\Big(\Big\lceil \frac{D^{(2)}_{\alpha,\beta}}{K_u} \Big\rceil\Big).
\end{multline}

Firstly, we list some useful properties regarding the unit Hamming distance as follows:

\smallskip\noindent{\bf Properties regarding unit Hamming distance}. For the unperturbed eigenstates $\ket{\Phi_{\alpha,j}^{(n)}}=\frac{1}{\sqrt{n}}\sum_{m=1}^n e^{i (\theta^{(n)}_{\alpha,m}+mj\frac{2\pi}{n}) } \ket{\chi^{(n)}_{\alpha,m}}$:
\begin{quote}
    Property \thesection.1. The unit Hamming distance $D_u(\ket{\chi_{\alpha,m}^{(n)}},\ket{\chi_{\alpha,m'}^{(n)}})$ between two distinct $Q$-basis components of the same eigenstate $\ket{\Phi_{\alpha,j}^{(n)}}$ is at least $l_\alpha$;\\
    
    \smallskip
    Property \thesection.2. The unit Hamming distance $D_u(\ket{\chi_{\beta,m'}^{(k)}},\ket{\chi_{\alpha,m}^{(n)}})$ between a smaller-period $Q$-basis state $\ket{\chi^{(k)}_{\beta,m'}}$ and any $Q$-basis component $\ket{\chi^{(n)}_{\alpha,m}}$ of the eigenstate $\ket{\Phi_{\alpha,j}^{(n)}}$ is at least $l_\alpha$;\\
    
    \smallskip Property \thesection.3. If the unit Hamming distance between two different $Q$-basis states $D_u(\ket{Q_{\alpha}},\ket{Q_\beta})=d$, then the unit Hamming distance for $D_u(U_F(0)\ket{Q_{\alpha}},U_F(0)\ket{Q_\beta})$ is also $d$, where $U_F(0)$ is the unperturbed Floquet operator. Namely, the unit Hamming distance between any two $Q$-basis states is invariant under the action of $U_F(0)$.
\end{quote}

Next, we prove the lemma by locality analysis for the $K_u$-local perturbation $V$. 

\smallskip\noindent{\bf Relations of diagonal elements in perturbation}. In the following discussion, we assume $l_\alpha \sim O(L)\gg K_u$. We can show the diagonal terms $V_{\alpha,j; \alpha,j}\equiv \braket{\Phi_{\alpha,j}^{(n)}|V|\Phi_{\alpha,j}^{(n)}}$ are always the same for different $j$ by simple calculation
\begin{equation}\label{eq:Perturbdiag}
V_{\alpha,j; \alpha,j}\equiv\braket{\Phi_{\alpha,j}^{(n)}|V|\Phi_{\alpha,j}^{(n)}}=\frac{1}{n}\sum_m \braket{\chi^{(n)}_{\alpha,m}|V|\chi^{(n)}_{\alpha,m}},
\end{equation}
where terms like $\braket{\chi^{(n)}_{\alpha,m}|V|\chi^{(n)}_{\alpha,{m'}}}$ with $m\neq m'$ vanish due to Property \thesection.1 and we see that $V_{\alpha,j ;\alpha,j}$ has no dependence on $j$. A more precise vanishing condition is
\begin{equation}\label{eq:vanishCondition}
    l_\alpha> K_u.
\end{equation}
The results can be extended to diagonal elements of $V^r$, i.e., we have
\begin{equation}\label{eq:Vndiag}
\braket{\Phi_{\alpha,j}^{(n)}|V^r|\Phi_{\alpha,j}^{(n)}}=\frac{1}{n}\sum_m \braket{\chi^{(n)}_{\alpha,m}|V^r|\chi^{(n)}_{\alpha,m}}.
\end{equation}
As $V$ is $K_u$-local, $V^r$ is at most $rK_u$-local. The relation holds for 
\begin{equation}\label{eq:vanishConditionRjj}
    l_\alpha> r K_u.
\end{equation}
Thus, for $r<\lceil l_\alpha/K_u \rceil$, terms like $\braket{\chi^{(n)}_{\alpha,m}|V^r|\chi^{(n)}_{\alpha,{m'}}}$ with $m\neq m'$ always vanish. Since $e^{-i\lambda V} = \sum_{m=0}^\infty\frac{(-i\lambda V)^m}{m!} $, we have
\begin{multline}\label{eq:VndiagTan}
\left|\braket{\Phi_{\alpha,j+c}^{(n)}|e^{-i\lambda V}|\Phi_{\alpha,j+c}^{(n)}}-\braket{\Phi_{\alpha,j}^{(n)}|e^{-i\lambda V}|\Phi_{\alpha,j}^{(n)}}\right|\\
\leq 2
R_{\lambda}\Big(\Big\lceil \frac{l_\alpha}{K_u} \Big\rceil\Big).
\end{multline}
where we define the function $R_\lambda(k)$ as
\begin{equation}
    R_\lambda(k)\equiv\frac{(\lambda L)^k}{k!},
\end{equation}
which is exponentially small if $k>e\lambda L$:
\begin{equation}
    R_\lambda(k)  \approx \frac{1}{\sqrt{2 \pi k}}\left(\frac{e \lambda L}{k}\right)^k.
\end{equation}
The function $R_\lambda(k)$ comes from the 2-norm of the remainder of the expansion $e^{-i\lambda V} = \sum_{m=0}^\infty\frac{(-i\lambda V)^m}{m!} $:
\begin{multline}
    \left\|\sum_{m=k}^\infty\frac{(-i\lambda V)^m}{m!}\right\|_2=\max _{j }\left|\sum_{m=k}^{\infty} \frac{\left(-i \mu_j\right)^m}{m!}\right|\\
    \leq\max_j\frac{|\mu_j |^k}{k!}\leq\frac{(\lambda L)^k}{k!},
\end{multline}
where $\mu_j$'s are eigenvalues of $\lambda V$. By our assumption $\|V\|_2\leq L$, we have $|\mu_j|\leq \lambda L$.

\smallskip\noindent{\bf Relations of off-diagonal elements in perturbation}. For the off-diagonal terms $V_{\alpha,j ; \beta,k}\equiv\braket{\Phi_{\alpha,j}^{(n)}|V|\Phi_{\beta,k}^{(n)}}$, we can also calculate its value explicitly,
\begin{equation}\label{eq:<phialphaVbeta>}
\begin{aligned}
\braket{\Phi_{\alpha,j}^{(n)}|V|\Phi_{\beta,k}^{(n)}}=\frac{1}{n}\sum_{m,m'} &e^{-i(\theta_{\alpha,m}^{(n)}-\theta_{\beta,{m'}}^{(n)}+(mj-m'k)\frac{2\pi}{n})}\\
&~~\boldsymbol{\cdot}\braket{\chi^{(n)}_{\alpha,m}|V|\chi^{(n)}_{\beta,{m'}}}.
\end{aligned}
\end{equation}
The leading order of the term is determined by the minimum unit Hamming distance $D_{\alpha,\beta}$ between all $\chi_{\alpha,m}$ and $\chi_{\beta,m'}$. Suppose one pair of the $Z$-basis terms satisfy the condition $D_u(\chi^{(n)}_{\alpha,m_1},\chi^{(n)}_{\beta,m_2})=D_{\alpha,\beta}\leq K_u$,  for $m'\neq m_2$, we have 
\begin{equation}
\begin{aligned}
D_u(\chi^{(n)}_{\alpha,m_1},\chi^{(n)}_{\beta,{m'}})&\geq D_u(\chi^{(n)}_{\beta,{m_2}},\chi^{(n)}_{\beta,{m'}})-D_u(\chi^{(n)}_{\alpha,m_1},\chi^{(n)}_{\beta,{m_2}})\\
&\geq l_\beta-D_{\alpha,\beta}\geq l_\beta-K_u,
\end{aligned}
\end{equation}
where we have used  Property \thesection.1 in the second inequality. We can also define the second minimum unit Hamming distance as $D_{\alpha,\beta}^{(2)}=\min_{m'\neq m_2}~D_u(\chi^{(n)}_{\alpha,m_1},\chi^{(n)}_{\beta,m'})$, which contributes to the leading order of the correction. One can also use Property \thesection.3 to get the relation $D_u(\chi^{(n)}_{\alpha,m_1},\chi^{(n)}_{\beta,{m'}})=D_u(\chi^{(n)}_{\alpha,m_1+m_2-m'},\chi^{(n)}_{\beta,{m_2}})$ and obtain
\begin{multline}
D_u(\chi^{(n)}_{\alpha,m_1},\chi^{(n)}_{\beta,{m'}})\geq D_u(\chi^{(n)}_{\alpha,{m_1+m_2-m'}},\chi^{(n)}_{\alpha,{m_1}})\\
-D_u(\chi^{(n)}_{\alpha,m_1},\chi^{(n)}_{\beta,{m_2}})
\geq l_\alpha-D_{\alpha,\beta}.
\end{multline}
Thus, in general, we have 
\begin{gather}\label{eq:D2>max-D}
    D_u(\chi^{(n)}_{\alpha,m_1},\chi^{(n)}_{\beta,{m'}})\geq D_{\alpha,\beta}^{(2)}\geq  \mathrm{max}(l_\alpha,l_\beta)-D_{\alpha,\beta}.
\end{gather} 
Together with the definition $D^{(2)}_{\alpha,\beta}\geq D_{\alpha,\beta}$, we have $D^{(2)}_{\alpha,\beta}\geq \lceil\frac{1}{2}\max(l_\alpha,l_\beta)\rceil$. Therefore, terms such as $\braket{\chi^{(n)}_{\alpha,m_1}|V|\chi^{(n)}_{\beta,{m'}}}$, vanish from $m'\neq m_2$ due to the large Hamming distance $l_\alpha -K_u \gg K_u$ between the two states.  The more precise vanishing condition is
\begin{equation}\label{eq:vanishCondition2}
   \mathrm{max}( l_\alpha,l_\beta) > 2 K_u.
\end{equation}
One can also verify that the minimum unit Hamming distance can be reached for different pairs $m$ and $m'$, i.e., $D_u(\chi^{(n)}_{\alpha,m},\chi^{(n)}_{\beta,m'})=D_{\alpha,\beta}$ \textit{iff} $m-m'=m_1-m_2$ by using the Property~\thesection.3. We can then define the invariant relative index difference as $d_{\alpha,\beta}\equiv m_1-m_2$. For arbitrary $m$ and $m'$, the term $\braket{\chi^{(n)}_{\alpha,m}|V|\chi^{(n)}_{\beta,{m'}}}$ is nonvanishing \textit{iff} $m-m'\equiv d_{\alpha, \beta} \pmod n$. From the definition, one sees that $d_{\alpha, \beta}=-d_{\beta, \alpha}$ and $d_{\alpha, \alpha}=0$.
Then Eq.~\eqref{eq:<phialphaVbeta>} can then be simplified to   
\begin{multline}
\label{eq:<phialphaVbetamm>}
\braket{\Phi_{\alpha,j}^{(n)}|V|\Phi_{\beta,k}^{(n)}}=\frac{1}{n}\sum_{m} \braket{\chi^{(n)}_{\alpha,m}|V|\chi^{(n)}_{\beta,{m-d_{\alpha,\beta}}}} \boldsymbol{\cdot} \\
e^{-i\left(\theta_{\alpha,m}^{(n)}-\theta_{\beta,{m-d_{\alpha,\beta}}}^{(n)}+\frac{2\pi}{n}[m(j-k)+d_{\alpha,\beta} k]\right)}.
\end{multline}
Here, we obtain another important relation: 
\begin{equation}\label{eq:<phii>=<phii+c>}
\braket{\Phi_{\alpha,j}^{(n)}|V|\Phi_{\beta,k}^{(n)}}=\braket{\Phi_{\alpha,{j+c}}^{(n)}|V|\Phi_{\beta,{k+c}}^{(n)}}\cdot e^{i  cd_{\alpha,\beta}\frac{2\pi }{n}},
\end{equation}
where the index addition $j+c$ means $(j+c \mod n)$. We note that this relation also holds for $\alpha=\beta$, and it reduces to Eq.~\eqref{eq:Perturbdiag} for the diagonal case: $\{\alpha,j\}=\{\beta,k\}$. We also note that there is a phase ambiguity for each eigenstate $\ket{\Phi_{\alpha,j}}\rightarrow e^{i\varphi}\ket{\Phi_{\alpha,j}}$, which may cause an additional phase difference for $\braket{\Phi_{\alpha,j}^{(n)}|V|\Phi_{\beta,k}^{(n)}}$ and $\braket{\Phi_{\alpha,{j+c}}^{(n)}|V|\Phi_{\beta,{k+c}}^{(n)}}$. This can be resolved by using the exact expression for eigenstates from Eq.~\eqref{eq:eStatesForSTDTC}, where the phase ambiguity is eliminated. 

With a similar derivation, we have
\begin{equation}\label{eq:<phii>=<phii+c>r}
\braket{\Phi_{\alpha,j}^{(n)}|V^r|\Phi_{\beta,k}^{(n)}}=\braket{\Phi_{\alpha,{j+c}}^{(n)}|V^r|\Phi_{\beta,{k+c}}^{(n)}}\cdot e^{i  cd_{\alpha,\beta}\frac{2\pi }{n}},
\end{equation}
where the relation holds for 
\begin{equation}
\mathrm{max}( l_\alpha,l_\beta) > (r+1) K_u.
\end{equation} 
Expanding the $e^{-i\lambda V}$ in power series, we can obtain the following relation
\begin{multline}\label{eq:<phii>=<phii+c>exp}
\left|\braket{\Phi_{\alpha,j}^{(n)}|e^{-i\lambda V}|\Phi_{\beta,k}^{(n)}}\cdot e^{-i  cd_{\alpha,\beta}\frac{2\pi }{n}}-\right.\\
\left.\braket{\Phi_{\alpha,{j+c}}^{(n)}|e^{-i\lambda V}|\Phi_{\beta,{k+c}}^{(n)}}\right| \leq 2R_\lambda\Big(\Big\lceil \frac{D^{(2)}_{\alpha,\beta}}{K_u} \Big\rceil\Big),
\end{multline}
where $D^{(2)}_{\alpha,\beta}$ is the second minimum unit Hamming distance between $Q$-basis components in $\ket{\Phi_{\alpha,j}}$ and $\ket{\Phi_{\beta,j'}}$. Note that the relation can also be applied to the diagonal cases, where $D^{(2)}_{\alpha,\alpha}=l_\alpha$. Thus, the lemma is proved.

\subsection{Proof of Lemma \thesection.2}
\label{app:proofOflemma2}
\smallskip\noindent{\bf Lemma \thesection.2}. Define $\tilde{A}_c\equiv U_c^\dagger A_c U_c$, with $U_c$ being a diagonal unitary matrix: $(U_c)_{\beta,j;\beta,j}=e^{ i c \frac{2\pi}{n}d_{\alpha,\beta}}$. Then the following relation is satisfied:
\begin{equation}
    A=e^{ic\frac{2\pi}{n}}\tilde{A}_c+\Delta A_c,
\end{equation}
with 
\begin{equation}
    \|\Delta A_c\|_2\leq 2N_{\mathbb{V}} R_\lambda\Big(\Big\lceil \frac{l_\alpha}{2K_u} \Big\rceil\Big).
\end{equation}

From Lemma \thesection.1, we know that the magnitudes of the matrix elements in $e^{-i\lambda V}$ and $\Pi_c^Te^{-i\lambda V}\Pi_c$ are almost the same, whereas there are additional relative phases in off-diagonal terms. We now show that the additional phases  can be eliminated by a diagonal unitary matrix $U_{c}$, whose diagonal terms are
\begin{equation}\label{eq:Uc}
    (U_c)_{\beta,m;\beta,m}=e^{ i c \frac{2\pi}{n}d_{\alpha,\beta}}.
\end{equation}

Specifically, the off-diagonal terms for sector $\beta_j$ and $\beta_{j'}$ differ from $e^{ic\frac{2\pi}{n}d_{\beta_j,\beta_{j'}}}$. By definition, $d_{\beta_j,\beta_{j'}}=m_j-m_{j'}$ if $D_u(\chi_{\beta_j,m_j},\chi_{\beta_{j'},m_{j'}})=D_{\beta_j,\beta_{j'}}$. Thus, if we choose $(U_c)_{\beta_j,m;\beta_j,m}=e^{ic\frac{2\pi}{n}m_j}$ and $(U_c)_{\beta_{j'},m;\beta_{j'},m}=e^{ic\frac{2\pi}{n}m_{j'}}$, then the additional phase can be eliminated. In such a way, we can further set $(U_c)_{\beta_{j''},m;\beta_{j''},m}=e^{ic\frac{2\pi}{n}m_{j''}}$ to eliminate the additional phase between sectors $\beta_{j'}$ and $\beta_{j''}$, where $D_u(\chi_{\beta_{j''},m_{j''}},\chi_{\beta_{j'},m_{j'}})=D_{\beta_{j''},\beta_{j'}}$. However, now that the diagonal elements for sectors $\beta_{j}$ and $\beta_{j''}$ are settled, the additional phase between the two sectors can be eliminated only if $D_u(\chi_{\beta_{j''},m_{j''}},\chi_{\beta_{j},m_{j}})=D_{\beta_{j''},\beta_{j}}$. In general, this condition may not hold. We now show that the condition holds if all sectors satisfy $D_{\alpha,\beta}<l_{\alpha}/4$.
In other words, the minimums of all $\min_{m,m'}D_u(\chi_{\beta_{j},m},\chi_{\beta_{j'},m'})$ can be simultaneously reached at some indexes $\{m_{j}\}$ for all $\beta_j$'s if $D_{\alpha,\beta}<l_{\alpha}/4$. 

Without loss of generality, we use $\ket{\chi_{\alpha,1}}$ as the reference for unit Hamming distance $D_u$'s. As we only consider $D_{\alpha,\beta}<\frac{l_{\alpha}}{4}$, the minimum of $\min_m D_u(\chi_{\alpha,1},\chi_{\beta_j,m})$ is reached at a unique index $m$ for all $j$, and we denote the index $m$ as $m_{j}$. For distances of other pairs $\chi_{\beta_j,m_{j}}$ and $\chi_{\beta_{j'},m_{{j'}}}$ with indexes that minimize the $D_u$ with $\chi_{\alpha,1}$, we have
\begin{multline}
    D_u(\chi_{\beta_j,m_{j}},\chi_{\beta_{j'},m_{{j'}}})\leq D_u(\chi_{\alpha,1},\chi_{\beta_{j},m_{{j}}})+\\
    D_u(\chi_{\alpha,1},\chi_{\beta_{j'},m_{{j'}}}) <\frac{l_{\alpha}}{2}.
\end{multline}
For another index $m\neq m_{j}$, we have
\begin{multline}
D_u(\chi_{\beta_j,m},\chi_{\beta_{j'},m_{{j'}}})\geq D_u(\chi_{\alpha,1},\chi_{\beta_{j},m})-\\
    D_u(\chi_{\alpha,1},\chi_{\beta_{j'},m_{{j'}}}) >\frac{l_{\alpha}}{2},
\end{multline}
where we have used $D_u(\chi_{\alpha,1+m_{j}-m},\chi_{\beta_{j},m_{j}})=D_u(\chi_{\alpha,1},\chi_{\beta_{j},m})$ so that
\begin{multline}
D_u(\chi_{\alpha,1},\chi_{\beta_{j},m})\geq D_u(\chi_{\alpha,1},\chi_{\alpha,1+m_{j}-m})-\\ D_u(\chi_{\alpha,1},\chi_{\beta_{j},m_{{j}}})>\frac{3 }{4}l_{\alpha}.
\end{multline}
Thus, all the minimums of $D_u$ between different subspaces are indeed simultaneously reached, which leads to
\begin{equation}
    d_{\beta_j,\beta_{j'}}=m_{j}-m_{{j'}}=d_{\beta_j,\alpha}-d_{\beta_{j'},\alpha}.
\end{equation}
Therefore, the $U_c$ cancels all the off-diagonal phase differences for $e^{-i\lambda V}$ and $\Pi_c^\dagger e^{-i\lambda V}\Pi_c$ in the block $\mathbb{V}\times \mathbb{V}$. Since $\Pi_c$ is defined in the unperturbed eigenbasis, we have
\begin{equation}
    \Pi_c^\dagger P_\mathbb{V}U_F(0)P_\mathbb{V}\Pi_c=e^{-ic\frac{2\pi}{n}}P_\mathbb{V}U_F(0)P_\mathbb{V}.
\end{equation}
As $U_F(0)$ is a diagonal unitary matrix in the unperturbed basis, both $P_\mathbb{V}$ and $U_c$ commute with $U_F(0)$. We can obtain
\begin{equation}
\begin{aligned}
    A&-e^{ic\frac{2\pi}{n}}U^\dagger_c A_c U_c\equiv \Delta A_c\sim \\
    &\begin{pmatrix}
        O(R_\lambda(\lceil\frac{l_{\alpha}}{K_u}\rceil)) & \dots & O(R_\lambda(\lceil\frac{D^{(2)}_{\alpha,\beta}}{K_u}\rceil)) \\
        \vdots & \ddots & \\
        O(R_\lambda(\lceil\frac{D^{(2)}_{\beta,\alpha}}{K_u}\rceil)) &        & O(R_\lambda(\lceil\frac{l_{\beta}}{K_u}\rceil))
    \end{pmatrix},
\end{aligned}
\end{equation}
where the bound of each entry can be obtained from Lemma \thesection.1. Then the operator 2-norm $||\cdot||_2$  of $\Delta A_c$ is bounded by the Frobenius norm $||\Delta A_c||_F$, whose value can be calculated by summation over the norm of entries $||\Delta A_c||_F\equiv\sqrt{\sum_{j,k}|(\Delta A_c)_{jk}|^2}$. From the above discussion, we have $D^{(2)}_{\beta_j,\beta_{j'}}>\frac{l_\alpha}{2}$ and $l_\beta>\frac{3}{4}l_\alpha$. Since $R_{\lambda}(k)$ is monotonically decreasing for $k>\frac{l_\alpha}{2K_u}>\lambda L$ when $\lambda$ is bounded by Eq.~\eqref{eq:lambdaBound=app}, all the matrix entries are bounded by $2 R_\lambda\big(\big\lceil \frac{l_\alpha}{2K_u} \big\rceil\big)$. Thus, the Frobenius norm $||\Delta A_c||_F$ is at most $2N_{\mathbb{V}} R_\lambda\big(\big\lceil \frac{l_\alpha}{2K_u} \big\rceil\big)$. Thus, the lemma is proved.

\section{Emergent symmetries and robust Hilbert space fragmentation}
\label{app:SymmetryS}
In this section, we will explain robust subharmonic oscillations for $U_F^{(n)}(\lambda)$ and $U_F^{(n_1\rightarrow n_2)}(s,\lambda)$ in the picture of approximate emergent symmetries. We call $G$ an approximate symmetry of an $L$-site system $U$ if $[G,U]\sim \lambda^{O(L)}$. When small perturbation breaks original symmetries in DTC models, the system does not get thermalized. Instead, the original symmetries are perturbed into approximate emergent symmetries, with lifetime growing exponentially with the system size $L$. Thus, the ergodicity remains broken when small finite-body period-$T$ perturbation is added.

\subsection{Symmetries and DTC-charges for $n$-DTC}

As all $Z$-basis states within the same sectors have the same quasi-energies in $[U_F^{(n)}(0)]^n$, one can identify a symmetry group $\mathbb{Z}_n$ for the system, with the group generator
\begin{eqnarray}\label{eq:SInPhi}
    S&=&\sum_{\alpha}\sum_{j=1}^k e^{-i j\frac{2\pi}{k}} \ket{\phi^{(k)}_{\alpha,j}}\bra{\phi^{(k)}_{\alpha,j}}\\
    &=&\sum_{\alpha}\sum_{m=1}^k e^{i (\theta^{(k)}_{\alpha,m+1}- \theta^{(k)}_{\alpha,m})}\ket{z_{\alpha,m+1}^{(k)}}\bra{z_{\alpha,m}^{(k)}},
\end{eqnarray}
where each sector contains $k$ irreducible representations of $\mathbb{Z}_n$ and $k|n$ is the corresponding dimension of the subspace. The second equation is from the relation in Eq.~\eqref{eq:eState}
\begin{equation}
    \ket{z_{\alpha,m}^{(k)}}=\frac{1}{\sqrt{k}}\sum_{j=1}^k e^{-i (\theta^{(k)}_{\alpha,m}+mj\frac{2\pi}{k}) } \ket{\phi^{(k)}_{\alpha,j}}.
\end{equation}
One can verify that $S^n=1$, and $S$ commutes with $U_F^{(n)}(0)$. Note that one can always construct such $\mathbb{Z}_n$ symmetry satisfying the two properties for an arbitrary unitary operator in its eigenbasis, according to the Eq.~\eqref{eq:SInPhi}. The non-trivial relation of the $U_F^{(n)}$ and $S$ is 
\begin{equation}\label{eq:UfSsameinAlpha}
    \left.U_F^{(n)}(0)\right|_\alpha= \left.e^{-i \frac{E_\alpha}{k}} S\right|_\alpha,
\end{equation}
where $E_\alpha\equiv \sum_{m=1}^k E_{\rm int}(z^{(k)}_{\alpha,m})$ and $U|_\alpha$ is the operator $U$ restricted in the subspace $\alpha$. This indicates that, when acting within $\alpha$-th subspace, the Floquet operator $U_F^{(n)}$ have exactly same effects as the $\mathbb{Z}_n$ symmetry generator $S$, up to a global phase $e^{-i E_\alpha/k}$.

The fact that $U_F^{(n)}$ acts like a $\mathbb{Z}_n$ symmetry generator in subspaces draws our attention to the operator $[U_F^{(n)}]^n$, and we identify some symmetries that only appear in $[U_{F}^{(n)}(0)]^n$, but not in $U_{F}^{(n)}(0)$. As they naturally exhibit subharmonic oscillations, we call them DTC-charges. For the unperturbed case, such DTC-charges are $\sigma^z$'s and most of their linear combinations. One can easily verify that all $\sigma^z$'s commute with $[U_{F}^{(n)}(0)]^n$, but none of them commute with $U_{F}^{(n)}(0)$. In fact, $U_{F}^{(n)}(0)$ and $S$ transform one $\sigma^z$ to another according to the rule specified in Appendix.~\ref{app:n-period}.
For convenience, we relabel the spins for each $n$-unit $\sigma_{i,1}\equiv\sigma_{(i-1)n+1}$ and
\begin{equation}
   \sigma^z_{i,j} \equiv \left(U_F^{(n)}\right)^{j-1}\circ (\sigma^z_{i,1}),
\end{equation}
where $U\circ \sigma\equiv U\sigma U^\dagger$. One can verify that
\begin{equation}
    S \circ (\sigma^z_{i,j})=U_F^{(n)} \circ (\sigma^z_{i,j})=\sigma^z_{i,j+1}.
\end{equation}
As all $\sigma^z$'s commute with $[U_{F}^{(n)}(0)]^n$, the $\sigma^z$ will be transformed back to itself after $n$ periods of evolution. 

Thus, if one prepares an initial state in $Z$-basis, one can observe the $nT$-period subharmonic oscillations by measuring $\sigma^z$'s. We remark that such enormous additional symmetries in $[U_F^{(n)}]^n$ are essential, as it divides the Hilbert space into exponentially many small fragments, prevents ergodicity in the whole Hilbert space, and offers oscillating observables so that time-crystalline structures can be experimentally measured.

When the perturbation is added, the symmetry $\mathbb{Z}_n$ and DTC-charges $\sigma^z$'s are perturbed into approximate emergent symmetries. Supposing the original eigenstates of $U_F^{(n)}(0)$ are $\{\ket{\phi_{\alpha,j}}\}$ and the perturbed eigenstates of $U_F^{(n)}(\lambda)$ are $\{\ket{\psi_{\alpha,j}(\lambda)}\}$, we define a unitary operator $\mathcal{V}$ that connects the perturbed eigenstates with the original one $\mathcal{V}(\lambda)\ket{\phi_{\alpha,j}}=\ket{\psi_{\alpha,j}(\lambda)}$. The $\mathbb{Z}_n$ symmetry is then perturbed to
\begin{equation}
\begin{aligned}
    \tilde{S}(\lambda)&\equiv\mathcal{V}(\lambda)S\mathcal{V}^\dagger(\lambda)=\sum_{\alpha}\sum_{j=1}^k e^{-i j\frac{2\pi}{k}} \ket{\psi^{(k)}_{\alpha,j}(\lambda)}\bra{\psi^{(k)}_{\alpha,j}(\lambda)}\\
    & =\sum_{\alpha}\sum_{m=1}^k e^{i (\theta^{(k)}_{\alpha,m+1}- \theta^{(k)}_{\alpha,m})}\ket{\tilde{z}_{\alpha,m+1}^{(k)}(\lambda)}\bra{\tilde{z}_{\alpha,m}^{(k)}(\lambda)},
\end{aligned}
\end{equation}
where $\ket{\tilde{z}(\lambda)}\equiv\mathcal{V}(\lambda)\ket{z}$ and we have used
\begin{equation}\label{eq:tildeZfromPsi}
    \ket{\tilde{z}_{\alpha,m}^{(k)}}=\frac{1}{\sqrt{k}}\sum_{j=1}^k e^{-i (\theta^{(k)}_{\alpha,m}+mj\frac{2\pi}{k}) } \ket{\psi^{(k)}_{\alpha,j}}.
\end{equation}
One can verify that $\tilde{S}(\lambda)$ is still an exact symmetry of $U_F^{(n)}(\lambda)$ and an exact $\mathbb{Z}_n$ symmetry generator satisfying $\tilde{S}^n=1$. However, the Eq.~\eqref{eq:UfSsameinAlpha} no longer holds. Instead, it becomes an approximate relation
\begin{equation}\label{eq:UfsameinAlphaWithLambda}
    \left.U_F^{(n)}(\lambda)\right|_\alpha= \left.e^{-i \frac{\tilde{E}_\alpha}{k}} \tilde{S}\right|_\alpha + \lambda^{O(l_\alpha)},
\end{equation}
where we use the conclusion we proved in Appendix.~\ref{app:perturbGap} that the $2\pi/k$ quasi-energy separation within $\alpha$-th sector is perturbed at the order of $\lambda^{O(l_\alpha)}$. Therefore, $U_F^{(n)}(\lambda)$ still acts like a $\mathbb{Z}_n$ symmetry generator in the new basis, up to a $\lambda^{O(l_\alpha)}$ correction.

This leads to the emergent approximate DTC-charges $\tau_i^z$'s of $U_F^{(n)}(\lambda)$, which can be obtained from similar transformations on $\sigma^z$'s. We can define $\tau_i^a\equiv\mathcal{V}(\lambda)\sigma_i^a\mathcal{V}^\dagger(\lambda)$ with $a \in \{x,y,z\}$. It is easy to check that $ \ket{\tilde{z}}$'s are eigenstates of $\tau_i^z$'s, with the same eigenvalues as corresponding $\sigma^z_i$'s. The quasi-symmetries $\tau^z$'s, becomes an approximate emergent symmetry of $[U_F^n(\lambda)]^n$,
\begin{equation}
    \left[\left(U_F^{(n)}(\lambda)\right)^n,\tau^z\right]\Bigr|_\alpha\sim \lambda^{O(l_\alpha)}.
\end{equation}
The relation is valid for the majority of the Hilbert space, with an exponentially small portion of exceptions. This can also be seen by applying $[U_F^n(\lambda)]^n$ to arbitrary $\tau^z$-basis state $\ket{\tilde{z}_{\alpha,m}^{(k)}}$
\begin{multline}
    \left(U_F^{(n)}(\lambda)\right)^n\ket{\tilde{z}_{\alpha,m}^{(k)}(\lambda)}\\=
    \frac{e^{-i n\frac{\tilde{E}_\alpha}{k}}}{\sqrt{k}}\sum_{j=1}^k e^{-i (\theta^{(k)}_{\alpha,m}+mj\frac{2\pi}{k}+\lambda^{O(l_\alpha)}) }
    \cdot\ket{\psi^{(k)}_{\alpha,j}(\lambda)}
    \\
    =e^{-i n\frac{\tilde{E}_\alpha}{k}}\ket{\tilde{z}_{\alpha,m}^{(k)}(\lambda)}+\sum_{j\neq m}\lambda^{O(l_\alpha)}\ket{\tilde{z}_{\alpha,j}^{(k)}(\lambda)},
\end{multline}
where we have used the $2\pi/k$ quasi-energy separation is perturbed at $\lambda^{O(l_\alpha)}$ order. In addition, we can see from the last line of the equation that the fragmentation structure of the Hilbert space remains the same, in a sense that, states in the original $\alpha$-th sector remain in the new $\alpha$-th sector after perturbations and the transformation $\mathcal{V}$, and different sectors of the perturbed Hilbert space remain dynamically disjoint. Thus, the small deviations $\lambda^{O(l_\alpha)}$ from the exact symmetry will only contain contributions from the same sector of subspace, and will not affect other sectors. That the fragmentation structure remains the same after perturbations is not surprising, because we are using the transformation $\mathcal{V}$ that relates eigenbases of the two Hamiltonians. Here, the non-trivial property of the perturbed Hamiltonian is from Eq.~\eqref{eq:UfsameinAlphaWithLambda}, that the $U_F^{(n)}(\lambda)$ robustly acts like a $\mathbb{Z}_n$ symmetry generator in each fragments, which is essentially the consequence of robust $2\pi/n$ quasi-spectral separations within each sector. 

Since $\tau^z$'s are approximate DTC-charges of $[U_F^{(n)}(\lambda)]^n$, we have
\begin{equation}
    U_F^{(n)}(\lambda) \circ (\tau^z_{i,j})=\tau^z_{i,j+1}+\lambda^{O(L)}.
\end{equation}
In addition, we expect $\mathcal{V}(\lambda)$ is close to identity when $\lambda$ is small. We can expand $\sigma^z$'s in $\tau$-basis
\begin{equation}
    \sigma^z_{i,j}=\mathcal{V}(\lambda)^\dagger\tau^z_{i,j}\mathcal{V}(\lambda)=[1-O(\lambda)]\tau^z_{i,j}+O(\lambda).
\end{equation}
Thus, by measuring $\sigma^z$'s, we can observe that their major components $\tau^z$'s have robust sub-harmonic oscillations, up to a time scale $e^{O(L)}$ going to infinity at thermodynamic limit.
 
\subsection{Symmetries and DTC-charges for ST-DTC}
For ST-DTC with Floquet operator $U_F^{(n_1\rightarrow n_2)}(s,\lambda=0)$ (we simplify the notation to $U_F(s,\lambda)$ in this section), in the unperturbed case, we can construct the $\mathbb{Z}_{n_G}$ symmetry and find the quasi-symmetries $Q_{i,j}$ in a similar way. The $\mathbb{Z}_{n_G}$ symmetry can be constructed as follows
\begin{eqnarray}\label{eq:SInPhiST}
    S&=&\sum_{\alpha}\sum_{j=1}^k e^{-i j\frac{2\pi}{k}} \ket{\Phi^{(k)}_{\alpha,j}}\bra{\Phi^{(k)}_{\alpha,j}}\\
    &=&\sum_{\alpha}\sum_{m=1}^k e^{i (\theta^{(k)}_{\alpha,m+1}- \theta^{(k)}_{\alpha,m})}\ket{\chi_{\alpha,m+1}^{(k)}}\bra{\chi_{\alpha,m}^{(k)}}.
\end{eqnarray}
The non-trivial relation of the $U_F^{(n)}$ and $S$ is 
\begin{equation}\label{eq:UfSsameinAlphaST}
    \left.U_F(s,\lambda=0)\right|_\alpha= \left.e^{-i \frac{E_\alpha}{k}} S\right|_\alpha.
\end{equation}
One can verify that the DTC-charges of $U_F(s,0)$ (or symmetries of $[U_F(s,0)]^{n_G}$) are $Q_{i,j}$'s, with the following oscillation relation
\begin{equation}
    S \circ (Q_{i,j})=U_F(s,0) \circ (Q_{i,j})=Q_{i,j+1}.
\end{equation}
As all $Q_{i,j}$'s are summations of $\sigma^z$'s, they can be directly measured. 

When perturbation is added, $U_F(s,\lambda)$ becomes an approximate $\mathbb{Z}_{n_G}$ symmetry generator
\begin{equation}\label{eq:UfSsameinAlphaWithLambdaInapp}
    \left.U_F(s,\lambda)\right|_\alpha= \left.e^{-i \frac{\tilde{E}_\alpha}{k}} \tilde{S}\right|_\alpha + \lambda^{O(l_\alpha)},
\end{equation}
where the relation holds for most sectors, with exponentially rare exceptions.

Similarly, one can find $\tilde{Q}_{i,j}\equiv\mathcal{V}(\lambda)Q_{i,j}\mathcal{V}^\dagger(\lambda)$ are emergent approximate symmetries of $[U_F(s,\lambda)]^{n_G}$, and $U_F(s,\lambda)$ approximately acts like a $\mathbb{Z}_{n_G}$ symmetry generator on $\tilde{Q}_{i,j}$ 
\begin{equation}
    U_F(s,\lambda) \circ (\tilde{Q}_{i,j})=\tilde{Q}_{i,j+1}+\lambda^{O(L)}.
\end{equation}
The quasi-symmetries $\tilde{Q}_{i,j}$'s, becomes an approximate emergent symmetry of $[U_F(s,\lambda)]^{n_G}$,
\begin{equation}
    \left[[U_F(s,\lambda)]^{n_G},\tilde{Q}_{i,j}\right]\Bigr|_\alpha\sim \lambda^{O(l_\alpha)}.
\end{equation}
Thus, the thermalizations are approximately confined within subspaces characterized by $\tilde{Q}_{i,j}$, within a time scale $e^{O(L)}$ going to infinity in the thermodynamic limit.
In addition, we expect $\mathcal{V}(\lambda)$ is close to identity when $\lambda$ is small. We can expand $Q_{i,j}$'s in $\tau$-basis
\begin{equation}
    Q_{i,j}=\mathcal{V}(\lambda)^\dagger \tilde{Q}_{i,j}\mathcal{V}(\lambda)=[1-O(\lambda)] \tilde{Q}_{i,j}+O(\lambda),
\end{equation}
where we use the fact that $\tilde{Q}_{i,j}$'s are combinations of $\tau^z$'s. Thus, by measuring $Q_{i,j}$'s, we can observe that their major components $\tilde{Q}_{i,j}$'s have robust sub-harmonic oscillations, up to a time scale $e^{O(L)}$ going to infinity in the thermodynamic limit.

\section{Susceptibility and spontaneous discrete time-translation symmetry breaking}\label{app:STauB}
To understand the spontaneous breaking of the temporal $T$ symmetry for our $n$-DTC, we can study the susceptibility of the system~\cite{khemani2019brief} from an infinitesimal $nT$-period perturbation $\epsilon V(t)$ on the original $U(\lambda,nT)=[U_F^{(n)}(\lambda)]^n$, which can be defined as 
\begin{equation}
\begin{aligned}
\hat{V}:=& \sum_{j,k} e^{-i k \frac{2\pi}{n}}\tau^z_{j,k}, \\
V(t):=& \sum_{N} e^{i N\frac{2\pi}{n}}\hat{V} \delta(t-NT).
\end{aligned}
\end{equation}
Using $U_F^{(n)}(\lambda) \circ \hat{V}=e^{i \frac{2\pi}{n}} \hat{V}$ , one can verify that the original $U(\lambda,nT)$ is perturbed to $U_{\epsilon}(\lambda,nT)=e^{-\epsilon n\hat{V}}U(\lambda,nT)$. Due to the robust $2\pi/n$ gap, eigenstates $\ket{\psi_{\alpha,m}(\lambda)}$ of $U(\lambda,nT)$ are exactly degenerate in the same $\alpha$-th subspace in the thermodynamic limit.  When perturbation $\epsilon V(t)$ is added, the perturbed eigenstates $\ket{\psi_{\alpha,m}(\lambda,\epsilon)}$ will be close to product states $\ket{\tilde{z}_{\alpha,m}}$ (see Eq.~\eqref{eq:tildeZfromPsi}) in $\tau^z$ basis, where the original $\ket{\psi_{\alpha,m}(\lambda)}$'s (without $\epsilon V(t)$ perturbation) are superpositions of $n$ $\tau^z$-basis states. Using the expansion in Eq.~\eqref{eq:tildeZfromPsi}, the expectation value of an observable $\hat{O}(NT)$ in the new eigenstate basis $\ket{\psi_{\alpha,m}(\lambda,\epsilon)}$ can be shown to have the form
\begin{multline}
  \lim_{\epsilon \rightarrow 0} \lim_{L\rightarrow \infty} \braket{\psi_{\alpha,m}(\lambda,\epsilon)|\hat{O}(NT)|\psi_{\alpha,m}(\lambda,\epsilon)}\\
   =\sum_{j,k} \braket{\psi_{\alpha,k}(\lambda)|\hat{O}|\psi_{\alpha,j}(\lambda)} e^{-i (N+m)(j-k) \frac{2\pi}{n}},
\end{multline}
where the phase $e^{-i (N+m)(j-k) \frac{2\pi}{n}}$ comes from the robust $\frac{2\pi}{n}$ gap of the quasi-spectrum in each $\alpha$-th subspace. Thus, with infinitesimal perturbation that breaks the $T$ time-translation symmetry, the observables can exhibit subharmonic oscillations in the thermodynamic limit, which indicates a spontaneous breaking of the discrete time-translation symmetry. We remark that the above analysis also works for the ST-DTC.

\section{Kicked Ising model in our analytical framework}
\label{app:KIsing}
To show our analytical framework can be easily applied to a broad range of disordered DTC, we use the well-studied kicked Ising model~\cite{von2016absolute,else2016floquet,yao2017discrete} as an example, whose evolution is governed by a period-$T$ Hamiltonian
\begin{gather}
    H(t)= \begin{cases}
    \begin{aligned}
    \sum_i J_i \sigma^z_i\sigma^z_{i+1}+\sum_i h^z_i \sigma^z_i+\sum_i \epsilon^x_i \sigma^x_i,\\
      \text { for } 0 \leq t<t_1;
    \end{aligned}\\ 
    (1-\epsilon)\frac{\pi}{2 t_2}\sum_i \sigma^x_i ,~~~~~\text { for } t_1 \leq t<T.
    \end{cases}
\end{gather}
where all perturbations $\epsilon\text{'s}\sim O(\lambda)$ are small and $T=t_1+t_2$. In the unperturbed case, the Floquet operator $U_F$ flips all sites for a $Z$-basis state
\begin{equation}
    U_F \ket{z}=e^{-i \phi}\ket{\bar{z}},
\end{equation}
where $\ket{\bar{z}}$ is from flipping all sites of $\ket{z}$. Thus, its eigenvalues can be written in the form of Eq.~\eqref{eq:eState}, and one can obtain an exact $\pi$ gap in the quasi-spectrum. As $\ket{z}$ and $\ket{\bar{z}}$ are globally different (with distance $L$ in the Hamming distance), using our analysis similar to Appendix~\ref{app:perturbGap}, one can obtain that the $\pi$-pairing of the quasi-spectrum is perturbed at $\lambda^{O(L)}$.

To probe the subharmonic oscillation, we need DTC-charges, which are symmetries in the $(U_F)^2$ but not in $U_F$. One can verify that  all $\sigma^z$'s are DTC-charges
\begin{equation}
\begin{aligned}
    (U_F)^2\circ \sigma^z_i= \sigma^z_i,\\
    U_F\circ \sigma^z_i= - \sigma^z_i.
\end{aligned}
\end{equation}
Thus, by measuring $\sigma^z$'s, one can observe the robust subharmonic oscillation. When perturbation is added, the DTC-charges become $\tau^z_i\equiv \mathcal{V}(\lambda)^\dagger \sigma_{i}^z\mathcal{V}(\lambda)$, where $\mathcal{V}(\lambda)$ transforms unperturbed eigenstates to perturbed ones. They still have robust subharmonic oscillations
\begin{equation}
\begin{aligned}
    \left[\left[U_F(\lambda)\right]^2,\tau^z_i\right] &\sim \lambda^{O(L)},\\
    U_F(\lambda) \circ (\tau^z_{i})&=-\tau^z_{i} +\lambda^{O(L)}.
\end{aligned}
\end{equation}
These relations are also obtained in previous works~\cite{von2016absolute,surace2019floquet}. Similarly, we have
\begin{equation}
    \sigma^z_{i}=\mathcal{V}(\lambda)^\dagger\tau^z_{i}\mathcal{V}(\lambda)=[1-O(\lambda)]\tau^z_{i}+O(\lambda).
\end{equation}
Thus, by measuring $\sigma^z$'s, we can observe that their major components $\tau^z$'s have robust sub-harmonic oscillations, up to a time scale $e^{O(L)}$ going to infinity in the thermodynamic limit.

We remark that although most of the arguments in this section are merely rephrasing of the existing results, they work for many existing models, including the kicked Ising DTC~\cite{von2016absolute,else2016floquet,yao2017discrete} and clock-model DTC~\cite{surace2019floquet}, with the latter having period $n$-tupling. The only different part is that we obtain the robust $\pi$-pairing from the proof in this work and use this as the stepping stone to obtain all the other time-crystalline features, whereas in the existing works, the robust $\pi$-pairing is from corollaries of MBL. The robustness of $\frac{2\pi}{n}$ gap obtained from the new perspective enables us to go beyond the MBL region to ST-DTC.

\section{Generalized to higher spin and higher dimensions}
\label{app:higherSandD}
Our $n$-DTC and ST-DTC models can be easily generalized to higher spins. To do this, one can simply replace $\sigma^{x,y,z}$ in $H_{\rm int}$ with higher-spin operators $S^{x,y,z}$, and replace the swap gates in $H_1^{(n)}$ and $H_2^{(n)}$ with higher-spin swap gates. Such a generalization for $(n=2)$-DTC in the spin-1 case has already been discussed in~\cite{gargiulo2024swapping}. As the swap gate becomes more complicated for higher spins, higher-spin $n$-DTC models are harder to implement, but the generalization is still immediate.

Our $n$-DTC and ST-DTC models can also be generalized to higher-$\mathcal{D}$ spatial dimensions. A straightforward way to do so is replacing the couplings in $H_{\rm int}$ with $\mathcal{D}$-dimensional disordered interactions, and defining the $H_1^{(n)}$ and $H_2^{(n)}$ on a 1-D chain that traverses all $\mathcal{D}$-dimensional sites of spins. Another way to define $H_1^{(n)}$ and $H_2^{(n)}$ is by replacing them with a $2\mathcal{D}$-layer permutation, with each 2-layer being the same permutation as $H_1^{(n)}$ and $H_2^{(n)}$ along one spatial dimension. In this case, an initial spin at $(1,1,..)$ will be permuted to $(3,3,...)$ and get back to $(1,1,..)$ after $n$ periods, with each permutation unit containing $n\times n\times ...$ sites.

\end{document}